\documentclass[epj, nopacs]{svjour}
\usepackage{epsfig}
\usepackage{epsf}
\usepackage{cite}

\newcommand{\bda}{\begin{\displaymath}\begin{array}{rl}}
\newcommand{\eda}{\end{array}\end{displaymath}}
\newcommand{\be}{\begin{equation}}
\newcommand{\ee}{\end{equation}}
\newcommand{\bdm}{\begin{displaymath}}
\newcommand{\edm}{\end{displaymath}}
\newcommand{\bea}{\begin{eqnarray}}
\newcommand{\eea}{\end{eqnarray}}
\newcommand{\no}{\nonumber \\}
\newcommand{\fs}{\,.}
\newcommand{\co}{\,,}
\newcommand{\al}{&}

\newcommand{\ind}{\scriptscriptstyle}
\newcommand{\indR}{\mbox{\scriptsize R}}

\newcommand{\indPW}{\mbox{\scriptsize PW}}

\newcommand{\lbar}{\,\overline{\rule[0.65em]{0.4em}{0em}}\hspace{-0.45em}\ell}

\newcommand{\betabar}{\,\overline{\rule[0.7em]{0.4em}{0em}}\hspace{-0.6em}\beta}
\newcommand{\Kbar}{\hspace{0.03cm}\overline{\rule[0.7em]{0.6em}{0em}}\hspace{-0.8em}K}
\newcommand{\Tbar}{\hspace{0.03cm}\overline{\rule[0.7em]{0.65em}{0em}}\hspace{-0.75em}T}
\newcommand{\OPWbar}{\hspace{0.03cm}\overline{\rule[0.7em]{0.65em}{0em}}\hspace{-0.7em}\mbox{O}_{\mbox{\scriptsize PW}}}

\newcommand{\gsim}{\,\mbox{\raisebox{0.3mm}{$>$}}\hspace{-2.8mm}\mbox{\raisebox{-1.4mm}{$\sim$}}\!}

\newcommand{\NN}{{\ind N\!N}}
\begin{document}

\title {\boldmath Regge analysis of the $\pi\pi$ scattering amplitude}

\author{I. Caprini\inst{1}, G. Colangelo\inst{2} and H. Leutwyler\inst{2}}
\institute{Horia Hulubei National Institute for Physics and Nuclear Engineering,
P.O.B. MG-6, 077125 Magurele, Romania \and
Albert Einstein Center for Fundamental Physics, Institute for Theoretical Physics, University of Bern,\\
Sidlerstrasse 5, CH-3012 Bern, Switzerland}

\date{December 21, 2011}

\abstract{The theoretical predictions for the subtraction constants lead to
  a very accurate dispersive representation of the $\pi\pi$ scattering
  amplitude below $0.8~\mbox{GeV}$. The extension of this representation up
  to the maximum energy of validity of the Roy equations
  ($1.15~\mbox{GeV}$) requires a more precise input at high energies.  In
  this paper we determine the trajectories and residues of the leading
  Regge contributions to the $\pi\pi$ amplitude (Pomeron, $f$ and $\rho$),
  using factorization, phenomenological parametrizations of the $\pi N$ and
  $NN$ total cross sections at high energy and a set of sum rules which
  connect the high and low energy properties of $\pi\pi$ scattering. We
  find that nonleading Regge terms are necessary in order to achieve a
  smooth transition from the partial waves to the Regge representation at
  or below 2 GeV. We obtain thus a Regge representation consistent both
  with the experimental information at high energies and the Roy equations
  for the partial waves with $\ell\leq 4$. The uncertainties in our result
  for the Regge parameters are sizable but in the solutions of the Roy
  equations, these only manifest themselves above $K\Kbar$ threshold.}
\authorrunning{I.~Caprini, G.~Colangelo and H.~Leutwyler}
\titlerunning{Regge analysis of $\pi\pi$ scattering}
\maketitle

\section{Motivation}
Low energy pion physics has become a precision laboratory: the chiral
symmetry properties of the Standard Model can now be compared with low
energy precision experiments, using Chiral Perturbation Theory ($\chi$PT),
dispersion theory and numerical simulations of QCD on a lattice. In
principle, as illustrated by the prediction for the magnetic moment of the
muon, physics beyond the Standard Model can show up at low energies,
provided the quantity of interest can not only be measured accurately, but
can also be calculated to sufficient precision. In this context, the
interaction among the pions and in particular, the elastic $\pi\pi$
scattering amplitude play a crucial role.

The Roy equations \cite{Roy} provide a suitable framework for the low
energy analysis of the $\pi\pi$ scattering amplitude, as they fully
incorporate the basic properties that follow from analyticity, unitarity
and crossing symmetry. They express the real parts of the partial waves as
integrals over the imaginary parts that extend over all energies. The high
energy behaviour of the imaginary parts thus enters the analysis, even if
the Roy equations are evaluated only at low energy. More specifically, the
contributions from the high energy region are contained in the so-called
{\it driving terms}. Near the $\pi\pi$ threshold, these terms are very
small. Accordingly, the results for the threshold parameters are not
sensitive to the high energy contributions. The driving terms, however,
grow with the energy. In \cite{ACGL,Descotes}, where the threshold
parameters and the coupling constants of the effective chiral
SU(2)$\times$SU(2) Lagrangian relevant for $\pi\pi$ scattering are
determined to high accuracy, the Roy equations are solved only below 0.8
GeV. On general grounds \cite{Roy} these equations are valid in the
region\footnote{The maximum energy can be increased by using, instead of
  fixed-$t$ dispersion relations as in the standard approach, more general
  dispersion relations along curves in the Mandelstam plane
  \cite{Mahoux,RoyWanders}.} $s\leq 68\,M_\pi^2$, i.e.~for $\sqrt{s}\leq
1.15\,\mbox{GeV}$, but in order to use them we need an accurate evaluation
of the driving terms. This calls for a better understanding of the
imaginary parts at high energies and motivates the present work.

For $\pi\pi$ scattering, several sets of data at high energies are
available from indirect experiments
\cite{Biswas,Hyams,Robertson,Cohen,Losty,Hanlon,Abramowicz}. These 
were analyzed in the late seventies, both in the frame of the
Lovelace-Shapiro-Veneziano (LSV) model \cite{LSV} and by means of various
dispersive sum rules
\cite{Olsson,Wanders:1969,Tryon,Basdevant-Schomblond,Pennington-Protopopescu,Basdevant:1973ru,Pennington-Annals,Froggatt:1977hu}.
While the 
determination of the structure in the resonance region gradually made
progress
\cite{Ochs:1991rw,Bugg,Kaminski,E852,Achasov:2003xn,WZLB,Zhou:2004ms,Bugg2007,Klempt:2007cp},
the Regge analysis of the $\pi\pi$ scattering amplitude was for a long time
practically abandoned.

The Regge parametrization used in \cite{ACGL,CGL,Descotes} is based on
Pennington's analysis \cite{Pennington-Annals}. This input was subject to a
criticism in \cite{PY2003}, where an alternative Regge parametrization was
proposed.  As shown in \cite{CCGL}, the driving terms and the $S$ and
$P$-phase shifts below 0.8 GeV are practically insensitive to the
difference between the Regge parametrization used in \cite{ACGL} and that
proposed in \cite{PY2003}. However, for extending the Roy analysis up to
1.15 GeV, an update of the Regge parametrization used as input is
necessary. Irrespective of the very modest impact it has on the region
below 0.8 GeV, credit is due to Pel\'aez and Yndur\'ain \cite{PY2003} for
pointing out the need of improving the Regge representation used in
\cite{ACGL}.  In the meantime, these authors and their collaborators
gradually improved their own Regge parametrization
\cite{PY2004-Regge,KPY2006,KPY2008,GarciaMartin:2011cn}. In the following we will
compare our results with the most recent version of this series,
\cite{GarciaMartin:2011cn}, more precisely with the variant referred to as
``CFD'', which appears to represent the net result of their work.

After a brief introduction to the Regge representation in the next section,
we review the phenomenology of the leading Regge trajectories and of the
total $NN$ and $\pi N$ cross sections and discuss the consequences of
factorization for the total $\pi\pi$ cross sections.  The Regge
representation used in the present work is specified in section
\ref{sec:regge pipi}. Section \ref{sec:partial waves} contains a short
description of the partial wave representation used in the present
analysis, and section \ref{sec:total cross sections} presents the resulting
picture for the total cross sections. In section \ref{sec:transition}, we
introduce the quantitative formulation of duality used in the present work
and section \ref{sec:sum rules} discusses the sum rules we are relying on
to calculate the $t$-dependence of the Regge residues.  Details concerning
the way in which our analysis is carried out are explained in section
\ref{sec:analysis}. Our results for the Regge residues at $t=0$ are
presented in section \ref{sec:sigmatot}, while those concerning their
$t$-dependence are discussed in section \ref{sec:nonzerot}.  A few
consistency checks are described in section \ref{sec:consistency}. Finally,
in section \ref{sec:summary}, we present a summary and draw some
conclusions.
 
The paper has four appendices: in appendix \ref{app:review} we give a short
review of the Regge parametrizations of the $\pi\pi$ amplitude proposed in
the literature. In appendix \ref{app:Roy} we specify the input used when
solving the Roy equations for the partial waves (a detailed presentation
will be given in a forthcoming paper \cite{paper-on-PW}). In appendix
\ref{app:SR} we provide the explicit expressions of the sum rules used in
this analysis and in appendix \ref{sec:details} we specify some details of
the minimization procedure used to impose our constraints.  Preliminary
results of this work were reported in
\cite{CCL,LeutwylerChDy,CapriniChDy,Colangelo-KAON-08,Colangelo-PANIC-08}.

\section{Regge poles}\label{sec:regge}
The high energy behaviour of soft hadronic and photon-induced reactions can
be described in terms of Regge poles and cuts in the angular-momentum
plane.  A Regge pole actually corresponds to the exchange of an infinite
series of particles or resonances with quantum numbers such that they can
be produced in the $t$-channel. For a detailed account of this framework
and its application to meson-nucleon and nucleon-nucleon scattering we
refer to the older literature
\cite{Frautschi,Serber,Barger,Rarita,Worden,Reggeon-calculus,ChRi,Kane-Seidl,Irving-Worden,Collins,Donnachie-Landshoff-1984}.
More recent studies 
\cite{Donnachie-Landshoff-1992,QCD-Pomeron,Donnachie-Landshoff-1998,PDG-2000,Desgrolard,Nachtmann,DDLN-book,Cudell2002,PDG-Regge,PDG-2011,Szczurek,Huang,BoSo,Godizov:2007dy,Godizov:2008zz,Sobol:2010mu,Cudell2006,Oko,Igi,Halzen,Sibirtsev}
consider the description 
of the Pomeron in QCD 
\cite{QCD-Pomeron} and applications of Regge theory to diffractive
phenomena and deep inelastic $e p$ scattering at low $x$ (see the excellent
review \cite{DDLN-book}).  For soft hadronic processes, global fits of the
$\pi N$, $KN$ and $NN$ high energy data sets based on the Regge
representation were performed in a systematic way \cite{Cudell2002}, as
quoted in the Review of Particle Physics \cite{PDG-Regge,PDG-2011}.  More
sophisticated representations are obtained by including Regge cuts to
describe absorption \cite{Szczurek,Huang}, or by the eikonalization of
Regge exchanges in the impact-parameter representations
\cite{BoSo,Godizov:2007dy,Godizov:2008zz,Sobol:2010mu}. These models extend
the validity of the framework to larger momentum transfers and to higher
energies, respectively.

The contribution of a Regge pole to a scattering amplitude at large c.m
energy squared $s$ and small momentum transfer $t\le 0$ has the form 
\be
\label{Regge} T(s,t)=- \beta(t)\,\frac{
  \exp[-i\pi\alpha(t)]+\tau}{\sin[\pi\alpha(t)]} \hspace{-0.1cm}
\left(\frac{s}{s_1}\right)^{\hspace{-0.1cm}\alpha(t)}\hspace{-0.1cm} \,,
\ee 
where $\alpha(t)$ and $\beta(t)$ denote the trajectory and the residue,
respectively, and $\tau$ is the signature, taking the value $1$ ($-1$) for
$C$-even (odd) trajectories. The scale factor $s_1$ is usually set equal to
$1\, \mbox{GeV}^2$ and we will adopt this convention. Like the scattering
amplitude $T(s,t)$ itself, the residue $\beta(t)$ of a Regge pole is a
dimensionless quantity.

The trajectories $\alpha(t)$ and the residues $\beta(t)$ are analytic
functions of $t$, with branch points at the unitarity thresholds in the
$t$-channel.  The trajectories are universal, {\it i.e.} a definite Regge
pole in the angular momentum plane is located at the same place
$J=\alpha(t)$ in all processes where it contributes. Information on the
trajectories is inferred from the plot of the spin of known resonances
versus the square of their mass (Chew-Frautschi diagrams).  Throughout this
paper, we assume that in the $t$ region of interest the trajectories are
approximately linear\footnote{Deviations from linearity were also
  investigated, see for instance \cite{Desgrolard}.}  \be\label{linear}
\alpha(t)= \alpha(0)+\alpha'(0)\,t\fs\ee

The residues $\beta(t)$ are much less known.  The zeros of the denominator
in (\ref{Regge}) give rise to poles of the amplitude -- unless the
numerator or the residue $\beta(t)$ vanish there.  For positive $t$, the
poles correspond to the physical resonances situated on the Chew-Frautschi
plots.  Poles in the spacelike region are unphysical, hence the residue
$\beta(t)$ must vanish if $\alpha(t)+\frac{1}{2}(1+\tau)$ is an odd integer
and $t<0$. In the literature, this condition is sometimes implemented with
factors depending explicitly on $\alpha(t)$ (see appendix
\ref{app:review}).

A remarkable property of the residues, which follows from unitarity, is
their factorization as a product of two vertices. For the exchange of a
Reggeon $R$ in the process $a+b\to c+d$, factorization reads
\be\label{factorization} \beta_R(t)= \gamma_{Rac}(t) \gamma_{Rbd}(t).\ee
For particles with spin, this relation holds separately for amplitudes of
definite helicity.
 
In the representation quoted by the Particle Data Group \cite{PDG-Regge},
the dominant contribution, the Pomeron, is described as a triple pole in
the angular momentum plane, which leads to an increase with energy as the
square of a logarithm. Also, at nonleading order, daughter poles as well as
Regge cuts arising from the simultaneous exchange of two or more poles
contribute \cite{Reggeon-calculus,Collins}. In fact, in exotic channels,
where resonances do not occur, the dominating contributions stem from such
cuts.
 
The position $\alpha_{cut}(t)$ of the first branch-point arising from the
exchange of two Regge poles is determined by the corresponding
trajectories. For the exchange of two identical poles, characterized by the
linear trajectory $\alpha(0)+\alpha'(0)\,t$, the expression reads
\cite{Reggeon-calculus,Collins} \be\label{alphac} \alpha_{cut}(t)=
2\alpha(0)-1 + \mbox{$\frac{1}{2}$}\alpha'(0)\, t.\ee At large $s$, a Regge
cut contributes as $ s^{\alpha_{cut}(t)}$, with additional logarithmic
factors like $\ln^\gamma (s/s_c)$, where the scale $s_c$ and the exponent
$\gamma$ of the correction are in general unknown \cite{DDLN-book}.
Factorization does not hold for cuts and generally fails for absorption
corrections \cite{Irving-Worden}, especially at large negative $t$.

\section{Phenomenology of leading trajectories and total cross sections}\label{sec:leading}

\subsection{\boldmath $\pi N$ and $NN$ total cross sections}\label{sec:piN}
During the last decade, a comprehensive compilation of the hadronic total
cross sections was developed by the Compete Collaboration
\cite{Cudell2002}. The Regge fit of the data above 5 GeV, quoted in PDG
2011 \cite{PDG-Regge}, is based on the expressions: \bea\label{sigmatot}
\sigma_{ab}(s) \al=\al B\,{\ln}^2(s/s_0) + Z_{ab}+ \\
\al\al Y_{1\,ab}\, (s_1/s)^{\eta_1}-Y_{2\,ab}\,(s_1/s)^{\eta_2}\co\no
\sigma_{\bar{a}b}(s) \al=\al B\,{\ln}^2(s/s_0) + Z_{ab}+ \no \al\al
Y_{1\,ab}\, (s_1/s)^{\eta_1}+Y_{2\,ab}\,(s_1/s)^{\eta_2}\,,\nonumber \eea
where the target particle $b$ is either a proton ($p$) or a neutron ($n$),
and the projectiles $a$ ($\bar a$) are $p$ or $\pi^+\!,$ or their
antiparticles.

The first two terms on the right hand side of (\ref{sigmatot}) represent
the contribution of the Pomeron. The coefficient $B$ and the scale $s_0$ of
the logarithmic term are assumed to be universal:
\be\label{Bs0}B=0.308\pm0.010\,\mbox{mb}\co\hspace{1em} \sqrt{s_0}=5.38\pm
0.50\,\mbox{GeV}\,. \ee The universality hypothesis was tested
independently in \cite{Igi,Halzen}.

The coefficients accounting for the energy independent part of the Pomeron
contribution to the cross section are \be\label{ZpiN} Z_{\ind NN}=35.63\pm
0.25\,\mbox{mb}, \quad Z_{\pi^+ p}=20.86\pm 0.03\,\mbox{mb}, \ee where the
first value is the mean of $Z_{pp}$ and $Z_{pn}$ (which agree within
errors, indicating that the isospin breaking effects are very small). The
numbers (\ref{Bs0})--(\ref{ZpiN}) show that -- unless the energy is taken
so high that $\ln(s/s_0)$ becomes large compared to 1 -- the Pomeron term
is approximately independent of the energy.

The remaining terms in (\ref{sigmatot}) account for the leading Regge
poles, which for $NN$ scattering are known to be $f$, $a_2$, $\rho$ and
$\omega$ \cite{Collins}. The connection with (\ref{Regge}) is provided by 
the optical theorem, which relates the total cross sections to the imaginary 
parts of the corresponding  forward amplitudes. At high energies, it yields 
\be \label{sigmapole}
\sigma_{\rm pole}(s) \approx \frac{\beta(0)}{s_1}
\left(\frac{s}{s_1}\right)^{\alpha(0)-1}\,.  \ee For $NN$ scattering, the
term proportional to $Y_1$ in (\ref{sigmatot}) collects the contributions
of the dominant $C$-even trajectories, $f$ and $a_2$, with the intercepts
taken to be the same, $\alpha_f(0)=\alpha_{a_2}(0)=1-\eta_1$, the
contributions from lower trajectories being neglected. Likewise, the
leading $C$-odd trajectories, $\rho$ and $\omega$, are represented by the
term proportional to $Y_2$, with
$\alpha_\rho(0)=\alpha_\omega(0)=1-\eta_2$. For completeness we quote the
values from \cite{PDG-Regge}: \bea\label{Y12}
Y_{1\,pp}&=&42.53 \pm 0.23\,\mbox{mb},\quad  Y_{2\,pp}=33.34\pm
0.33\,\mbox{mb}\rule{0.5cm}{0cm}\\ 
~Y_{1\,pn}&=&40.15 \pm 1.59\,\mbox{mb},\quad Y_{2\,pn}=30.00\pm
0.96\,\mbox{mb}.\nonumber \eea

Using (\ref{sigmatot}) for the collisions of protons or antiprotons on
protons and neutrons, we can separate the contributions from the individual
trajectories, $f$, $a_2$, $\rho$ and $\omega$:
$Y_{1\,pp}=Y_{f\NN}+Y_{a_2\NN}$ and $Y_{2\,pp}=Y_{\rho\NN}+Y_{\omega\NN}$.
If the proton target is replaced by a neutron, the isospin odd
contributions from $\rho$ and $a_2$ change sign. Hence, the combinations
relevant for the scattering on neutrons are given by $Y_{1\,pn}=Y_{f\NN}-
Y_{a_2\NN}$ and $Y_{2\,pn}=-Y_{\rho \NN}+Y_{\omega\NN}$.

At high energies, the cross sections barely show any difference between
proton and neutron targets, indicating that $Y_{a_2\NN}$ and $Y_{\rho\NN}$
are small. In early works \cite{Barger}, the $\rho$ and $a_2$ contributions
to $pp$ and $\bar{p}p$ scattering were found to be negligible and in some
analyses they were simply dropped \cite{Rarita}. The recent Regge analysis
of nucleon-nucleon scattering above 3 GeV reported in \cite{Sibirtsev}
confirms that $f$ and $\omega$ dominate over $a_2$ and $\rho$,
respectively. Indeed, the values obtained from (\ref{Y12}),
\bea\label{YNN}  Y_{f\NN} &=& 41.3\pm 0.8\,\mbox{mb}\co\;  Y_{a_2 \NN}=1.2\pm0.8\,\mbox{mb}\co\\
Y_{\rho\NN}&=& 1.7\pm 0.5\,\mbox{mb}\co\hspace{0.39cm} Y_{\omega
  \NN}=31.7\pm0.5\,\mbox{mb}\,\nonumber\eea show that the couplings of
$a_2$ and $\rho$ to the nucleons are too small to clearly stick out.

If isospin breaking effects are neglected, only the $f$ and $\rho$
trajectories are present in $\pi N $ scattering, in accord with the fact
that two pions in a configuration with even (odd) isospin must carry even
(odd) angular momentum, so cannot couple to $a_2$ or $\omega$.  The
parameters $Y$ given in \cite{PDG-Regge} for this process are:
\be\label{YpiN} Y_{f\pi \ind N} =19.24 \pm 0.18\, \mbox{mb}\co\; \;
Y_{\rho\hspace{0.2mm}\pi \ind N} = 6.03\pm 0.09\,\mbox{mb}.  \ee

\subsection{Leading trajectories}\label{sec:trajectories}
The intercepts of the leading trajectories can be read off from the powers
of $s$ occurring in the representation (\ref{sigmatot}) for the total cross
sections. As mentioned above, this representation describes the Pomeron as
a triple pole with  
\be\label{alpha0P}\alpha_P(0)=1\fs
\ee
Using the values  $\eta_1=0.458\pm 0.017$ and  $\eta_2=0.545\pm 0.007$ from
\cite{PDG-Regge} and enlarging the errors to take into account other
results in the literature (see appendix \ref{app:review}) we take:  
\be \label{alpha0frho} 
\alpha_f(0)=0.54\pm 0.05\co\hspace{1em}
\alpha_\rho(0)=0.45\pm 0.02\fs
\ee 

As stated above, we assume that in the $t$ region of interest, the
trajectories are adequately approximated by the linear formula
(\ref{linear}). For the slope of the Pomeron trajectory, we use the value 
\be
\label{alpha1P} \alpha'_P(0)= 0.25 \pm 0.05\, \mbox{GeV}^{-2}\co
\ee 
which is consistent with the information on the shape of diffraction peak
at high energies (see appendix \ref{app:review}). 

For $\alpha_f(t)$ and $\alpha_\rho(t)$, we fix the slopes with the
requirement that the trajectories pass through the masses squared:
$\alpha_f(M_{f_2}^2)=2$ and $\alpha_\rho(M_\rho^2)=1$, respectively.  Using
the experimental values for the masses and adding errors to cover other
results quoted in appendix \ref{app:review}, we take
\bea\label{alpha1frho} \alpha'_f(0)\al=\al 0.90\pm 0.05\, \mbox{GeV}^{-2}\,,\\
\alpha'_\rho(0)\al=\al 0.91\pm 0.02\, \mbox{GeV}^{-2}\,.\nonumber \eea
\subsection{Factorization}\label{sec:factorization}
We rely on the values derived in section \ref{sec:piN} in order to extract the
Regge parameters of the $\pi\pi$ amplitude through factorization, using 
equations (\ref{factorization}) and (\ref{sigmapole}). As
remarked in section \ref{sec:regge}, if the particles carry spin,
factorization applies to amplitudes of definite helicity. In the forward
direction, only the non-flip coupling survives, and factorization holds for
each Regge pole contributing to the $\pi\pi$, $\pi N$ and $NN$ cross
sections \cite{Collins}.

In the case of $\pi\pi$ scattering the representation (\ref{sigmatot}) reads
\bea\label{sigmatot pipi} 
\sigma_{\pi^\pm\pi^+}(s)&=& 
B\,{\ln}^2(s/s_0) +Z_{\pi\pi} +\\  &&Y_{f\,\pi\pi}\,
(s_1/s)^{\eta_1} \mp Y_{\rho\hspace{0.2mm}\pi\pi}\,(s_1/s)^{\eta_2}\fs
\nonumber
\eea
The expression involves the leading Regge terms  $P$, $f$ and $\rho$. 
For the Pomeron, factorization gives the universality of the parameters $B$
and  $s_0$ and 
\be\label{Zpipi} 
Z_{\pi\pi}=  \frac{Z_{\pi \ind N}^2}{Z{_{\NN}}}
= 12.2\pm0.1\,\mbox{mb}\fs\ee 
This is to be compared with the number obtained from $Z_{\pi \ind N}$ with the quark
counting rule: $Z_{\pi \pi}\approx \frac{2}{3}\,Z_{\pi\ind N}\approx 14\,\mbox{mb}$. 
For the $Y$ parameters, the values in (\ref{YNN}) and (\ref{YpiN}) give 
\bea\label{Ypipi} 
Y_{f\hspace{0.2mm}\pi\pi}&=& \frac{Y_{f\pi N}^2}{Y_{f \NN}} = 8.95\pm
0.24\,\mbox{mb}\co\\
Y_{\rho\hspace{0.2mm}\pi\pi}&=&\frac{Y_{\rho\hspace{0.2mm}\pi\ind
    N}^2}{Y_{\rho\ind NN}}=  21.8\pm 9.0\,\mbox{mb}\nonumber\fs
\eea  
The phenomenological information about the coefficients $Z$ and $Y$ will be
converted into standard Regge terminology in section \ref{sec:regge pipi}.
\subsection{\boldmath Data on $\pi\pi$  cross sections}\label{sec:comparison} 
We complete this phenomenological discussion with a brief review of the
experimental results for the $\pi\pi$ total cross sections obtained
indirectly, from reactions like $\pi^- p\rightarrow\pi^+\pi^- n$, $\pi^-
p\rightarrow\pi^-\pi^- \Delta^{++}$ \cite{Biswas,Robertson,Cohen} and
$\pi^\pm p\to \Delta^{++}X$, $\pi^\pm n\to pX$ \cite{Hanlon,Abramowicz}.  

\begin{figure}[tbh]
\begin{center} 
\includegraphics[width=8.7cm]{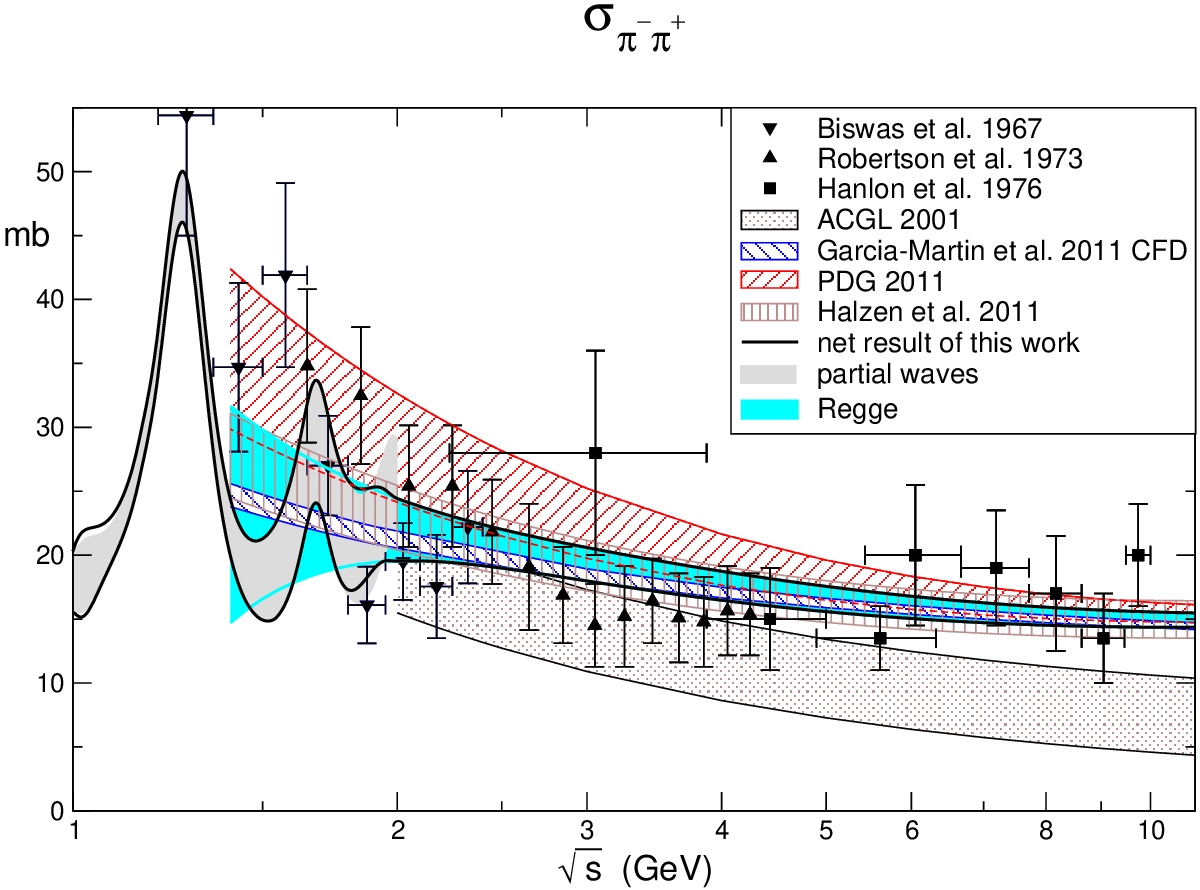} 
\end{center} 
\caption{\label{fig:pimpip} Total $\pi^-\pi^+$ cross section} 
\end{figure}

Fig.\,\ref{fig:pimpip} shows that in the $\pi^-\pi^+$ channel, the three
available data sets \cite{Biswas,Robertson,Hanlon} are more or less
consistent, not only with one another, but also with the cross section that
follows by applying factorization to the Regge representation in PDG 2011.
The figure also indicates the cross section corresponding to some of the
Regge representations available in the literature, as well as the outcome
of our own analysis. These entries will be discussed in section
\ref{sec:discussion sigmatot}.   

In the channel $\pi^-\pi^0$, we are aware of only one set of data
\cite{Biswas}. As these concern the region below 2 GeV, they do not contain
information about the behaviour at high energies. 
\begin{figure}[tbh]
\begin{center}
\includegraphics[width=8.7cm]{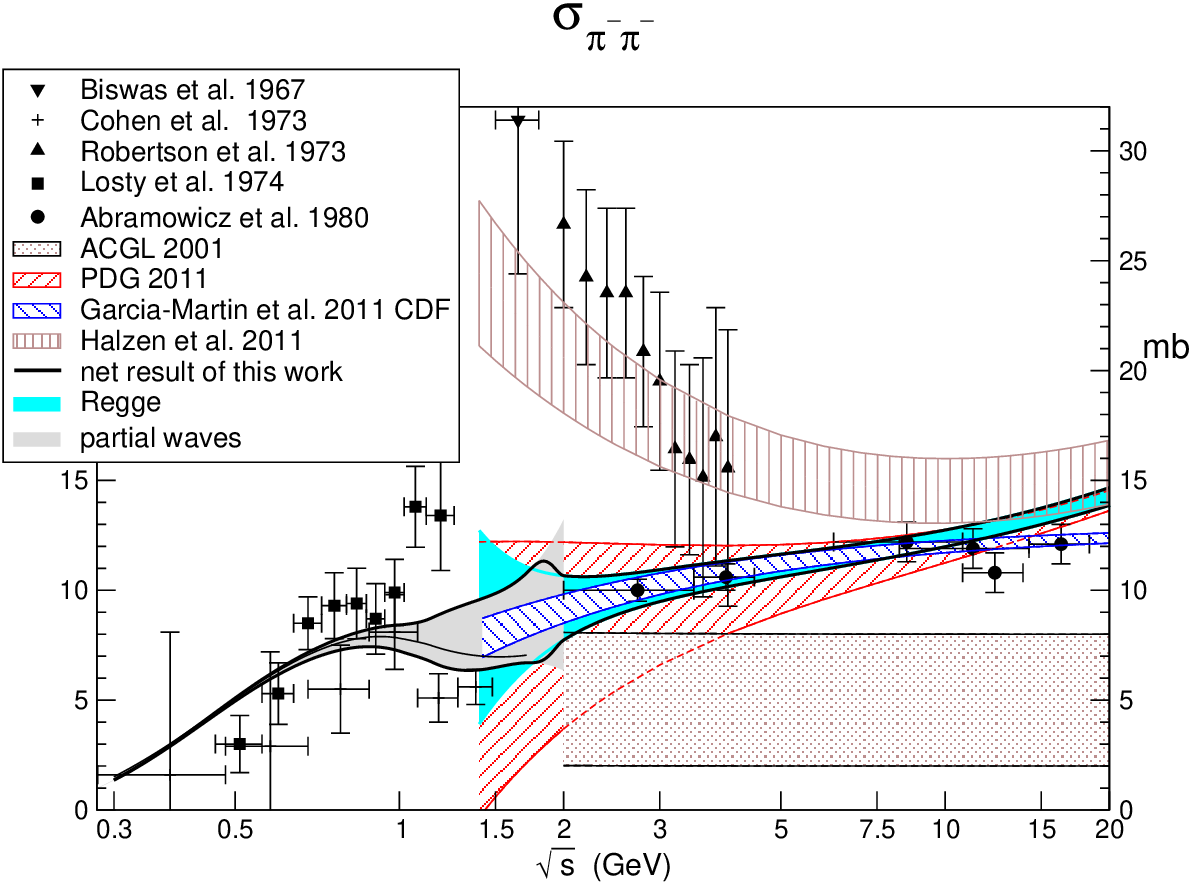}
\end{center}
\caption{\label{fig:pimpim} Total $\pi^-\pi^-$ cross section}
\end{figure}  

The $\pi^-\pi^-$ channel does not contain resonances. At low energies, the
cross section is small and notoriously difficult to measure. Various data
are shown in Fig.\,\ref{fig:pimpim}. Those of Cohen et al.~\cite{Cohen}
concern the elastic rather than the total cross section. At low energies,
however, the inelasticity in this channel is very small, so that the
elastic cross section barely differs from the total one. In particular, the
main source of inelasticity in $\pi^-\pi^+$ collisions -- transitions into
$K\Kbar$ states -- is forbidden here. This implies that the data of Biswas
et al.~\cite{Biswas} and Losty et al.~\cite{Losty} are in conflict with
those of Cohen et al.~\cite{Cohen}. Concerning the behaviour at higher
energies, there is a clash between the data of Robertson et
al.~\cite{Robertson} and Abra\-mo\-wicz et al.~\cite{Abramowicz}. Note that
the energy range of this plot extends to 20 GeV, far beyond the region
which matters in our context.

In the Regge parametrization of the $\pi^-\pi^-$ cross section, the $\rho$
contributes with a sign opposite to Pomeron and $f$. Since factorization
determines the latter rather sharply, the uncertainty of the result for
$\sigma_{\pi^-\pi^-}$ is dominated by the one in the residue of the Regge
pole associated with the $\rho$. The red shaded area indicates the result
obtained from $\pi N$ and $NN$ scattering with factorization, using
equations (\ref{Zpipi}) and (\ref{Ypipi}). The uncertainty in this result
is large, because (a) the $\rho$ contribution to $NN$ scattering is small
and (b) the extraction of the $\rho$ residue requires data on $p\,n$
scattering, which are less precise. The other entries shown in the figure
will be discussed in section \ref{sec:discussion sigmatot}.

\section{\boldmath Regge representation of the $\pi\pi$ amplitude}\label{sec:regge pipi}
We now specify the parametrization of the $\pi\pi$ scattering amplitude we
are going to use at high energies. The dominant Regge poles with $C=1$ are
the Pomeron and the $f$, while the channel with $C=-1$ is dominated by the
$\rho$-pole. What counts is the isospin in the $t$-channel. We use the
notation\footnote{In \cite{PY2003,GarciaMartin:2011cn} the
  scattering amplitude is instead denoted by $F(s,t)$. The normalization
  differs from ours by a numerical factor: $T(s,t)=4\pi^2F(s,t)$.}  of
reference \cite{ACGL}, where the component of the scattering amplitude with
$t$-channel isospin $ I_t$ is denoted by $T^{(I_t)}(s,t)$. In this
notation, the optical theorem takes the form \be\label{optical
  theorem}\sigma^{(I_t)}(s) = \frac{\mbox{Im}\,
  T^{(I_t)}(s,0)}{\sqrt{s\,(s-4M_\pi^2)}}\fs\ee The Pomeron and the $f$
contribute to $T^{(0)}$ and $\rho$ contributes to $T^{(1)}$, while the
exotic amplitude $T^{(2)}$ only receives subdominant contributions from the
exchange of two or more Regge poles. In the dispersive framework we are
using, the real parts are represented as dispersion integrals over the
imaginary parts. As we are evaluating these integrals only at low energies,
it suffices to specify the Regge representation of the imaginary parts --
for the analysis of the Roy equations and sum rules discussed in the
present paper, the high energy behaviour of the real parts is not relevant.

The quantities $\sigma_{\pi^\pm\pi^+}$, $\sigma_{\pi^\pm\pi^0}(s)$ and
$\sigma_{\pi^0\pi^0}(s)$ represent linear combinations of the total cross
sections of definite isospin in the $t$ channel: \bea\label{isospin
  decomposition} \sigma_{\pi^\pm \pi^+}(s)\al=\al \mbox{$\frac{1}{3}$}\,
\sigma^{(0)}(s)\mp \mbox{$\frac{1}{2}$}\,\sigma^{(1)}(s)+ \mbox{$\frac{1}{6}$}\,\sigma^{(2)}(s)\co\\
\sigma_{\pi^\pm\pi^0}(s)\al=\al \mbox{$\frac{1}{3}$}\,\sigma^{(0)}(s)-
\mbox{$\frac{1}{3}$}\,\sigma^{(2)}(s)\co\no \sigma_{\pi^0\pi^0}(s)\al=\al
\mbox{$\frac{1}{3}$}\,\sigma^{(0)}(s)+\mbox{$\frac{2}{3}$}\,
\sigma^{(2)}(s)\fs\nonumber \eea

\subsection{\boldmath  Parametrization of ${\rm  Im}\,T^{(0)}(s,t)$}\label{sec:ImT0t}
Our explicit representation for the imaginary part of the component with
$I_t=0$ reads: \bea\label{Regge
  ImT0t}&&\hspace{-0.5cm}\mbox{Im}\,T^{(0)}(s,t)_{\indR}=\beta_P(t)
\,\left(\frac{s}{s_1}\right)^{\alpha_P(t)}\!\times\\
&&\hspace{0.2cm}\left\{1+\bar{B}\,\ln^2\left(\frac{s}{s_0}\right)+
  p_1\,\frac{s_1}{s}\right\}
+\beta_f(t)\,\left(\frac{s}{s_1}\right)^{\alpha_f(t)}\hspace{-0.2cm},\nonumber\eea
It contains the trajectories of Pomeron and $f$,
\bea\label{alphaPf} \alpha_P(t)\al=\al \alpha_P(0)+\alpha_P'(0)\,t\co\\
\alpha_f(t)\al=\al\alpha_f(0)+\alpha_f'(0)\,t \fs\nonumber\eea The range
used for the intercepts and slopes is specified in equations
(\ref{alpha0P})--(\ref{alpha1frho}).

We first comment on the coefficients relevant at $t=0$, which govern the
total cross section $\sigma^{(0)}(s)$.  Using the optical theorem
(\ref{optical theorem}) on the right hand side of the first equation in
(\ref{isospin decomposition}) and comparing the result with (\ref{sigmatot
  pipi}), we obtain: \bea \label{betaPf PDG Regge} &&\beta_P(0)= 3\,s_1
Z_{\pi\pi}(\hbar c)^{-2}
=  94\pm 1\,,\\
&&\bar{B}= B/Z_{\pi\pi}=0.025 \pm 0.001, \no &&\beta_f(0)= 3\,s_1
Y_{f\hspace{0.2mm}\pi\pi}(\hbar c)^{-2} =69\pm 2\,,\nonumber \eea where we
have explicitly indicated the conversion factor $(\hbar c)^2
=0.389\,\mbox{mb}\,\mbox{GeV}^2$ needed to arrive at dimensionless
residues.  As discussed in appendix \ref{app:review}, the values of
$\beta_P(0)$ and $\beta_f(0)$ proposed in the literature cover a large
range. The relative weight of the contributions from the Pomeron and from
the $f$ depends on the para\-me\-tri\-zation chosen for the Pomeron.  For
our analysis, this does not present a significant source of uncertainty:
the relative weight of the two contributions barely matters. It is
important, however, that in the energy range of interest, factorization
yields an unambiguous prediction for the sum of the two, that is for the
total cross section with $I_t=0$.
  
In (\ref{Regge ImT0t}), we have allowed for a pre-asymptotic contribution,
whose size is determined by the coefficient $p_1$. The reason for including
such a term is that we are making use of the parametrization in terms of
Regge poles down to an energy of 1.7 GeV. The uncertainties in the residues
given in (\ref{betaPf PDG Regge}) stem from an analysis of the observed
$\pi N$ and $NN$ cross sections above 5 GeV. They do not cover the
contributions generated by nonleading Regge poles, which become
increasingly important as the energy is lowered. In the representation
(\ref{Regge ImT0t}), these contributions are modeled by a single term,
which drops off more rapidly with the energy than the Pomeron, by one power
of $s$.

We now turn to the $t$-dependence. While the term associated with the $f$
is of the standard form of a Regge pole given in equation (\ref{Regge}),
the one describing the Pomeron contains the square of a logarithm.
Following \cite{PDG-Regge}, we made the assumption that the Pomeron is a
triple pole in the angular momentum plane, located at $J=\alpha_P(t)$. This
is legitimate only if the energy is not too high. As pointed out in
\cite{MacDowell:1964zz,Auberson:1977jz}, unitarity imposes a bound on the
shape of the diffraction peak. The parametrization in (\ref{Regge ImT0t})
violates this bound at very high energies, for the following reason. In our
normalization of the scattering amplitude, the differential cross section
for elastic scattering reads \be\label{eq:differential cross
  section}\frac{d\sigma}{dt}=\frac{|T(s,t)|^2}{16\pi\, s\,(s-4M_\pi^2)}\;.
\ee The contribution to this cross section from $|\mbox{Im}\,T(s,t)|^2$
alone grows in proportion to $\ln^4\hspace{-0.1cm} s\, s^{2\alpha_P'(0)
  t}$, so that the integral over $t$ -- the total elastic cross section --
increases with the third power of $\ln s$ and thus eventually becomes
larger than the total cross section. The same problem also occurs if the
intercept of the Pomeron is taken above unity, $\alpha_P(0)>1$. There are
well-known ways to avoid the clash with unitarity, for instance in
Regge-eikonal models \cite{BoSo,Godizov:2007dy,Godizov:2008zz,Sobol:2010mu}
valid at very high energies.

In the context of the present paper, the behaviour at very high energies is
an academic issue, because the integrals over the Regge representation
occurring in the evaluation of the various quantities of interest pick up
significant contributions only from modest energies, where the logarithmic
term represents a small, roughly energy independent correction. It could
just as well be absorbed in the residue $\beta_P(t)$ and in the sub-leading
coefficient $p_1$, so that the problem would disappear. Nevertheless, a
parametrization of the Pomeron that (i) reduces to the one used in the
Review of Particle Properties \cite{PDG-Regge} when $t\rightarrow0$ and
(ii) is consistent with general principles also for large $s$ and $t\neq 0$
would be of considerable interest -- we did not find such a representation
in the literature.

We give the term proportional to $p_1$ the same $t$-depen\-dence as the
Pomeron. This has the advantage that the pre-asymptotic contributions in
the channel with $I_t=0$ are accounted for in terms of a single parameter,
but it is evident that a simple parametrization of this sort can account
for the nonleading terms only in a very crude way. Nevertheless, we will
show that this parametrization allows us to make contact between the region
where the first few partial waves represent a reliable approximation and
the one where the amplitude can be represented in terms of a few Regge
poles. The $t$-dependence of the residues $\beta_P(t)$ and $\beta_f(t)$ is
discussed in section \ref{sec:nonzerot}.

\subsection{\boldmath Parametrization of ${\rm  Im}\,T^{(1)}(s,t)$}\label{sec:ImT1t}
For the component with $I_t=1$, we work with the representation
\be\label{Regge ImT1t}\mbox{Im}\,T^{(1)}(s,t)_{\indR}=\beta_\rho(t)
\left\{1+r_{1}\frac{s_1}{s}\right\}
\left(\frac{s}{s_1}\right)^{\alpha_\rho(t)}\fs\ee It involves the
$\rho$-trajectory, \bea \alpha_\rho(t)\al=\al\alpha_\rho(0)+
\alpha_\rho'(0)\,t\co \nonumber\eea whose parameters are specified in
(\ref{alpha0frho}) and (\ref{alpha1frho}), as well as a pre-asymptotic
contribution, determined by the parameter $r_1$.  For the intepretation of
this term, we refer the reader to the last paragraph of the previous
section -- what we wrote for $p_1$ also applies to $r_1$.

The result for the coefficient $Y_{\rho\hspace{0.2mm}\pi\pi}$ in
(\ref{Ypipi}), which follows from the Regge fits to the $\pi N$ and $NN$
data in \cite{PDG-Regge}, comes with a large uncertainty. The same applies
to the corresponding result for the $\rho$ residue at $t=0$: \be
\label{eq:betarho PDG} \beta_\rho(0)= 2\,s_1
Y_{\rho\hspace{0.2mm}\pi\pi}(\hbar c)^{-2} = 112\pm 46\fs \ee In the
following, we do not make use of this crude estimate.  Instead, we rely on
the determination of $\beta_\rho(0)$ based on the Olsson sum rule
\cite{Olsson}, which will be discussed in section \ref{sec:Olsson}. Our
determination is consistent with the above number, but somewhat more
precise. The behaviour of $\beta_\rho(t)$ at nonzero values of $t$ is
discussed in section \ref{sec:nonzerot}.
\subsection{\boldmath Parametrization of ${\rm  Im}\,T^{(2)}(s,t)$}\label{sec:ImT2t}
At the energies considered in the analysis of the high energy $\pi N$ and
$NN$ data, contributions from exotic Regge channels are neglected. In the
energy range discussed in the present paper, however, this is not
justified. Indeed, as we will see, some of the sum rules obeyed by the
scattering amplitude do require a small component with $I_t=2$. In this
channel, the dominating contributions stem from the exchange of two Regge
poles with $I_t=1$, so that the leading singularity in the angular momentum
plane is a branch cut \cite{Pennington-Annals}. The leading contributions
stem from $\rho\!-\!\rho$ and $a_2\!-\!a_2$ exchange. In the Regge
representation we are using, which is based on \cite{PDG-Regge}, the $a_2$
trajectory is taken degenerate with the $f$. Moreover, equations
(\ref{alpha0frho}), (\ref{alpha1frho}) show that the $f$ and $\rho$
trajectories are rather close to one another: for the cut generated by
$\rho\!-\!\rho$ exchange, the formula (\ref{alphac}) gives
$\alpha_{\rho\rho}(0)=-\,0.10\pm 0.04$ and $\alpha_{\rho\rho}'(0)=0.46\pm
0.01$, while the cut from $f\!-\!f$ exchange is characterized by
$\alpha_{f\hspace{-0.05cm}f}(0)=+\,0.08\pm 0.10$,
$\alpha_{f\hspace{-0.05cm}f}'(0)=0.50\pm 0.08$. Also, at the relatively low
energies considered here, the neglected logarithms cannot be distinguished
from the difference between $\alpha_{\rho\rho}(0)$ and
$\alpha_{f\hspace{-0.05cm}f}(0)$.  The numerical values indicate that,
compared to the dominating contribution from the Pome\-ron, the exotic
component of the amplitude roughly falls off in proportion to $1/s$. The
$I_t=2$ component of the scattering amplitude may thus be viewed like a
pre-asymptotic contribution. We approximate it in the same manner as the
one occurring in the amplitude with $I_t=0$, parametrizing it with a simple
Regge pole, \be\label{Regge ImT2t} {\rm Im}\, T^{(2)}(s,t)_{\indR} =
\beta_{e} (t) \left(\frac{s}{s_1}\right)^{\alpha_{e}(t)},\ee where
$\beta_{e}(t)$ represents an effective residue and
\be\label{alphae}\alpha_{e}(t)=\alpha_e(0)+\alpha_e'(0)\,t\ee is an
effective trajectory with $\alpha_e(0)=0$.
 
The formula (\ref{alphac}) implies that the slope of the effective exotic
trajectory is smaller than the slopes of the $\rho$ or $f$-trajectories, by
a factor of 2. This indicates that, compared to $\rho$ and $f$, the
relative importance of the exotic amplitude increases as one goes away from
$t=0$, towards spacelike values \cite{DDLN-book}. On the other hand,
according to (\ref{alpha1P}), the slope of the Pomeron trajectory is even
smaller, so that, compared to the Pomeron, the exotic component loses
weight if $t$ becomes negative. We fix the exotic trajectory with
\be\label{alphae01} \alpha_{e}(0)=0\co\quad \alpha_e'(0)=0.5\pm 0.1 \,
\mbox{GeV}^{-2}\fs\ee The properties of the residue $\beta_e(t)$ as a
function of $t$ are discussed in section \ref{sec:nonzerot}.

\section{\boldmath Partial waves}\label{sec:partial waves}
The representation of the scattering amplitude in terms of Regge poles and
cuts is dual to the one in terms of partial waves: the amplitude can be
represented with either one of the two. We denote the partial wave
amplitudes by $t^I_\ell(s)$, where $I=0,1,2$ stands for the $s$-channel
isospin and $\ell=0,1,2\ldots$ denotes the angular momentum. We use the
normalization \bea \label{eq:T and t}
t^I_\ell(s)\al=\al\frac{1}{2i\rho(s)}\left\{\eta_\ell^I(s)\,e^{2i\delta_\ell^I(s)}-1\right\}\,, \\
\rho(s)\al=\al\sqrt{1-4M_\pi^2/s}\,,\nonumber\eea where $\eta^I_\ell(s)$
and $\delta^I_\ell(s)$ denote the elasticity and the phase shift,
respectively. The partial wave decomposition of the scattering amplitude
then reads
\bea\label{eq:partial waves} T^I(s,t_z) \al=\al 32 \pi\sum_\ell(2\ell+1)P_\ell(z)\,t^I_\ell(s)\\
t^I_\ell(s)\al=\al \frac{1}{64\pi}\int_{-1}^{+1}\!dz\, P_\ell(z)\,
T^I(s,t_z)\,,\no t_z\al=\al
\mbox{$\frac{1}{2}$}(4M_\pi^2-s)(1-z)\,,\nonumber\eea where $P_\ell(z)$ is
the Legendre polynomial of order $\ell$.  For the even angular momenta,
Bose statistics permits $I=0$ and $I=2$, while waves of odd angular
momentum carry $I=1$. We use the symbols S$^0$ and S$^2$ to distinguish the
two S-waves, write P$^1$ for the P-wave and label the higher partial waves
analogously. In the region where the partial waves are small, the real and
imaginary parts are approximately related by \be\label{eq:t small}
\mbox{Im}\,t^I_\ell(s)\simeq \rho(s)\,
\mbox{Re}\,t^I_\ell(s)^{2}+\frac{1-\eta^I_\ell(s)}{2\,\rho(s)}\fs\ee In
this region, unitarity ($0\leq \eta^I_\ell(s)\leq1$) sets a lower bound to
the imaginary part, given by the first term in the above relation.
 
At low energies, the angular momentum barrier suppresses the contributions
from high angular momenta. The\-re, the decomposition of the amplitude into
partial waves is preferable, because the first few terms in this
decomposition suffice to obtain a decent approximation. At high energies,
on the other hand, the Regge representation is more efficient, because the
high energy behaviour of the amplitudes is dominated by the leading Regge
poles. In the context of the present paper, which aims at an improved Regge
parametrization, the partial wave representation plays an important role
for two reasons: (i) We are making use of sum rules to analyze the
$t$-dependence of the Regge residues. Since these involve integrals
extending from threshold to infinity, we need an adequate representation
also at low energies. (ii) The leading Regge poles yield a good
approximation for the scattering amplitude only at very high energies. We
need to bridge the gap between the region where the leading Regge poles
dominate and the one where the Roy equations provide a reliable
representation for the partial waves. Since the behaviour of the amplitudes
between these two domains represents the main source of uncertainty in our
analysis, we need to discuss it in some detail.

In the region where the Roy equations are valid, we determine the partial
waves by solving these equations. The {\it input} of the calculation
consists of four parts:
\begin{enumerate}
\item subtraction constants 
\item elasticities below $s_{max}$ 
\item partial wave imaginary parts above $s_{max}$ 
\item driving terms
\end{enumerate} 
where $\sqrt{s_{max}}$ is the upper end of the energy interval over which
the equations are solved. In \cite{ACGL,CGL}, the upper end was taken at
$\sqrt{s_{max}}=0.8\,\mbox{GeV}$ and the analysis was restricted to the S-
and P-waves. In that case, the input unambiguously determines the solution.
In the present paper, we consider the entire range where the Roy equations
are valid, i.e.\,set $\sqrt{s_{max}}=\sqrt{68}\,M_\pi$ $\simeq
1.15\,\mbox{GeV}$, and also calculate the D-, F- and G-waves. As discussed
in detail in \cite{ACGL}, the input listed above does then not determine
the solution uniquely: on the extended interval, the Roy equations admit a
3-parameter family of solutions. We find it convenient to identify two of
these degrees of freedom with the value of the S$^0$ phase shift at two
points in the interior of the interval on which we solve the Roy equations:
$\delta_0^0(s_A)$, $\delta_0^0(4M_K^2)$, with $\sqrt{s_A}=0.8\,\mbox{GeV}$.
The third one is fixed with the P$^1$ phase shift: $ \delta_1^1(s_A)$. The
phenomenological estimates used for these quantities as well as those for
the elasticities and for the imaginary parts of the partial waves are
discussed in appendix \ref{app:input}. The calculation of the driving terms
is sketched in section \ref{sec:analysis}.

Near threshold, the subtraction constants control the behaviour of the
scattering amplitude. In the case of $\pi\pi$ scattering, the dispersive
representation involves two subtraction constants, which may be expressed
in terms of the two S-wave scattering lengths, $a_0^0$, $a_0^2$. At leading
order of the chiral expansion, their values can be predicted in terms of
the pion decay constant \cite{Weinberg:1966}. The higher order corrections
have been calculated in the framework of $\chi$PT \cite{GL:1983/84,BCEGS}.
As pointed out in \cite{CGL}, the dispersive analysis of the scalar pion
form factor \cite{DGL} can be used to pin down one of the low energy
constants occurring therein, so that a remarkably sharp theoretical
prediction for the scattering lengths follows: \be\label{a00
  a20}a_0^0=0.220\pm0.005\,,\quad a^2_0=-0.0444\pm 0.0010\,.\ee In the
meantime, this prediction has been tested in a series of beautiful low
energy precision experiments\footnote{For recent reviews, we refer to
  \cite{CD09:experiment}.}  concerning the decays
$K\rightarrow\ell\nu\pi\pi$ \cite{NA48Kl4,Pislak}, $K\rightarrow \pi\pi\pi$
\cite{NA48K3pi} and pionic atoms \cite{DIRAC}. For the combination
$a_0^0-a_0^2$, where the experimental precision is comparable to the
accuracy of the prediction, theory and experiment agree within errors. 
The data do not pin down the two individual scattering lengths to the same accuracy, but in combination with the fact that the dispersive evaluation of the scalar form factor leads to a sharp correlation between $a_0^0$ and $a_0^2$, the experimental results imply  $a_0^0= 0.2198 (46)_{stat} (16)_{syst} (64)_{th}$,
$a^2_0=-0.0445(11)_{stat} (4)_{syst}(8)_{th}$, thus confirming the predictions to a remarkable degree of accuracy. Alternatively, as pointed out by Stern and collaborators \cite{Stern}, the two scattering lengths may be disentangled by invoking comparatively crude experimental results extracted from $\pi N$ scattering. The recent update of this analysis by Garc\'ia-Martin et al.\,\cite{GarciaMartin:2011cn} leads to $a_0^0=0.220(8)$, $a_0^2=-0.042(4)$, also consistent with the predictions, albeit somewhat less precise. 

Moreover, the recent
progress made on the lattice now allows a calculation of $a_0^2$ from first
principles, analyzing the volume dependence of the energy levels of the
system in a box of finite size \cite{Beane,Feng-Jansen-Renner,Yagi:2011jn}.
Also, the low energy constants $\lbar_3,\lbar_4$, which dominate the
uncertainties in the theoretical prediction for $a_0^0$ and $a_0^2$, can
now be determined on the lattice, from the dependence of $M_\pi$ and
$F_\pi$ on the quark masses. The results are consistent with the estimates
used in \cite{CGL} and the uncertainties in $\lbar_3$ are reduced
significantly \cite{FLAG}. All of this corroborates the conclusion that the
subtraction constants are reliably and accurately known.  The remaining
three parts of the input will be discussed below.
\section{Partial wave representation of the total cross sections}\label{sec:total cross sections}
\begin{figure}[thb]
\begin{center}
\includegraphics[width=7.4cm]{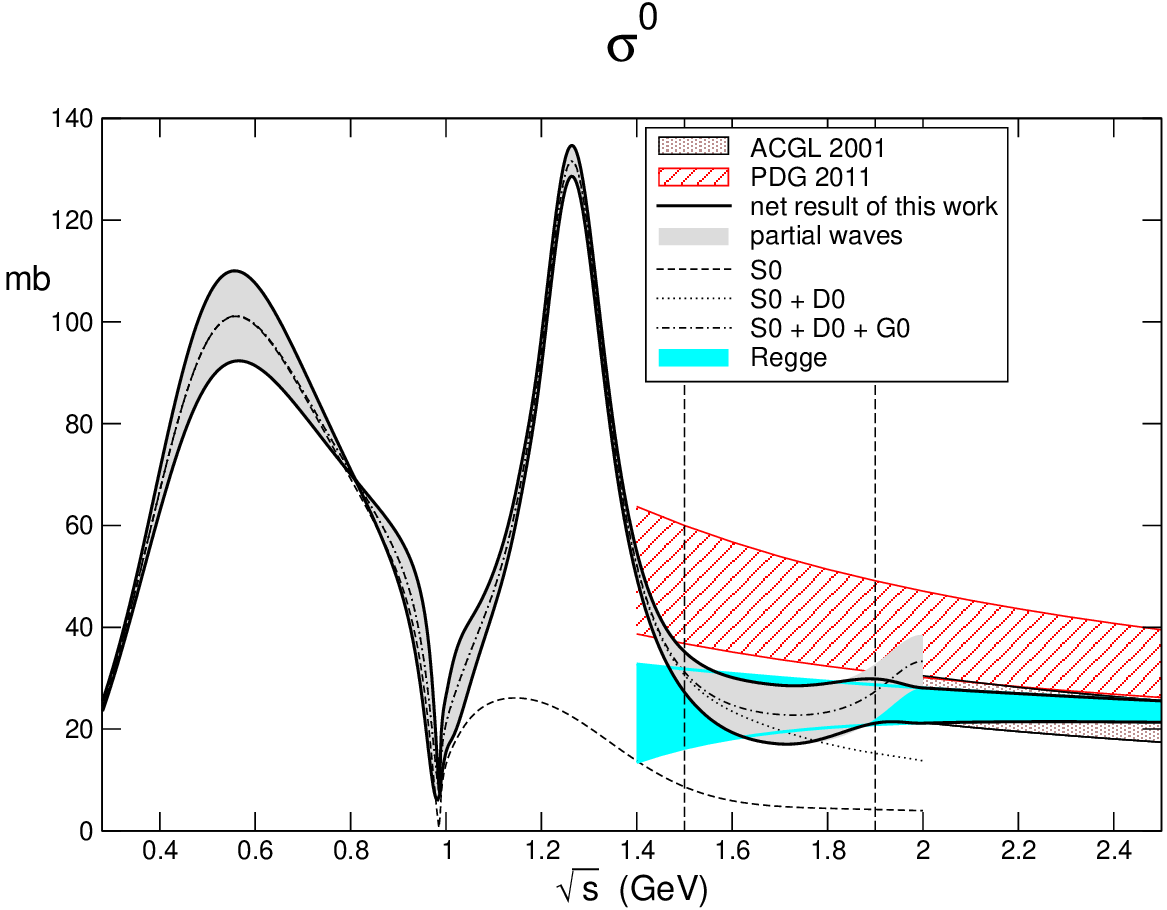}
 
\vspace{0.2cm}
{\bf Fig.\,3a} Total $\pi\pi$ cross section with $I_s=0$.

\vspace{0.3cm}
\includegraphics[width=7.4cm]{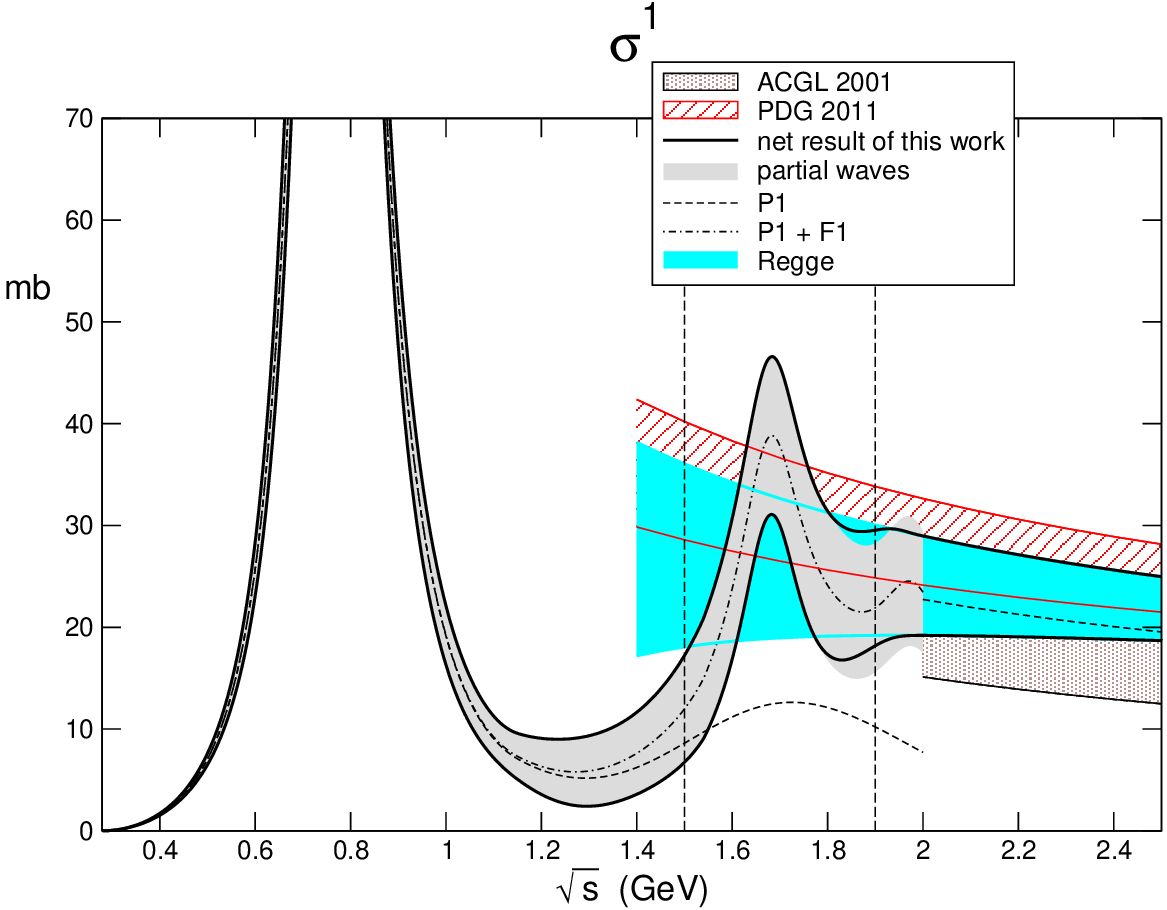}

\vspace{0.2cm}
{\bf Fig.\,3b} Total $\pi\pi$ cross section with $I_s=1$.

\vspace{0.3cm}
\includegraphics[width=7.4cm]{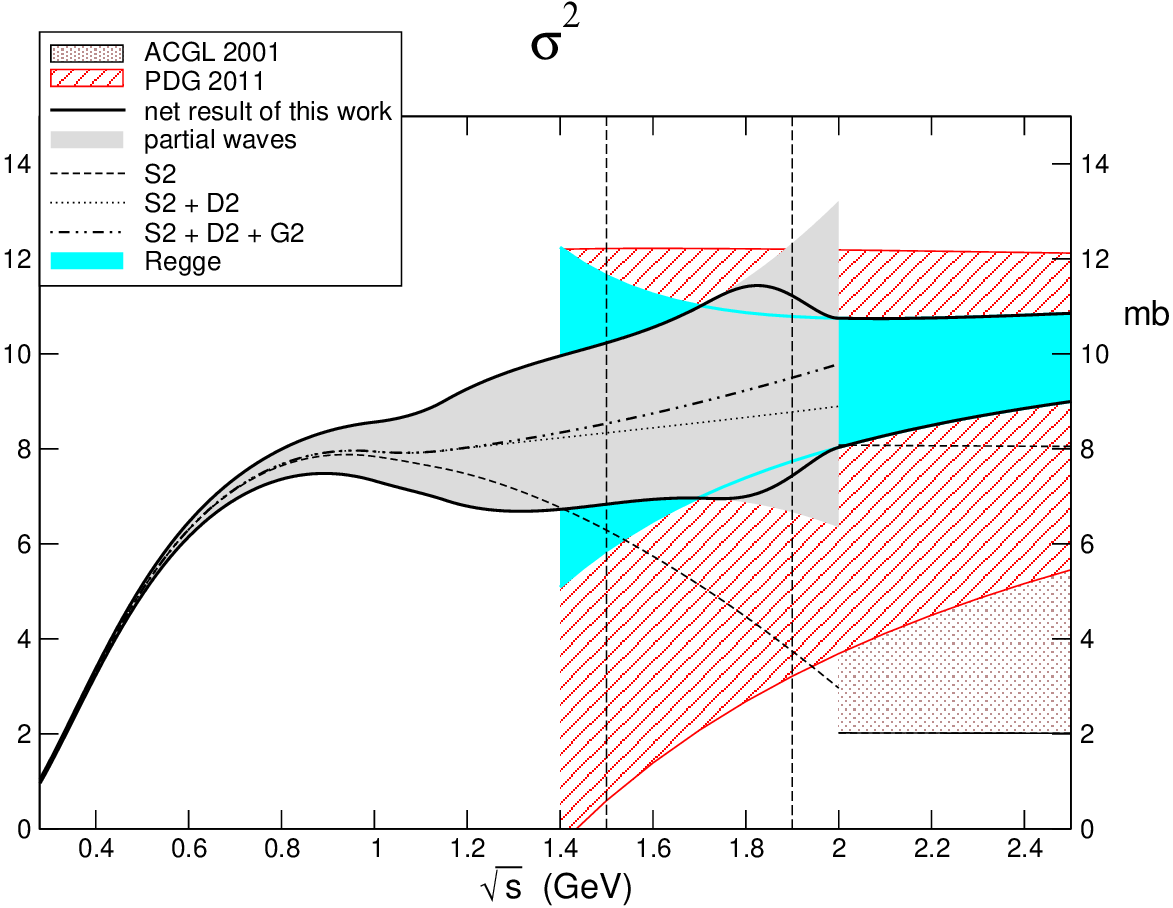}

\vspace{0.2cm}
{\bf Fig.\,3c} Total $\pi\pi$ cross section with $I_s=2$. 
\end{center}\vspace{-0.5cm} \end{figure}

In the present section, we briefly describe the qualitative picture
obtained for the total cross sections, in order to set the stage for our
analysis. The method used to derive this picture is outlined in section
\ref{sec:analysis}.

We denote the total cross sections with $s$-channel iso\-spin $I_s=0,1,2$
by $\sigma^0(s)$, $\sigma^1(s)$, $\sigma^2(s)$, respectively. In this
notation, the total cross sections for pions of definite charge are given
by
\bea\label{eq:sigmapipi}\sigma\rule[-1mm]{0mm}{0mm}_{\!\pi^+\pi^+}(s)\al=\al\sigma\rule[-1mm]{0mm}{0mm}_{\!\pi^-\!\pi^-}(s)= \sigma^2(s)\,,\\
\sigma\rule[-1mm]{0mm}{0mm}_{\!\pi^-\pi^+}(s)\al=\al\mbox{$\frac{1}{3}$}\sigma^0(s)+\mbox{$\frac{1}{2}$}\sigma^1(s)+\mbox{$\frac{1}{6}$}\sigma^2(s)\,,\no
\sigma\rule[-1mm]{0mm}{0mm}_{\!\pi^\pm\pi^0}(s)\al=\al
\mbox{$\frac{1}{2}$}\sigma^0(s)-\mbox{$\frac{1}{2}$}\sigma^2(s)\,, \no
\sigma\rule[-1mm]{0mm}{0mm}_{\!\pi^0\pi^0}(s)\al=\al
\mbox{$\frac{1}{3}$}\sigma^0(s)+\mbox{$\frac{2}{3}$}\sigma^2(s) \,.
\nonumber\eea

The grey bands in Fig.\,3 indicate our net result for the total cross
sections obtained from the sum over the partial waves. The dashed black
lines represent the central solutions of the Roy equations for the S- and
P-waves. The contributions from these waves dominate the total cross
sections below 1 GeV, reach a maximum and then drop with the energy. The
dotted black curves are obtained by adding the imaginary parts of the
central Roy solution for D$^0$ and D$^2$, respectively, while the
dash-dotted ones include the F- and G-waves.

In the total cross section, the resonance peaks occurring in the imaginary
parts of the individual partial waves generate an oscillatory behaviour,
which only disappears gradually with increasing energy.  Fig.\,3a shows
that, below 1 GeV, $\sigma^0(s)$ is dominated by the contribution from the
S$^0$-wave, which exhibits a broad bump around 600 MeV. The phenomenon
reflects the presence of a pole on the second sheet of the partial wave
amplitude $t^0_0(s)$, the well-known $f_0(600)$ or $\sigma$. Slightly above
0.8 GeV, the phase shift passes through 90$^\circ$. There, the inelasticity
is very small, so that the S$^0$-wave nearly reaches the unitarity limit.
This explains why the uncertainties in the total cross section are very
small in that region. While the destructive interference with the
$f_0(980)$ generates a pronounced dip in the vicinity of $K\Kbar$
threshold, the behaviour between 1.15 and 1.5 GeV is dominated by the
$f_2(1275)$. The contribution from the G$^0$-wave becomes visible only
above 1.6 GeV. The first resonance with these quantum numbers, the
$f_4(2018)$, can be seen, but it does not produce a marked peak, because
many decay channels are open and the branching fraction
$\Gamma(f_4\rightarrow\pi\pi)/\Gamma(f_4\rightarrow \mbox{all})$ is below
$20\%$.

The most prominent resonance, the $\rho(770)$, manifests itself in
$\sigma^1(s)$. Since an appreciable inelasticity sets in only at
$M_\omega+M_\pi\simeq 0.92$ GeV, the contribution from the P$^1$-wave
nearly saturates the unitarity limit at the peak of this resonance. The
cross section $\sigma^1(s)$ exceeds 220 mb there -- the upper two thirds of
the peak are chopped off, to make the behaviour of the cross section above
1 GeV more visible. The spin 3 resonance $\rho_3(1690)$ produces a more
modest peak, followed by an even weaker structure around 2 GeV. In the PDG
listings, the lowest resonance with $\ell=5$ is the $\rho_5(2330)$.  We
expect partial waves with $\ell>4$ to play a significant role only above 2
GeV.

The exotic cross section $
\sigma^2(s)=\sigma\rule[-1mm]{0mm}{0mm}_{\!\pi^-\!\pi^-}(s)$, which does
not receive contributions from resonances, is shown in Fig.\,3c, a
magnified version of Fig.\,2. Below 1.15 GeV, the Roy equations provide
good information about this cross section: they imply that the real parts
of the exotic partial waves are dominated by the contributions from the
imaginary parts of the non-exotic waves, which are known comparatively
well.  The uncertainties in $\sigma^2(s)$, however, grow with the energy --
at $\sqrt{s}=2$ GeV, they are of order 3 mb.

Fig.\,3 shows that, in the region between 1.5 and 2 GeV, our partial wave
representations for the total cross sections with $I_s=0$ and $I_s = 2$ are
roughly energy independent, while the one with $I_s=1$ clearly reveals the
presence of a resonance.  Our quantitative estimates for the imaginary
parts of the partial waves, which are outlined in appendix \ref{app:input},
lead to substantial uncertainties in the total cross sections, which,
moreover, grow with the energy -- beyond 2 GeV, the partial wave
decomposition is barely of practical use in our context. In the channels
with $I_s=0$ or 1, the uncertainties mainly stem from the background
underneath the resonances. In this language, the cross section in the
channel with $I_s=2$ represents pure background. Indeed, Fig.\,3 shows that
our estimate for the uncertainties in $\sigma^0(s)$ and $\sigma^1(s)$ are
comparable to the exotic total cross section $\sigma^2(s)$.

\section{Duality}\label{sec:transition}
In the preceding sections, we discussed the Regge and partial wave
representations by themselves. We now merge the two and first consider the
forward direction, $t=0$.
 
A simple way to join the two parametrizations of the imaginary parts is to
use the partial wave representation below some energy $E_t$ and the Regge
representation above that energy. In order for the total cross sections not
to make a jump at $\sqrt{s}=E_t$, the imaginary parts must be continuous
there:

$\bullet\;$In the case of $\mbox{Im}\,T^{(0)}(s,0)$, this requirement can
be used to fix the coefficient $p_1$ of the pre-asymptotic term, so that,
at $t=0$, the Regge representation (\ref{Regge ImT0t}) does then not
contain any unknowns: the total cross section $\sigma^{(0)}(s)$ is
determined also above the transition energy $E_t$.

$\bullet\;$The Olsson sum rule subjects the cross section $\sigma^{(1)}(s)$
to a constraint, which correlates the value of $\beta_\rho(0)$ with the
pre-asymptotic coefficient $r_1$ (see section \ref{sec:Olsson}). Since the
continuity requirement yields a second constraint of this type, we can
determine $\beta_\rho(0)$ as well as $r_1$. As a result, the total cross
section is determined above $E_t$, also in the channel with $I_t=1$.

$\bullet\;$Finally, in the exotic $t$-channel, continuity determines the
residue $\beta_e(0)$ in terms of the partial wave representation at $E_t$,
so that the Regge representation of $\sigma^{(2)}(s)$ does then not contain
any free parameters, either.

The choice of the transition energy represents a compromise: if $E_t$ is
taken too low, then our Regge representation does not provide a decent
approximation, if it is taken too high, then our neglect of the partial
waves with $\ell>4$ is not justified.

In \cite{ACGL,CGL,Descotes}, the transition from the partial wave
representation to the Regge representation is made at 2 GeV, while in
\cite{PY2003,PY2004-Regge,KPY2006,KPY2008,GarciaMartin:2011cn} the Regge parametrization is
assumed to be valid down to 1.42 GeV. The problem with the former analysis
is that the phenomenological information used for the partial waves above
1.4 GeV leaves much to be desired, while the problem with the latter choice
is that Regge asymptotics does not set in that early -- see Fig.\,3b.

Continuity may be viewed as a local version of duality: it requires the
partial wave and Regge representations to agree at one particular energy.
In the following, we use a weaker form of this requirement: we assume that
the partial wave and Regge representations agree with one another only {\it
  in the mean}. For short, we refer to the resulting constraints as {\it
  duality conditions}.
 
More precisely, we replace the continuity conditions discussed above by the
following constraints.  Fig.\,4 shows that the total cross sections with
fixed $t$-channel isospin
\begin{figure}[t]
\begin{center}
\vspace{0.15cm}
\includegraphics[width=7.2cm]{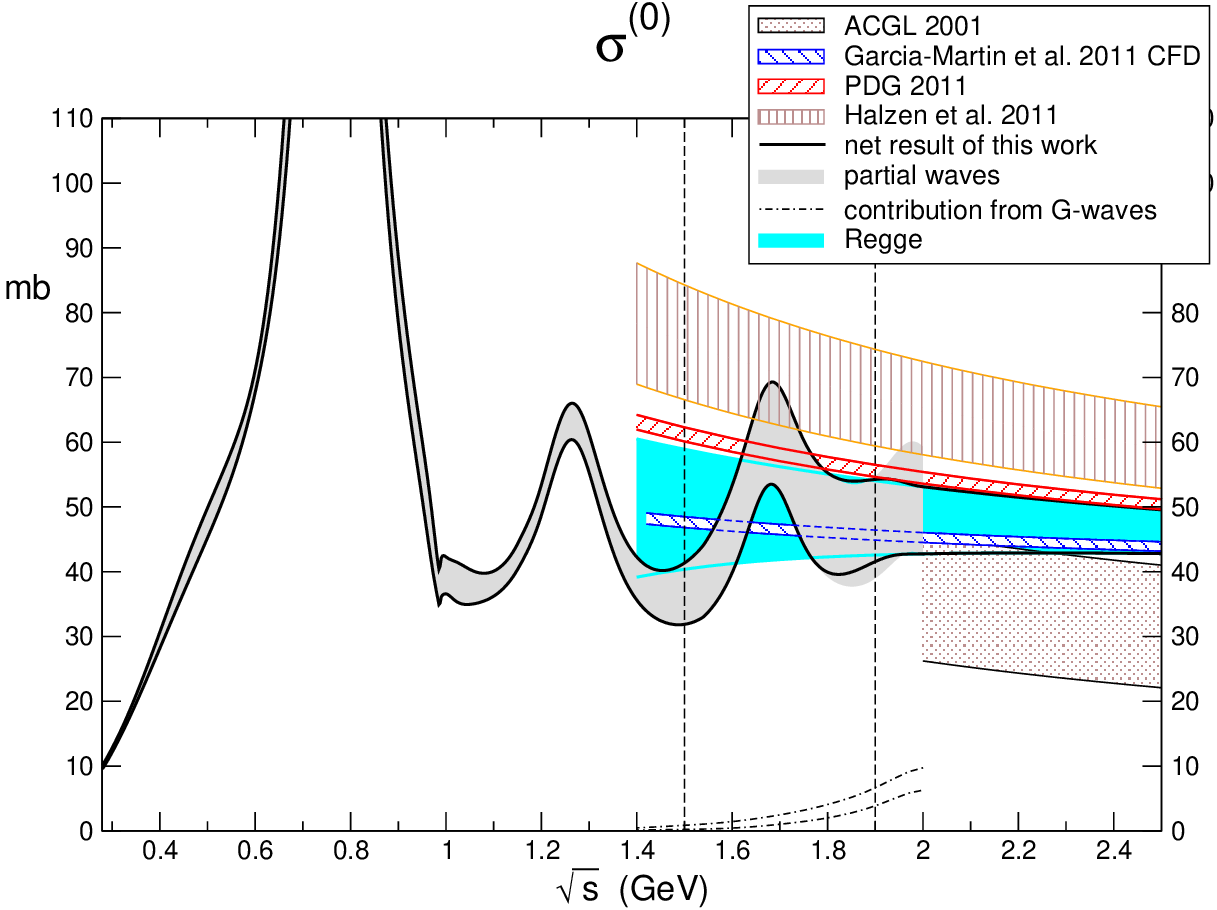}
 
\vspace{0.2cm}
{\bf Fig.\,4a} Total $\pi\pi$ cross section with $I_t=0$.

\vspace{0.35cm}
\includegraphics[width=7.3cm]{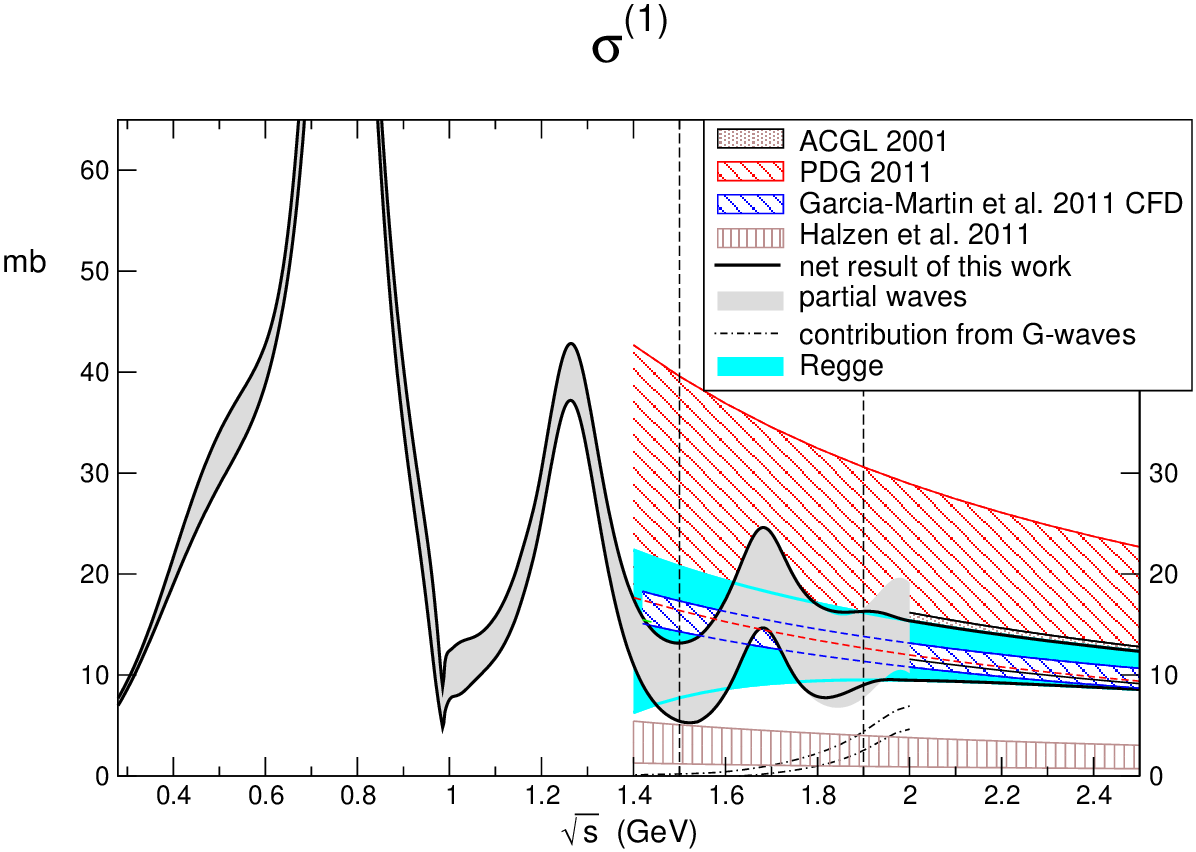}

\vspace{0.2cm}
{\bf Fig.\,4b} Total $\pi\pi$ cross section with $I_t=1$.

\vspace{0.3cm}
\includegraphics[width=7.5cm]{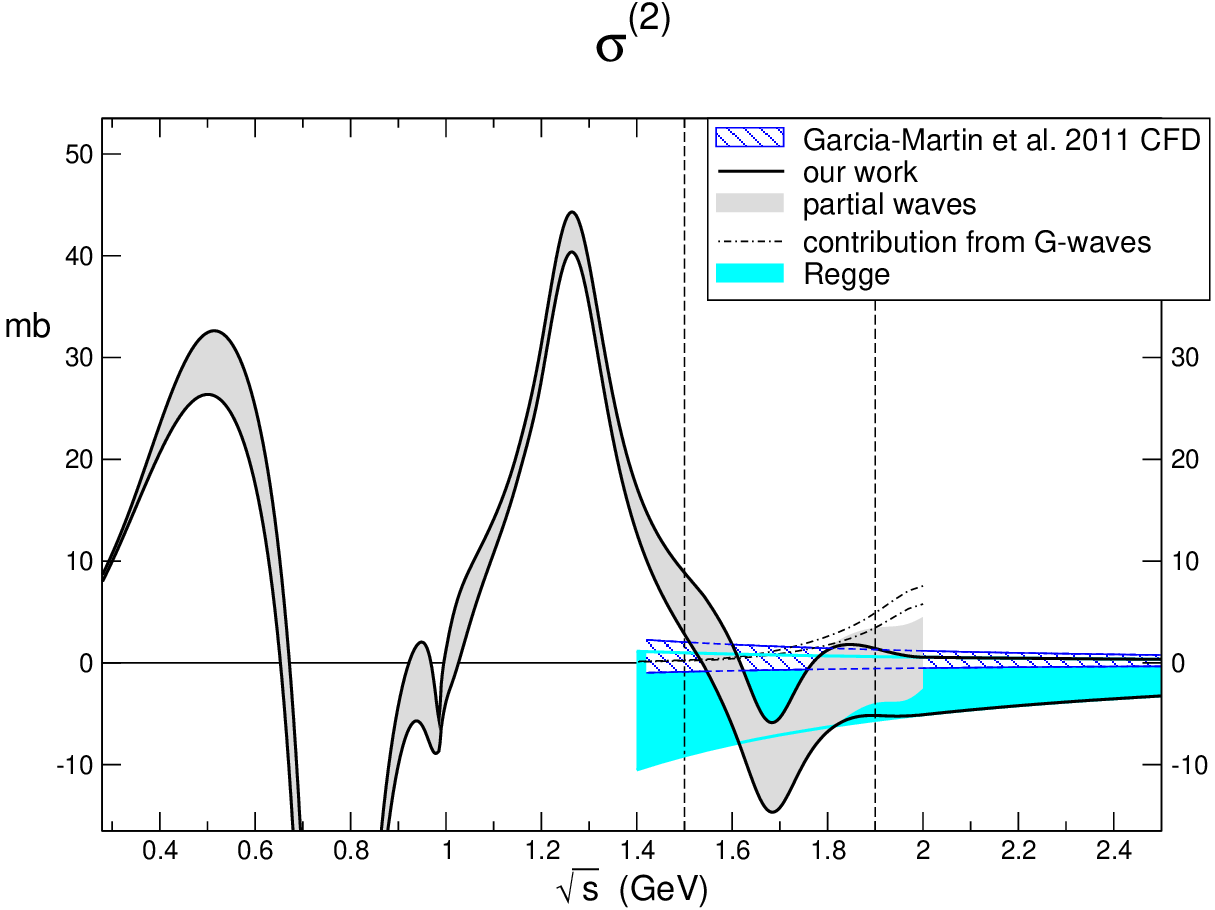}

\vspace{0.2cm}
{\bf Fig.\,4c} Total $\pi\pi$ cross section with $I_t=2$. 
\end{center}\vspace{-0.5cm}\setcounter{figure}{4} \end{figure}
\bea \sigma^{(0)}(s)\al=\al \mbox{$\frac{1}{3}$}\sigma^0(s)+\sigma^1(s)+ \mbox{$\frac{5}{3}$}\sigma^2(s)\,,\\
\sigma^{(1)}(s)\al=\al \mbox{$\frac{1}{3}$}\sigma^0(s)+
\mbox{$\frac{1}{2}$}\sigma^1(s)- \mbox{$\frac{5}{6}$}\sigma^2(s)\,,\no
\sigma^{(2)}(s)\al=\al \mbox{$\frac{1}{3}$}\sigma^0(s)-
\mbox{$\frac{1}{2}$}\sigma^1(s)+
\mbox{$\frac{1}{6}$}\sigma^2(s)\,,\nonumber\eea either exhibit a peak or a
dip, generated by the $\rho(1690)$. The dashed vertical lines indicate the
adjacent valleys at 1.5 and 1.9 GeV, respectively. The bands labeled
``Regge'' are obtained by assuming that, on the interval spanned by these
lines, 1.5 GeV $<\sqrt{s}<1.9$ GeV, the average over the partial wave
representation of the total cross sections agrees with the average over the
Regge representation. In other words, we assume that the inclusion of the
pre-asymptotic terms allows us to extrapolate the Regge representation down
to the above interval, not locally, but in the mean.  Roughly, this amounts
to imposing continuity at $E_t$, varying the value of $E_t$ over the above
range and averaging the result.

This version of duality lends itself to a natural generalization as well:
it suffices to replace the total cross sections $\sigma^{(k)}(s)$ by the
functions $\rho(s)\,\mbox{Im}\,T^{(k)}(s,t)$. The duality conditions used
in the present paper represent the requirement that the mean value of the
Regge representation for these functions agrees with the partial wave
representation thereof. Duality is useful in our framework, because it
provides us with an estimate for the pre-asymptotic terms
$p_1,r_1,\beta_e(0)$, much like the continuity conditions discussed above.
Moreover, its generalization provides a constraint on the $t$-dependence of
the Regge residues.

Whenever we calculate integrals over the imaginary parts of the $\pi \pi$
amplitude we need to specify how the transition between the partial-wave
and the Regge representation takes place. We assume that the partial wave
representation provides an adequate approximation below 1.7 GeV and that
the Regge representation yields an adequate approximation above 2 GeV. In
the transition region, we use an interpolation, \bea\label{interpolation}
\mbox{Im}\,T^{(I_t)}(s,t)\al=\al \{1-h(s)\}\,
\mbox{Im}\,T^{(I_t)}(s,t)_{\indPW}\\\al+\al h(s)\,
\mbox{Im}\,T^{(I_t)}(s,t)_{\indR}\co\hspace{0.8cm}s_a<s<s_b\co\nonumber\eea
with $\sqrt{s_a}=1.7$ GeV, $\sqrt{s_b}=2$ GeV. The function $h(s)$
interpolates between $h(s_a)=0$ and $h(s_b)=1$. For definiteness, we take a
cubic polynomial and fix the coefficients with the requirement that the
first derivative is continuous at the lower and upper ends of the
interpolation.\footnote{The corresponding explicit expression for the
  interpolating function reads
  $h(s)=(s-s_a)^2(3\,s_b-s_a-2\,s)/(s_b-s_a)^3$. } The interpolation
(\ref{interpolation}) automatically ensures continuity, also at non-zero
values of $t$.

\section{Sum rules}\label{sec:sum rules}
The Olsson sum rule \cite{Olsson} originates in the fact that the
$t$-channel $I=1$ amplitude does not receive a Pomeron contribution, and
thus grows only in proportion to $s^{\alpha_\rho(t)}$ for $s\rightarrow
\infty$. The fixed-$t$ dispersion relation obeyed by this amplitude,
however, does contain terms that grow linearly with $s$.  For the relation
to be consistent with Regge asymptotics, the contribution from the
subtraction term must cancel the one from the dispersion integral.  At
$t=0$, this condition reduces to a sum rule which expresses the combination
$2a_0^0-5a_0^2$ of the $S$-wave scattering lengths as an integral over the
cross section $\sigma^{(1)}(s)$ \cite{Olsson}:
\bea\label{Olsson}2a_0^0-5a_0^2=\frac{3 M_\pi^2}{4\pi^2}
\int_{4M_\pi^2}^\infty
ds\,\frac{\sigma^{(1)}(s)}{\sqrt{s(s-4M_\pi^2)}}\,.\eea

The requirement that the amplitude $ T^{(1)}(s,t)$ has the proper high
energy behaviour also for $t\neq0$ implies a further constraint. This
amplitude obeys the fixed-$t$ dispersion relation in equation
(\ref{fixedt}) of appendix \ref{app:dispersion relation}. The constraint
can be derived by evaluating the coefficient of the term on the r.h.s\, of
this equation that grows linearly with $s$ and requiring that the term has
the same value as for $t=0$.  This condition subjects the function
$\mbox{Im}\,T^{(1)}(s,t)$ to an entire family of sum rules of the form \be
S(t)=0\co\ee where $t$ is a free parameter and the function $S(t)$ is given
by: \bea\label{St} S(t)&\equiv&\frac{1}{\pi}\int_{4M_\pi^2}^\infty ds\,
\frac{\mbox{Im}\,\Tbar^{\,(1)}(s,t)}{s\,(s+t-4M_\pi^2)}\\
&-&\frac{1}{\pi}\int_{4M_\pi^2}^\infty ds\,\frac{2(s-2M_\pi^2)\,
  \mbox{Im}\,T^1(s,0)}{s\,(s-4M_\pi^2)\,(s-t)\,(s+t-4M_\pi^2)}\fs\nonumber\eea
The first integral involves only the $I_t=1$ component of the imaginary
part.  The bar indicates that the value at $t=0$ is subtracted:
\be\label{ImTbar} \Tbar (s,t)\equiv\frac{T(s,t)-T(s,0)}{t}\fs\ee The
function $T^1(s,0)$ in the second integral is the $I_s=1$ component at
$t=0$, which exclusively picks up contributions from partial waves of odd
angular momentum.  The integrand in (\ref{St}) displays a fictitious pole
for $s= 4M_\pi^2- t$. However, for a crossing symmetric integrand, the
residue of the pole vanishes. The sum rule $S(t)=0$ was written down
already in \cite{Wanders:1969} and was applied for the determination of the
Regge residues in \cite{Tryon,Basdevant-Schomblond}.  For a detailed
discussion, we refer to appendix C in \cite{ACGL}.

The Roy equations are manifestly symmetric with respect to
$s\leftrightarrow u$, but they do not ensure crossing symmetry with respect
to $s\leftrightarrow t$, which requires the scattering amplitude
$\vec{T}=(T^0,T^1,T^2)$ to obey the relation \be\label{st}
\vec{T}(s,t)-C_{st}\vec{T}(t,s) =0\,.\ee On the other hand, any solution of
the Roy equations that obeys this constraint is then automatically also
crossing symmetric with respect to $t\leftrightarrow u$. Note that the
subtraction constants as well as the contributions from the S- and P-waves
drop out here: the relation (\ref{st}) amounts to a family of sum rules
that relate integrals over the imaginary parts of the partial waves with
$\ell\geq 2$ to integrals involving the Regge residues. We exploit these
additional constraints by evaluating the derivative of (\ref{st}) with
respect to $s$ at $s=0$. As shown in appendix \ref{app:SR}, this leads to a
new family of sum rules of the same type as the one above. We write these
in the form $C_k(t)=0$, with $k=0,1,2$. Explicit expressions for the three
functions $C_0(t)$, $C_1(t)$, $C_2(t)$ are given in equation (\ref{eq:Ck}).
As we will see, these sum rules provide us with a good tool to determine
the $t$-dependence of the residues.
\section{Method of analysis}\label{sec:analysis}
Our analysis runs as follows:

\vspace{0.1cm}\noindent 1. We start with a complete representation of the
amplitude, which specifies the imaginary parts of the partial waves below 2
GeV as well as the parameters occurring in the Regge representation.

\vspace{0.1cm}\noindent 2. As a first step, we calculate the driving terms
for the partial waves with $\ell\leq 4$. The driving terms of a given wave
consist of two parts: contribution from the imaginary parts of the other
partial waves with $\ell \leq 4$ below 2 GeV and contribution from all
waves above 1.7 GeV. The former is smoothly cut off with the factor
$\{1-h(s)\}$, while the latter is gradually turned on with the factor
$h(s)$, in accord with equation (\ref{interpolation}).

\vspace{0.1cm}\noindent 3. Next, we specify the input for which we wish to
determine the full amplitude. Some of the input variables concern the
partial wave representation, others concern the Regge para\-metrization. In
fact, neither one of the two representations is fully specified by the
input variables alone (see appendix \ref{app:input}). We refer to the
remaining parameters as output variables. The Roy equations determine the
partial wave output, while the sum rules and duality conditions serve to
determine the Regge output.

\vspace{0.1cm}\noindent 4. Together with the driving terms, the partial
wave input uniquely determines the solution of the Roy equations. We solve
these equations iteratively. The result is a new complete representation of
the amplitude.

\vspace{0.1cm}\noindent 5. We then evaluate the various sum rules and
duality conditions, which amount to constraints imposed on the Regge output
variables. Since some of the sum rules and the duality conditions depend on
$t$, we are in principle dealing with an infinite number of equations for a
finite number of parameters and can therefore not expect to find an exact
solution. Instead, we determine the minimum of a discrepancy function,
obtained by adding up the various equations in quadrature -- for details,
we refer to appendix \ref{sec:details}. The minimum fixes the output
variables of the Regge parametrization. In combination with the input
specified in step 3, we thus arrive at a new Regge representation.

\vspace{0.1cm}\noindent 6. If the result of the preceding step differs
significantly from the Regge representation we started with, we must return
to step 2 and iterate the entire procedure. We find that the process
converges rapidly -- two or three iterations suffice to arrive at a
self-consistent solution of good quality.

\vspace{0.1cm}\noindent 7.  For a given choice of the input, the above
procedure provides us with a complete representation of the amplitude. We
finally need to determine the response of the output to changes in the
input. As the variations of interest are small, the result for the partial
wave and Regge representations of the scattering amplitude is approximately
linear in these variations. We treat the input parameters as independent
Gaussian variables, with one exception: we account for the
correlation\footnote{This correlation cannot be ignored; the error in the
  combination $2a_0^0-5a_0^2$, for instance, which plays a prominent role
  in the Olsson sum rule, is significantly smaller than what is obtained if
  the two terms are treated as independent.}  between the two scattering
lengths $a_0^0$ and $a_0^2$. The error analysis is then straightforward --
the quantities of interest are linear functions of the input variables and
we can calculate the uncertainties therein as well as the correlations
between them by averaging over the Gaussian fluctuations in these
variables.

\section{\boldmath Regge analysis of the total cross sections}\label{sec:sigmatot}
\subsection{\boldmath Total cross section with $I_t=0$} \label{sec:sigma0t}
We first illustrate the method with the determination of the pre-asymptotic
coefficient $p_1$. As discussed in section \ref{sec:leading}, factorization
yields a rather sharp prediction for the asymptotic behaviour of the total
$\pi\pi$ cross section in the channel with $I_t=0$. The prediction is shown
as a narrow red band in Fig.\,4a, obtained by inserting the values
(\ref{betaPf PDG Regge}) for $\beta_P(0)$ and $\beta_f(0)$ in the formula
(\ref{Regge ImT0t}), setting the pre-asymptotic coefficient $p_1$ equal to
zero and calculating the corresponding cross section with the optical
theorem (\ref{optical theorem}).

Since the grey band in Fig.\,4a, which represents the partial wave
representation of the cross section $\sigma^{(0)}(s)$, passes below the
extrapolation of the contributions from $\beta_P(0)$ and $\beta_f(0)$, the
pre-asymptotic term must be negative. The numerical evaluation of the
duality condition formulated above implies that the coefficient $p_1$ is in
the range \be\label{p1} p_1=-0.8\pm0.6\fs\ee
\subsection{\boldmath Olsson sum rule}\label{sec:Olsson}
The analysis of the total cross section in the channel with $I_t=1$ is more
involved, because we do not make use of the result obtained from the $\pi
N$, $NN$ and $\bar{N}N$ data with factorization. As discussed in section
\ref{sec:ImT1t}, the error bar in that result is rather large.  We instead
rely on the Olsson sum rule (\ref{Olsson}), which states that an integral
over the cross section $\sigma^{(1)}(s)$ is determined by a combination of
S-wave scattering lengths. The l.h.s.~is known from $\chi$PT \cite{CGL}:
\be\label{low energy} 2a_0^0-5a_0^2=0.663\pm 0.007\,.  \ee The prediction
is very sharp, because the NLO corrections to Weinberg's low energy theorem
\cite{Weinberg:1966} exclusively involve the low energy constant $\ell_4$
\cite{GL:1983/84}. Chiral symmetry implies that the same constant also
governs the chiral expansion of the scalar radius of the pion. The value
quoted in (\ref{low energy}) relies on a dispersive calculation of the
scalar pion form factor \cite{DGL} and accounts for all contributions up to
and including NNLO of the chiral expansion \cite{BCEGS}.  As discussed in
section \ref{sec:partial waves}, the predictions for the scattering lengths
have in the meantime been corroborated, both experimentally and on the
lattice. Concerning the specific combination relevant for the Olsson sum
rule, the ``best value'' obtained on the basis of the available
experimental information in \cite{GarciaMartin:2011cn} is
$2a^0_0-5a^2_0=0.650 \pm 0.015$, thus confirming the prediction (\ref{low
  energy}) to an accuracy of 2.5\%.

The integral on the r.h.s.~of (\ref{Olsson}) can be evaluated with the
optical theorem (\ref{optical theorem}). Below 1.7 GeV, the integrand is
represented as a sum over the partial waves discussed in section
\ref{sec:partial waves}, while above 2 GeV, the Regge representation in
(\ref{Regge ImT1t}) is relevant. In the transition region, we are using the
interpolation (\ref{interpolation}). The Olsson integral is then given by
the sum of two terms: $\mbox{O}=\mbox{O}_{\indPW}+\mbox{O}_{\indR}$, with
\bea\label{eq:OPW OR} \mbox{O}_{\indPW}\al\equiv\al \frac{3
  M_\pi^2}{4\pi^2} \int_{4 M_\pi^2}^\infty\hspace{-0.3cm}
ds\,\frac{1-h(s)}{\sqrt{s(s-4M_\pi^2)}}\;\sigma^{(1)}_{\indPW}(s)\co\\
\mbox{O}_{\indR}\al\equiv\al \frac{3 M_\pi^2}{4\pi^2} \int_{4
  M_\pi^2}^\infty\hspace{-0.3cm} ds\,
\frac{h(s)}{\sqrt{s(s-4M_\pi^2)}}\;\sigma^{(1)}_{\indR}(s)\nonumber \fs\eea
 
Below 1.15 GeV, we solve the Roy equations for the S, P, D, F and G partial
waves. From there to 2 GeV, we use a parametrization based on phenomenology
(see appendix \ref{app:input}). Table \ref{tab:Olsson} lists the numerical
results obtained for the contributions from the partial waves considered.
\begin{table}[thb]\begin{center}
\begin{tabular}{|c|l|}\hline
 S$^0$ &$\hspace{0.8em}0.315^{+0.019}_{-0.011}$\rule[-0.1cm]{0cm}{0.5cm} \\
S$^2$ &$-0.078^{+0.007}_{-0.005}\rule[-0.1cm]{0cm}{0.5cm}$\\
P$^1$ &$\hspace{0.8em}0.246^{+0.017}_{-0.019} \rule[-0.2cm]{0cm}{0.6cm}$ \\
D$^0$ &$\hspace{0.8em}0.061\pm 0.003 $\\
D$^2$ &$-0.007\pm 0.003 $\\
F$^1$ &$\hspace{0.8em}0.013\pm 0.001 $\\
G$^0$ &$\hspace{0.8em}0.0021\pm 0.0003$\\
G$^2$ & $-0.0007\pm 0.0007$\\
\hline
$\mbox{O}_{\indPW}$&$\hspace{0.8em}0.552^{+0.022}_{-0.016}$\rule[-0.2cm]{0cm}{0.6cm}\\ \hline
\end{tabular}\vspace{0.2cm}
\caption{\label{tab:Olsson} Partial wave contributions to the Olsson sum
  rule.}\end{center}\vspace{-0.5cm}\end{table} 
The table shows that the result is dominated by the contributions from
S$^0$ and P$^1$. The uncertainties in the contributions from the partial
waves are not symmetric, because the input used for the value of the phase
shift $\delta^0_0$ at 0.8 GeV is asymmetric (see appendix \ref{app:input}).
  
Note that the errors are correlated -- the uncertainty attached to the sum
given at the bottom of the table is smaller than what is obtained by
summing the individual errors in quadrature, because it accounts for the
correlations. These also need to be accounted for when evaluating the
combination
\be\label{eq:OPWbar}\OPWbar\equiv\mbox{O}_{\indPW}-2a_0^0+5a_0^2\,:\ee
since the values of the scattering lengths enter the Roy solutions, the
term $\mbox{O}_{\indPW}$ depends on these. The net result reads
\be\label{eq:OPWbar num} \OPWbar=-0.110^{+0.022}_{-0.015}\fs\ee
\subsection{\boldmath Total cross section with $I_t=1$}\label{sec:sigma1t}
In the above notation, the Olsson sum rule amounts to the relation
$\OPWbar+ \mbox{O}_{\indR}=0$. Since the representation (\ref{Regge ImT1t})
for $\mbox{Im}\,T^{(1)}(s,t)$ is proportional to the residue of the Regge
pole with the quantum numbers of the $\rho$, the cross section
$\sigma^{(1)}_{\indR}(s)$ is proportional to $\beta_\rho(0)$ and in
addition involves the coefficient $r_1$ of the pre-asymptotic term, as well
as the intercept of the $\rho$-trajectory, $\alpha_\rho(0)$. Actually, in
the narrow range (\ref{alpha0frho}) we adopt for this intercept, the
integral is not sensitive to $\alpha_\rho(0)$. Evaluating it with the
central value, the result (\ref{eq:OPWbar num}) implies \be
\label{eq:Olsson band} \beta_\rho(0)(1+0.104\, r_1) =80^{+11}_{-16}\fs\ee
\begin{figure}[tbh]  
\begin{center} 
\includegraphics[width=8.5cm]{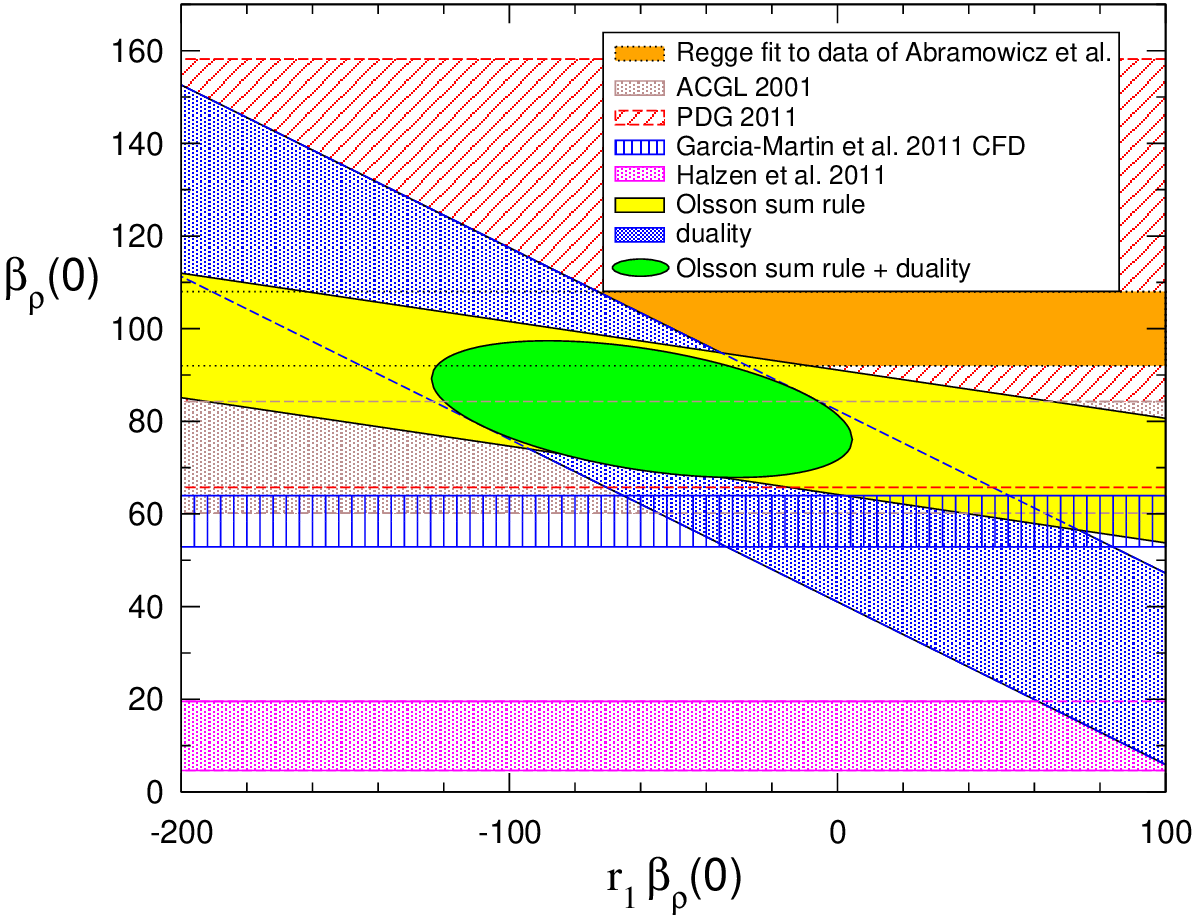} 
\end{center} 
\caption{\label{fig:betarho} Regge term with the quantum numbers of the
  $\rho$. The constraints imposed by the Olsson sum rule and by dual\-ity
  are compared with the results for $\beta_\rho(s)$ obtained from other
  sources: Regge parametrization quoted in the Review of Particle
  Properties \cite{PDG-Regge} plus factorization; fit to the data of
  Abramowicz et al.\,\cite{Abramowicz}; various values taken from the
  literature \cite{ACGL}, \cite{GarciaMartin:2011cn}, \cite{Halzen}. The
  ellipse represents the result of our analysis in equation
  (\ref{eq:betarho0 and r1}).}\end{figure}
The Olsson sum rule thus leads to a linear correlation between
$\beta_\rho(0)$ and $r_1 \beta_\rho(0)$, which is indicated by the yellow
band in Fig.\,\ref{fig:betarho}. Because of the linearity, the error
analysis simplifies considerably if the variable $r_1$ is replaced by
$r_1\beta_\rho(0)$ and we will do that.

The duality condition, which requires that, on the interval
$1.5\,\mbox{GeV}<\sqrt{s}<1.9\,\mbox{GeV}$, the average over the partial
wave representation agrees with the average over the Regge representation
yields a second relation of this form: \be\label{eq:Duality band}
\beta_\rho(0)(1+0.351\, r_1) =61^{+10}_{-9} \fs\ee In
Fig.\,\ref{fig:betarho}, this constraint is shown as a blue band.

The right hand sides of (\ref{eq:Olsson band}) and (\ref{eq:Duality band})
are obtained by averaging over the uncertainties attached to the Roy
solutions and are therefore correlated. In order to account for these
correlations as well as for the uncertainties in $\alpha_\rho(0)$, we use
the method outlined in section \ref{sec:analysis} and calculate the central
values and the correlation matrix of the two quantities $\beta_\rho(0)$ and
$r_1\beta_\rho(0) $. The ellipse shown in Fig.\,\ref{fig:betarho} indicates
the corresponding 1 $\sigma$ contour and represents our final result for
these two quantities. While the bands do not account for the correlation
between the r.h.s.\,of (\ref{eq:Olsson band}) and (\ref{eq:Duality band}),
the ellipse does -- this is the main reason why the bands are not exactly
tangent to the ellipse (the remaining difference stems from the fact that
the ellipse also accounts for the uncertainty in $\alpha_\rho(0)$, while
the bands do not). Expressed in terms of the two original parameters
$\beta_\rho(0)$ and $r_1$, our result reads: \be\label{eq:betarho0 and r1}
\beta_\rho(0)=84^{+13}_{-16}\,,\quad r_1=-0.8^{+0.8}_{-0.7}\,.\ee There is
a modest correlation between these two variables; the correlation
coefficient is $-0.33$.  The result for $\beta_\rho(0)$ will be compared
with values found in the literature in section \ref{sec:discussion
  sigmatot}.
 
\subsection{\boldmath Total cross section with $I_t=2$} \label{sec:sigma2t}
The analysis of the total cross section with isospin 2 in the $t$-channel
closely parallels the one in section \ref{sec:sigma0t}.  The grey band in
Fig.\,4c indicates the sum of the partial waves for the corresponding total
cross section. The resonances are seen clearly, because $\sigma^{(2)}(s)$
is not positive and the resonance contributions alternate in sign. The last
resonance seen is the $\rho_3(1690)$ which generates a small dip.

The high energy behaviour is parametrized with the expression (\ref{Regge
  ImT2t}). According to (\ref{alphae01}), the total cross section falls
approximately in inverse proportion to $s$. In this channel, the
requirement that, on the interval between 1.5 and 1.9 GeV, the average over
the partial wave representation agrees with the average over the Regge
representation implies \be\label{betae0} \beta_e(0)= -24 \pm 29 \fs\ee

\subsection{Discussion: total cross sections}\label{sec:discussion
  sigmatot}
The Regge parameters emerging from our analysis at $t=0$ are given in
equations (\ref{betaPf PDG Regge}), (\ref{p1}) for $I_t=0$, in
(\ref{eq:betarho0 and r1}) for $I_t=1$, and in (\ref{betae0}) for $I_t=2$.
The final result for the total cross sections, obtained with the
interpolation described in section \ref{sec:transition}, is shown as a
solid line in Figs. \ref{fig:pimpip}--4.

An important feature of our representation is the presence of
pre-asymptotic terms. As mentioned above, such terms become relevant if the
Regge representation is used at relatively low energies. In the following
sections, we will show that the semi-local form of duality invoked in the
present paper does lead to a coherent picture, provided the pre-asymptotic
contributions are accounted for. We emphasize, however, that the simple
parametrization used for these terms can only give a very rough description
of the contributions occurring underneath the leading Regge poles.
 
In the cross section with $I_t=0$, the estimate (\ref{p1}) for the
pre-asymptotic term reduces the contribution from Pomeron and $f$ at 2 GeV
by $20\pm 15\%$. Fig.\,4a shows that, towards lower energies, this term
widens the band spanned by the Regge representation, leaving room for
resonances. Towards higher energies, the pre-asymptotic contribution
shrinks. At 5 GeV (lower end of the energy range used for the Regge
analysis of the $\pi N$ and $NN$ data in \cite{PDG-Regge}), it amounts to
$3\pm2\%$.

Our estimate (\ref{eq:betarho0 and r1}) for the pre-asymptotic term in the
channel with $I_t=1$ is consistent with zero. The main effect of this term
is to increase the uncertainties in the result obtained for the Regge
parametrization at low energies. Non-leading contributions of this size
generate sufficient broadening, so that the extrapolation below 2 GeV
becomes consistent with the representation in terms of partial waves, at
least in the average sense. Note that in the channel with $I_t=1$, the
resonances generate more pronounced fluctuations than for $I_t=0$, because
the constant background related to the Pomeron is missing here: between 1.3
and 1.5 GeV, the sum of the partial wave contributions drops from $34\pm 2$
to $7\pm 3$ mb and then again increases to $14 \pm 4$ mb when the energy
reaches 1.7 GeV.

The entire component of the Regge representation with $I_t=2$ amounts to a
pre-asymptotic contribution. The result given in (\ref{betae0}) shows that,
at $t=0$, this term vanishes within errors. The corresponding total cross
section is in the range $\sigma^{(2)}(2\,\mbox{GeV})=-2\pm 3$ mb, small
compared to the leading cross section $\sigma^{(0)}(2\,\mbox{GeV})=48\pm 5$
mb, which is dominated by the contribution from the Pomeron.

We conclude that the Regge representation exhibits the expected qualitative
features already at 2 GeV.  Our results are in general agreement with the
total $\pi^-\pi^+$ cross section data shown in Fig.\,\ref{fig:pimpip},
while among the $\pi^-\pi^-$ data above 1.5 GeV shown in
Fig.\,\ref{fig:pimpim}, only those of Abramowicz et al. are consistent with
our framework. In fact, these data provide an independent estimate for the
size of the coefficient $Y_{\rho\hspace{0.2mm}\pi\pi}$ in the Regge formula
(\ref{sigmatot pipi}) for the total $\pi\pi$ cross section. Taking the
values of the coefficients $B,s_0,Z_{\pi\pi},Y_{f\pi\pi}$ from
factorization, neglecting contributions with $I_t=2$ as well as
pre-asymptotic terms, and treating $Y_{\rho\hspace{0.2mm}\pi\pi}$ as a free
parameter, we obtain an acceptable fit to the data of Abramowicz et
al.\,\cite{Abramowicz}, with $\chi^2=7.7$, for 5 degrees of freedom and
$Y_{\rho\hspace{0.2mm}\pi\pi} =19.5\pm 1.6\,\mbox{mb}$. This corresponds to
$\beta_\rho(0)=100\pm 8$, consistent with (\ref{eq:betarho0 and r1}), as
seen also from Fig.\,\ref{fig:betarho}.

It can be further checked that the pre-asymptotic terms do not distort the
description at the energies where the data of Abramowicz et al. were taken.
Evaluating $\chi^2$ with the results obtained for $p_1$, $\beta_e(0)$,
$\beta_\rho(0)$ and $r_1$ in (\ref{p1}), (\ref{betae0}), (\ref{eq:betarho0
  and r1}), and accounting for the correlations among these variables, we
obtain $\chi^2=6.8$ for 6 data points, no free parameters. As compared to
the fit mentioned above, the quality thus even improves a little. We
conclude that the results obtained above for the non-leading Regge
contributions at $t=0$ are perfectly consistent with the data of Abramowicz
et al.

Figs.\,1--4 also show several other Regge parametrizations of the $\pi\pi$
amplitude at high energies.  The red bands indicated as PDG 2011 in Figs.
3a--c and 4a, 4b are based on the results given in equations (\ref{betaPf
  PDG Regge}) and (\ref{eq:betarho PDG}), which are obtained from the $\pi
N$, $NN$ and $\bar{N}N$ data above 5 GeV with factorization. As discussed
in section \ref{sec:ImT1t}, the large width of these bands in Figs. 3 and
4b is due to the poor extraction of $\beta_\rho(0)$.  The central value of
the factorization estimate for $\beta_\rho(0)$ is on the high side, but in
view of its large error bar it is consistent with the value
(\ref{eq:betarho0 and r1}), extracted from totally independent information.

The Roy equation analysis performed in \cite{ACGL,CGL} is based on
Pennington's representation \cite{Pennington-Annals}. The bands denoted as
ACGL 2001 in Figs.\,3 and 4 show that the Pomeron estimate used in
\cite{ACGL}, $\beta_P(0)=46 \pm 38$, although equipped with a large error,
is too low. On the other hand, our analysis does confirm the value
$\beta_\rho(0)=72$ given in \cite{Pennington-Annals} and adopted in
\cite{ACGL}. As noted in \cite{CGL}, this result is perfectly consistent
with the theoretical prediction for the combination $2a_0^0-5a_0^2$ of
scattering lengths that is relevant for the Olsson sum rule.

The hatched bands that extend down to 1.42 GeV indicate the Regge
representation labeled CFD in \cite{GarciaMartin:2011cn}, which corresponds
to $\beta_P(0)=98.7 \pm 1.6$, $\beta_f(0)=31.6 \pm 1.9$,
$\beta_\rho(0)=58.4 \pm 5.5$ and $\beta_e(0)=3.15\pm 7.9$. The
corresponding trajectories, as well as the results for the residues
obtained in the previous works of the same authors, are quoted in appendix
\ref{app:review}.  We note that, although the values of the $f$ and $\rho$
residues are smaller than our results in (\ref{betaPf PDG Regge}) and
(\ref{eq:betarho0 and r1}), the Regge representation for the total cross
section with $I_t=0$ and $I_t=1$ given in \cite{GarciaMartin:2011cn} is
consistent with ours. The reason is the presence of the pre-asymptotic
terms in our amplitudes, which are negative and in effect reduce the
contributions from the dominant residues.  The larger intercepts of the $f$
and $\rho$ trajectories adopted in \cite{GarciaMartin:2011cn} also count:
at 2 GeV, for instance, with the value $\alpha_ \rho(0)=0.53$ adopted in
\cite{GarciaMartin:2011cn}, the factor $s^{\alpha_\rho}$ is larger by 10\%
than the same factor calculated with our value (\ref{alpha0frho}). As
concerns the overall behaviour of the cross sections above 1.7 GeV, the
representation given in \cite{GarciaMartin:2011cn} is consistent with ours.

Factorization is exploited also in a very recent analysis\footnote{
{\it Footnote added in August 2012.} The comments made in this
paragraph concern a preliminary version of reference \cite{Halzen}, available at
{\tt http://arxiv.org/abs/1110.1479v1}. In the published version of that paper,
which recently appeared in Phys. Rev. D85 (2012) 074020, the deficiency is
taken care of. The new values for the residues $Y_{f\pi\pi} = (11.5 \pm
0.9) {\rm mb}$ and $Y_{\rho\pi\pi} = (19\pm 5){\rm mb}$   are consistent
with our estimates.} of the hadronic
cross sections by Halzen et al.\,\cite{Halzen}, which leads to a value
$Z_{\pi \pi}=(12.7\pm 1.4)\,\mbox{mb}$ consistent with (\ref{Zpipi}), but a
higher $Y_{f\hspace{0.2mm}\pi\pi}=(16.0\pm 3.9)\,\mbox{mb}$ and a lower
$Y_{\rho\hspace{0.2mm}\pi\pi}=1.9\,^{+ 1.9}_{- 1.0}\,\mbox{mb}$, compared
with the numbers in (\ref{Ypipi}).  These discrepancies originate in the
fact that the authors of \cite{Halzen} use a small coupling $Y_{f \NN}$,
and a large coupling $Y_{\rho\NN}$ in the denominator of the factorization
relations (\ref{Ypipi}).  Actually, the values of $Y_{1\,pp}$ and
$Y_{2\,pp}$ obtained in \cite{Halzen} by independent fits of $pp$ forward
scattering data, including the total cross section at 7 TeV measured
recently at LHC, are not very different from those given in PDG 2011 and
quoted in equation (\ref{Y12}). But the individual contributions of the
trajectories are then separated assuming the equalities $Y_{f
  \NN}=Y_{a_2\NN}$ and $Y_{\rho\NN}=Y_{\omega\NN}$.  The values reported in
equation (\ref{YNN}) and the discussion below it show that this assumption
is strongly violated. As seen from Figs. 4a, 4b and \ref{fig:betarho}, the
unusual Regge picture of the $\pi\pi$ cross sections proposed in
\cite{Halzen} is in conflict with our analysis.

Finally, we consider the so-called exchange degeneracy, which requires
$\alpha_f(t)=\alpha_\rho(t)$ and $\beta_f(t)= 3/2\, \beta_\rho(t)$. In the
Lovelace-Shapiro-Veneziano model \cite{LSV}, these relations indeed follow
from the absence of exotic resonances. In contrast to other analyses,
\cite{ACGL} for instance, we do not assume exchange degeneracy to hold, and
it turns out that our results for $\beta_f(0)$ and $\beta_\rho(0)$ given in
(\ref{betaPf PDG Regge}) and (\ref{eq:betarho0 and r1}), respectively, do
not have this property. However, as emphasized below equation (\ref{betaPf
  PDG Regge}), only the sum of the contributions from Pomeron and $f$
counts, so that the value of $\beta_f(0)$ is sensitive to the chosen
parametrization. To illustrate this point, we mention that the
parametrization of the total $NN$ and $\pi N$ cross section adopted by PDG
before 2002 \cite{PDG-2000} leads via factorization to the values
$\beta_P(0) = 57 \pm 1$, $\beta_f(0) = 111 \pm 2$, $\beta_\rho(0) = 82 \pm
15$, which happen to be consistent with exchange degeneracy. We conclude
that, within our framework, the residue $\beta_f(t)$ is too sensitive to
the parametrization used for the Pomeron to give an unambiguous answer to
the question of whether or not the residues of $\rho$ and $f$ are
approximately exchange degenerate.

\section{\boldmath Regge analysis for $t\neq0$}\label{sec:nonzerot} 
\subsection{\boldmath $t$-dependence of the
  residues}\label{sec:t-dependence} 
The status of the Regge representation for amplitudes at nonzero momentum
transfer is more uncertain than that for the total cross sections. Several
global Regge fits of data on soft hadronic processes, available in the
early literature \cite{Barger}, \cite{Rarita}, \cite{Collins} are in rough
agreement, but disagree on details. The absorption corrections, which are
large for nonforward scattering, are expected to change the residues of the
pure Regge poles. Since these corrections are in general process-dependent,
they can violate factorization \cite{Pennington-Annals}, \cite{Kane-Seidl}.
We mention also the more recent analysis reported in \cite{Cudell2006}.
However, a systematic treatment of all the data sets for $t\neq 0$ is not
yet available.
 
The representations (\ref{Regge ImT0t}), (\ref{Regge ImT1t}), (\ref{Regge
  ImT2t}) leave the $t$-depen\-dence of the Regge representation for the
imaginary parts open. In order to complete our parametrization, we need to
specify the dependence of the residues $\beta_P(t)$, $\beta_f(t)$,
$\beta_\rho(t)$, $\beta_e(t)$ on the variable $t$. It is convenient to
factor out the value at $t=0$: \be\label{eq:profile}
\beta_k(t)=\beta_k(0)\,\betabar_k(t)\,, \quad\quad\quad k= P, f, \rho,
e.\ee We refer to the functions $\betabar_k(t)$ as {\it profiles}.

In the framework of the Roy equations, we are interested in the range
$-32M_\pi^2<t<4M_\pi^2$, where Mandelstam analyticity implies that fixed
$t$ dispersion relations hold in the interval
$-4M_\pi^2<s<68M_\pi^2\simeq(1.15\,\mbox{GeV})^2$. Numerically, the range
of interest for the variable $t$ corresponds to
$-0.62\,\mbox{GeV}^2<t<0.08\,\mbox{GeV}^2$.

In the following, we exploit the fact that the sum rules strongly correlate
the $t$-dependence of the Regge residues with the $t$-depen\-den\-ce of the
partial waves. In fact, these sum rules can be solved for $\betabar_P(t)$,
$\betabar_\rho(t)$, $\betabar_e(t)$. The results are of the same accuracy
as those for the residues at $t=0$, because the uncertainties in the
$t$-dependence originate in the same sources.

The fourth profile, $\betabar_f(t)$, cannot be determined in this way,
because the sum rules only concern the sum of Pome\-ron and $f$ and do not
allow us to clearly separate the two contributions. For the same reason,
however, the specific form used for $\betabar_f(t)$ is not crucial.

The discussion below equation (\ref{linear}) imposes an important
constraint on $\betabar_f(t)$. The $f$ is even under charge conjugation,
$\tau_f=1$. Hence the absence of ghosts implies that $\betabar_f(t)$ must
vanish at the place where the trajectory $\alpha_f(t)$ passes through zero.
This happens around $t\simeq-0.6\,\mbox{GeV}^2$, that is, within the
$t$-interval of interest, close to the lower end. The condition leaves
little freedom in the $t$-dependence of this residue. We neglect the
curvature and approximate it with a straight line,
$\betabar_f(t)\propto\alpha_f(t)$, so that the profile is given by:
\be\label{eq:betaf}\betabar_f(t) = 1+b_f\, t\,,\quad\quad
b_f=\frac{\alpha'_f(0)}{\alpha_f(0)}\fs\ee
\begin{figure}[thb]
\begin{center}
\includegraphics[width=8.5cm]{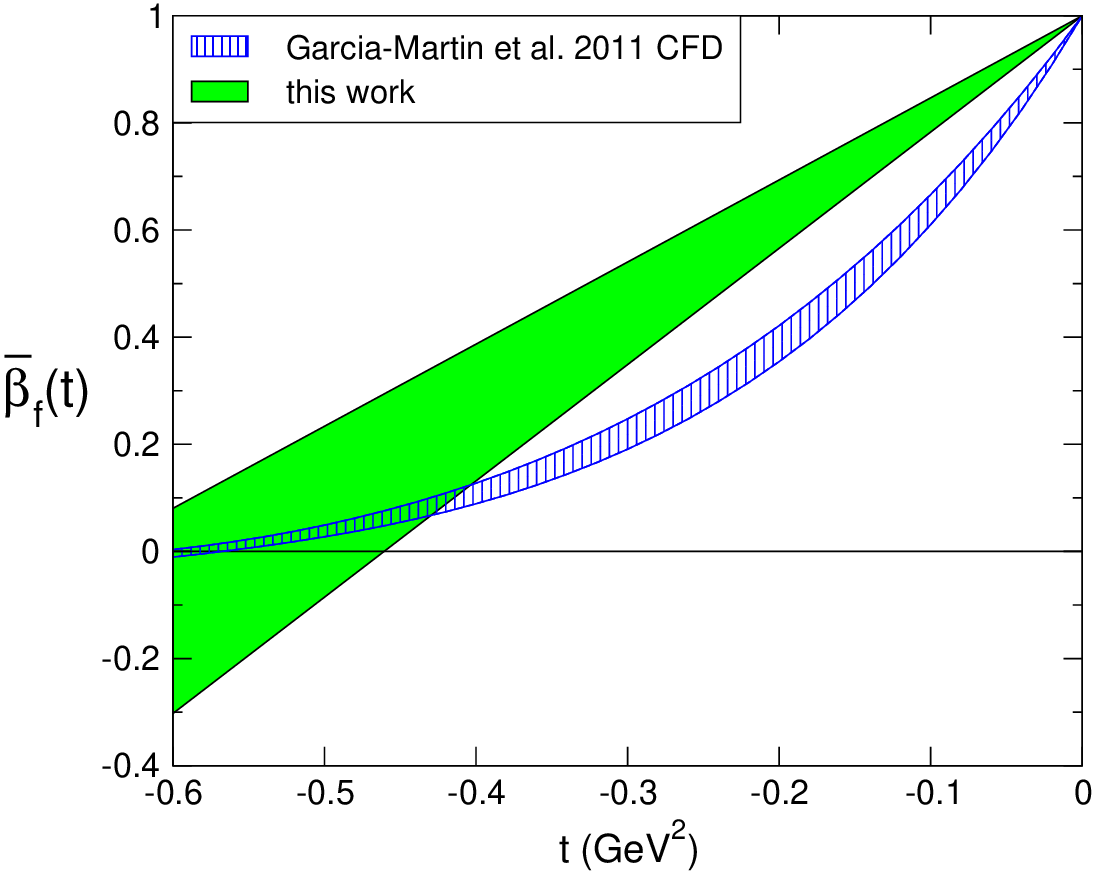}
\caption{\label{fig:profilebetaf} Profile
  of the $f$ residue.
  }  \end{center} 
\end{figure}
Fig.~\ref{fig:profilebetaf} visualizes this approximation: the area spanned
by the full band corresponds to the uncertainty range in (\ref{alpha0frho})
and (\ref{alpha1frho}).  We can allow for a curvature term -- the sum rules
then yield a somewhat different profile for the Pomeron. We have checked
that, as long as the curvature is not larger than in the other residues, it
barely affects our results.  For comparison, the figure also shows the
profile of the analogous term in the Regge parametrization of
\cite{GarciaMartin:2011cn}, where the symbol P$^\prime$ is used instead of
f.

Concerning the approximation used for the remaining three profiles, the
situation is the same as with the parametrization of the phase shifts used
when solving the Roy equations: since the sum rules allow us to calculate
these, the parametrization of the profiles is a mere matter of convenience.
We find that, on the domain of interest to us, $-0.62\,\mbox{GeV}^2 $ $\le
t\le 0$, a parabolic parametrization is adequate: \bea\label{eq:parabola}
\betabar_k(t)\al=\al 1+ b_k\, t +c_k\, t^2 \,, \quad\quad\quad k= P, f,
\rho, e. \eea The accuracy to which the sum rules and duality conditions
are obeyed does improve a little if we allow for cubic or quartic terms,
but the results can barely be distinguished from those obtained with
(\ref{eq:parabola}).  In this notation, the representation (\ref{eq:betaf})
for $\betabar_f(t)$ amounts to $c_f=0$ and the numerical values for
intercept and slope of the $f$-trajectory in (\ref{alpha0frho}) and
(\ref{alpha1frho}) imply \be b_f=1.7\pm0.3\;\mbox{GeV}^{-2}\fs\ee

As we will see, the central solution for $\betabar_\rho(t)$ also contains a
zero, not far from the one in $\betabar_f(t)$. This phenomenon, however, is
not related to the absence of a ghost. Since the $\rho$ is odd under charge
conjugation, $\tau_\rho=-1$, the absence of ghosts calls for a zero in
$\betabar_\rho(t)$ at the value of $t$ where $\alpha_\rho(t)$ passes
through $-1$. As this happens near $t\simeq-1.6\,\mbox{GeV}^2$, far outside
the range if interest, the condition that $\betabar_\rho(t)$ must vanish
there does not impose a significant constraint on our analysis
(incidentally, the parabolic parametrization (\ref{eq:parabola}) of our
central solution does contain a second zero at large negative values of
$t$, but it occurs outside the region where that parametrization is
relevant).
 
\subsection{Profile of the Pomeron}\label{sec:profile pomeron}
\begin{figure}[thb]
\begin{center} 
\includegraphics[width=9cm]{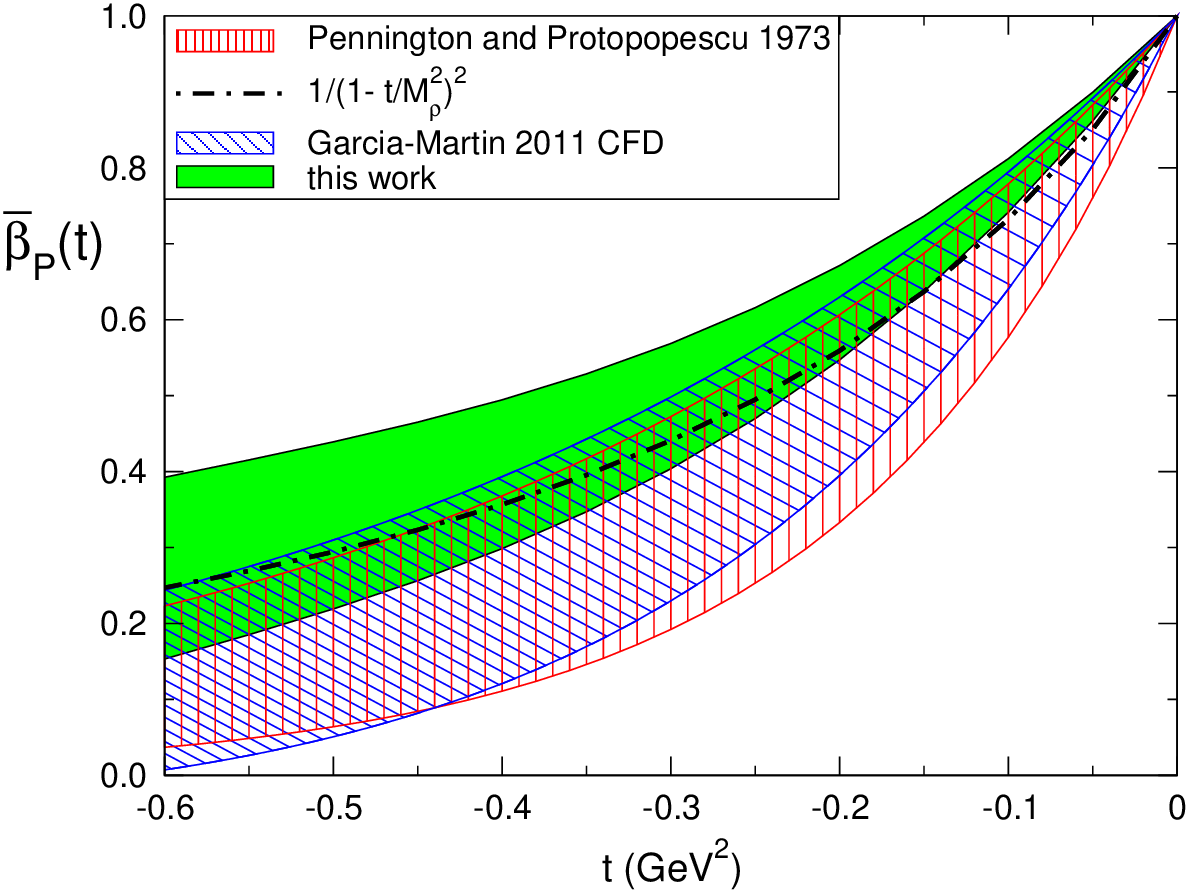} 
\caption{\label{fig:profilebetaP} Profile of the Pomeron residue.}
   \end{center}   
\end{figure}

The green band in Fig.\,\ref{fig:profilebetaP} shows the outcome of our
calculation for the $t$-dependence of the Pomeron residue. Numerically, the
upper and lower edges of the band can be represented as $\betabar_P(t)\pm
\delta\betabar_P(t)$, with\footnote{Throughout, the coefficients occurring
  in the polynomial representations of the residues as well as the slope
  parameters $b_P,b_f,b_\rho,b_e$ and $\beta_e'(0)$ are given in GeV
  units.}
\bea\label{eq:betaP num} \betabar_P(t)\al=\al 1 + 2.24\,  t + 1.76\, t^2\,,\\
\delta\betabar_P(t)\al=\al 0.36\, t + 0.26\, t^2\,.\nonumber\eea For the
slope at $t=0$, the calculation yields:\footnote{The value given for $b_P$
  agrees with $\betabar_P'(0)\pm \delta \betabar_P'(0)$ only roughly: the
  former involves the square root of a sum of squares of the derivative
  while the latter is obtained by taking the derivative of a square root of
  a sum of squares.}  \be\label{eq:bP}b_P=2.5\pm 0.4\;\mbox{GeV}^{-2}\fs\ee

The figure shows that the sum rule analysis confirms the description of the
Pomeron profile in terms of the e.m.\,form factor of the pion,
\begin{equation}\label{eq:form factor}
\betabar_P(t)=F_\pi(t)^2\,,
\end{equation}
which was introduced by Donnachie and Landshoff
\cite{Donnachie-Landshoff-1984} and adopted in
\cite{Nachtmann,DDLN-book,BoSo}: the dash-dotted 
line shows the profile obtained from this formula with vector meson
dominance, which predicts $F_\pi(t)=1/(1-t/M^2)$, $M=M_\rho$. A
parametrization of this form is used also by Cudell et
al.\,\cite{Cudell2006}, but the resulting picture is qualitatively
different from the one in \cite{Donnachie-Landshoff-1984}: the estimate for
the parameter $M$ resulting from the fits constructed in \cite{Cudell2006}
is more than twice as large as $M_\rho$, so that the Pomeron profile falls
off much less rapidly with $|t|$ than the square of the e.m.\,form factor.

Results for the structure of the diffraction peak in elastic $\pi\pi$
scattering would be of considerable interest, in particular in view of
possible direct pion-pion collisions at high energies in the near future
(for a recent discussion of these prospects see \cite{Halzen}).  The TOTEM
experiment at the LHC \cite{Totem}, which reaches down to very small
momentum transfer in elastic $NN$ scattering, renewed theoretical interest
in topics like the ratio of the real to the imaginary part of the forward
scattering amplitude, the comparison of the total elastic and inelastic
cross sections, the picture that represents hadrons as black disks,
etc.\,\cite{Halzen,Block:2011uy,Block-Halzen-black-disk}. However, for the
reasons given in section \ref{sec:ImT0t}, these questions are beyond the
reach of our framework.

 \subsection{\boldmath Profile of the $\rho$ residue}\label{sec:profile
   rho} 
 For the profile of the $\rho$, the outcome of our analysis is well
 described by the numerical representation $\betabar_\rho(t) \pm
 \delta\betabar_\rho(t)$, with
 \bea\label{eq:betarho num} 
\betabar_\rho(t)\al=\al 1 + 3.00\, t + 1.37\, t^2\,,\\
 \delta\betabar_\rho(t)\al=\al -0.44\, t + 0.21\, t^2\,.\nonumber\eea 
In this case, the value of the slope is given by 
\be \label{eq:brho}
b_\rho=3.4\pm 0.5\;\mbox{GeV}^{-2}\fs
\ee 
Our result for this profile is  indicated by
\begin{figure}[thb] 
\begin{center} 
\includegraphics[width=8cm]{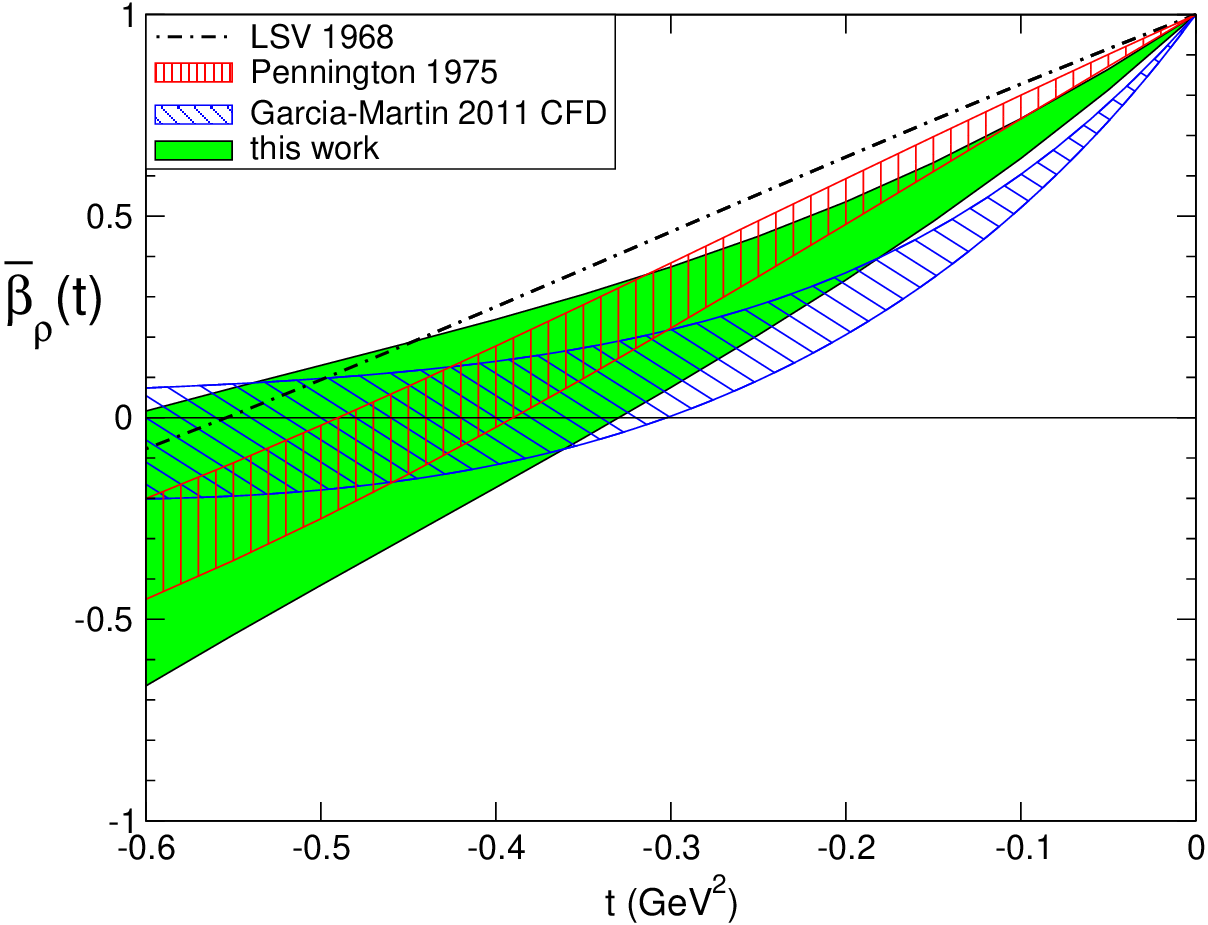} 
\caption{\label{fig:profilebetarho} Profile of the $\rho$ residue,
  } \end{center}   
\end{figure} 
the green band in Fig.~\ref{fig:profilebetarho}.  The residue has a zero at
$t=-0.42^{+0.09}_{-0.24}\,\mbox{GeV}^2$.  The uncertainty band partly
overlaps with the one describing the $f$ residue, which is shown in
Fig.~\ref{fig:profilebetaf}. Despite the different origin of the two zeros
emphasized at the end of section \ref{sec:t-dependence}, the two profiles
are consistent with exchange degeneracy.

The recent Regge analysis of $pp$ and $\pi p$ data in \cite{Cudell2006}
leads to a rather different picture for the $t$-dependence of the residues.
The authors assume factorization, but neglect the difference between the
spin-flip and non-flip amplitudes and do not consider data on the charge
exchange process.  Their representation for $\beta_\rho(t)$ contains a zero
at $t=-0.153\pm0.003\,\mbox{GeV}^2$ or at $t=-0.148\pm
0.003\,\mbox{GeV}^2,$ depending on whether or not a hard Pomeron component
is included. We recall that a zero at small momentum transfer was assumed
also in other works, in order to account for the so-called cross-over
phenomenon, {\em i.e.} the equality of the $\pi^+p$ and $\pi^-p$
differential cross sections around $t\sim -0.2\,\mbox{GeV}^2$. The Regge
representation of \cite{PY2003}, for instance, contains a zero at
$t=-0.22\pm0.02\,\mbox{GeV}^2$. The analysis of the $\rho$-dominated
charge-exchange process $\pi^-p\to\pi^0 n$ in \cite{Huang}, however, does
not confirm the presence of such a zero. It appears that the cross-over can
naturally be explained in the framework of a Regge model with strong
absorption for $\pi N$ scattering \cite{Pennington-Annals,Kane-Seidl},
without the $\rho\,\pi\pi$ residue passing through zero there. At any rate,
our results are not consistent with the presence of a zero in
$\beta_\rho(t)$ in the vicinity of the cross-over point.

Fig.~\ref{fig:profilebetarho} contains three other results for the $\rho$
profile.  The one obtained with the LSV model \cite{LSV} is shown as a
dash-dotted line. The zero predicted by this model is compatible with our
result.  A similar zero position, at $-0.44 \pm 0.05\, \mbox{GeV}^2$, was
found in the analysis of Pennington \cite{Pennington-Annals}, which was
also based on the Roy equations and a set of sum rules (the resulting
representation for $\beta_\rho(t)$ was used as an input in \cite{ACGL}).
The figure shows that the corresponding band is in very good agreement with
the outcome of the present analysis. The figure also shows the most recent
version of the $\rho$-profile used by the Madrid-Krakow collaboration
\cite{GarciaMartin:2011cn}. Despite some qualitative differences (it is
steeper close to the origin and has a larger curvature), their result is in
fair agreement with ours.

The uncertainties in $\betabar_\rho(t)$ rapidly grow with $|t|$: note
however that what counts in the applications is not $\betabar_\rho(t)$, but
the imaginary part of the amplitude which is proportional to
$\betabar_\rho(t) (s/s_1)^{\,\alpha_\rho(t)}$. At an energy of 2 GeV, the
term $(s/s_1)^{\alpha_\rho(t)}$ drops roughly by a factor of 2 as $t$ runs
from the upper to the lower end of the interval relevant for the partial
wave projections, $0\geq t\gsim -0.6 \,\mbox{GeV}^2 $.

\subsection{\boldmath Effective residue in the channel with
  $I_t=2$}\label{sec:profile e} 
   
\begin{figure}[thb]
\begin{center}
\includegraphics[width=8cm]{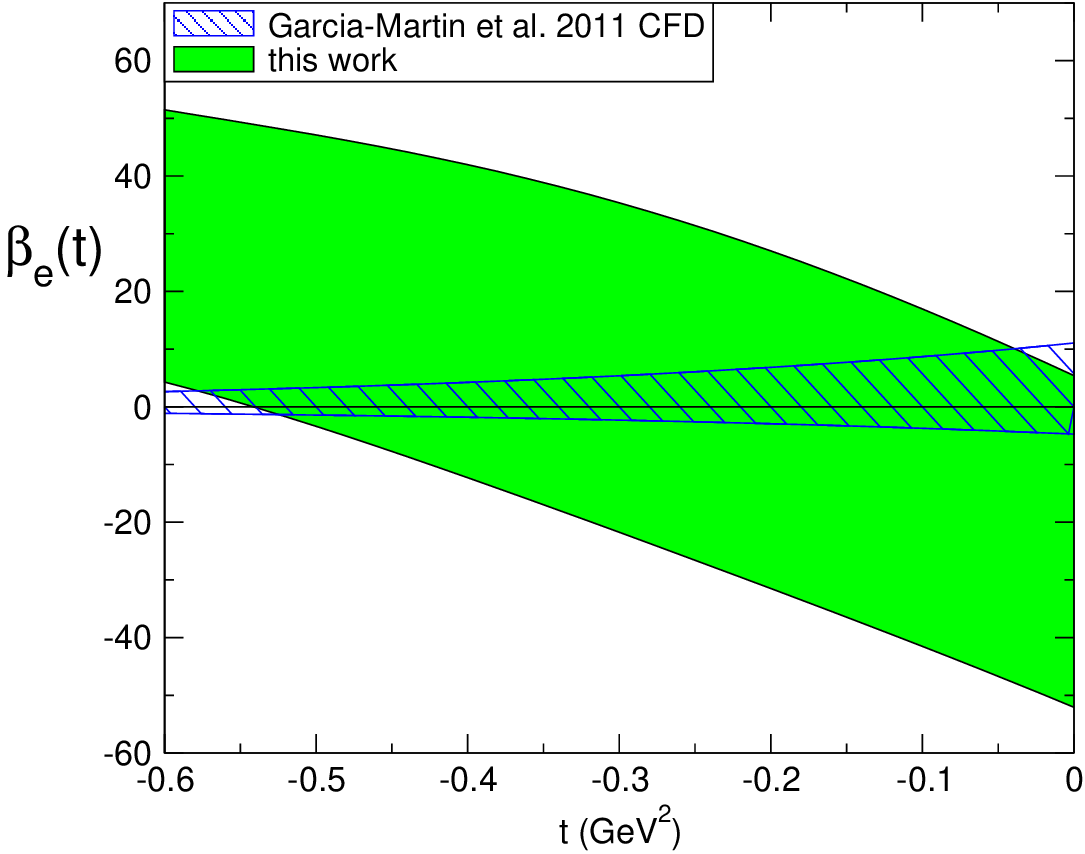} 
\caption{\label{fig:betae}$t$-dependence of the effective residue with
  $I_t=2$ } \end{center}  
\end{figure}

Within errors, the effective residue in the exotic channel, $\beta_e(t)$,
vanishes at $t=0$. The corresponding profile $\betabar_e(t)$ does therefore
not represent a meaningful quantity. Instead, Fig.\,\ref{fig:betae} shows
the $t$-dependence of the residue itself.  The upper and lower edges of the
band shown can be represented as $\beta_e(t)\pm\delta \beta_e(t)$, with
\bea\label{eq:betae num} \beta_e(t)\al=\al -24 - 139\, t - 94\,  t^2\,,\\
\delta\beta_e(t)\al=\al 29 - 2\, t - 19\, t^2\,,\nonumber\eea and for the
slope at $t=0$, we find \be\label{eq:be}\beta_e'(0)=-120\pm
40\;\mbox{GeV}^{-2}\fs\ee
In the entire range considered here, this residue is small.  The
uncertainty is approximately independent of $t$.  The value at $t=0$ is
related to the size of the total cross section in the channel with $I_t=2$,
which is shown in Fig.\,4c. At 2 GeV, this cross section amounts to
$\sigma^{(2)}(s)=-2\pm3$ mb, small compared to the leading term from the
Pomeron. At negative values of $t$, the residue passes through zero and
then becomes positive.  In view of the fact that the amplitude is
proportional to $\beta_e(t)\,s^{\alpha_e(t)}$, the uncertainty even
decreases a little towards the lower end of the range of $t$-values where
our analysis applies.

Fig.\,\ref{fig:betae} also shows the residue used in the analysis of
\cite{KPY2008,GarciaMartin:2011cn}, where the $t$-dependence of the
residues is taken from \cite{Froggatt:1977hu,Rarita}.  In the latter
references, the Regge parameters for $\pi \pi$ scattering are determined on
the basis of factorization. There is, however, barely any direct
information on the component of the Regge representation with $I_t=2$. For
what concerns $\beta_e(t)$, the expectation expressed in
\cite{KPY2008,GarciaMartin:2011cn} that their Regge parametrization
describes experimental data in the region $1.42 \,\mbox{GeV}\leq \sqrt{s}
\leq 20 $ GeV, $ -0.4$ GeV $ \leq t \leq 4 M_\pi^2$ merely represents an
educated guess.  Our quantitative analysis does confirm that $\beta_e(t)$
is small, but it does not support the uncertainties attached to their
parametrization, which in our opinion are grossly underestimated.
 
\section{Consistency checks\label{sec:consistency}} 
The present section concerns the internal consistency of our results: we
compare the contributions to the sum rule and duality relations arising
from the low energy region, where we are using the partial wave
representation, with those from high energies, for which we use the Regge
representation. For the relations to be approximately satisfied, the two
contributions must approximately cancel.
\subsection{\label{sec:Olsson check}Checking the Olsson sum rule} 
Consider first the Olsson sum rule. The contribution from the integral over
the partial waves and from the subtraction constants was analyzed in
section \ref{sec:Olsson}, with the result
$\OPWbar=-0.110^{+0.022}_{-0.015}$. Evaluating the remainder with the Regge
represention constructed in the preceding sections, we obtain
$\mbox{O}_{\indR}=0.106^{+0.016}_{-0.018}$. The two terms thus indeed
cancel within errors. This does not come as a surprise, because the Olsson
sum rule occurs among the conditions imposed when constructing the Regge
representation. What the agreement mainly checks is that the Olsson sum
rule did receive enough weight among the many constraints considered, so
that the outcome for the Regge representation does respect it.
\subsection{\label{sec:S check}\boldmath Verifying the sum rule $S(t)=0$} 
\begin{figure}[thb] 
\begin{center} 
\includegraphics[width=8cm]{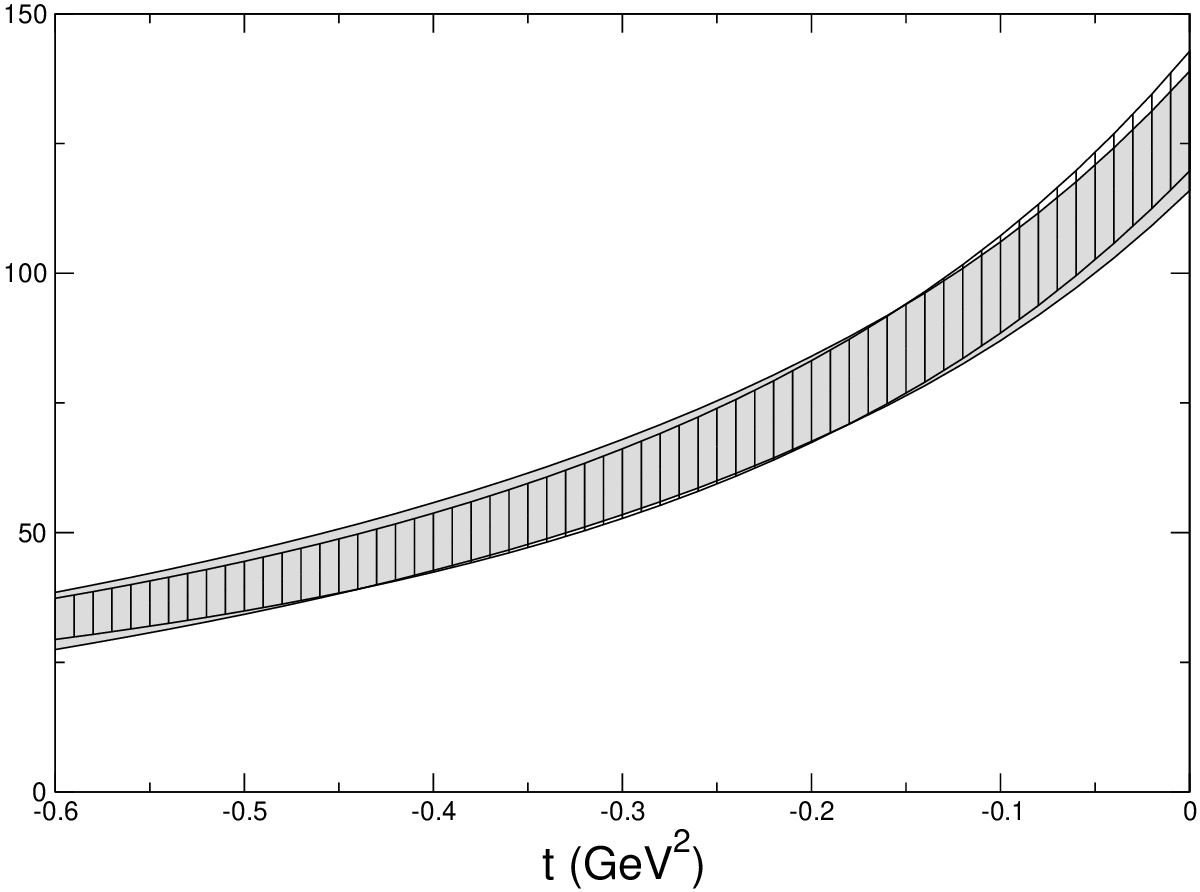} 
\caption{\label{fig:sum rule S} Validity of the sum rule $S(t)=0$.}
\end{center}   
\end{figure}
In Fig.\,\ref{fig:sum rule S}, the low energy contributions to the function
$S(t)$ defined in (\ref{St}) are shown as a grey band (numerical values in
GeV units). To check whether the high energy contributions do approximately
cancel these, we flip their sign, so that instead of canceling, the two
contributions would become equal if the sum rule was obeyed exactly. The
result is indicated by a hatched band. Evidently, the sum rule is very well
obeyed within errors.

\subsection{\label{sec:Ck check}Checking crossing symmetry} 
As discussed in section \ref{sec:sum rules}, the fact that the crossed
channels of elastic $\pi\pi$ scattering are identical imposes strong
constraints on the scattering amplitude. In particular, these imply a set
of sum rules of the form $C_k(t)=0$, where $k=0,1,2$ and $t$ is a free
parameter (see appendix \ref{app:SR}). In Fig.\,\ref{fig:sum rule DSR}, the
red and blue bands for $C_0(t)$ and $C_1(t)$ show the partial wave
contributions to the integrals occurring in the definition (\ref{eq:Ck}) of
these functions, while the hatched ones indicate the corresponding Regge
contributions, changed in sign (numerical values in GeV units). In order to
disentangle the bands for $C_1(t)$ and $C_2(t)$, the sign of the low energy
contributions to $C_2(t)$ is flipped instead of the one from the Regge
region. The fact that the partial wave and Regge bands are consistent with
one another demonstrates that the sum rules $C_k(t)=0$ are well obeyed
within errors. 
\begin{figure}[thb] 
\begin{center} 
\includegraphics[width=8cm]{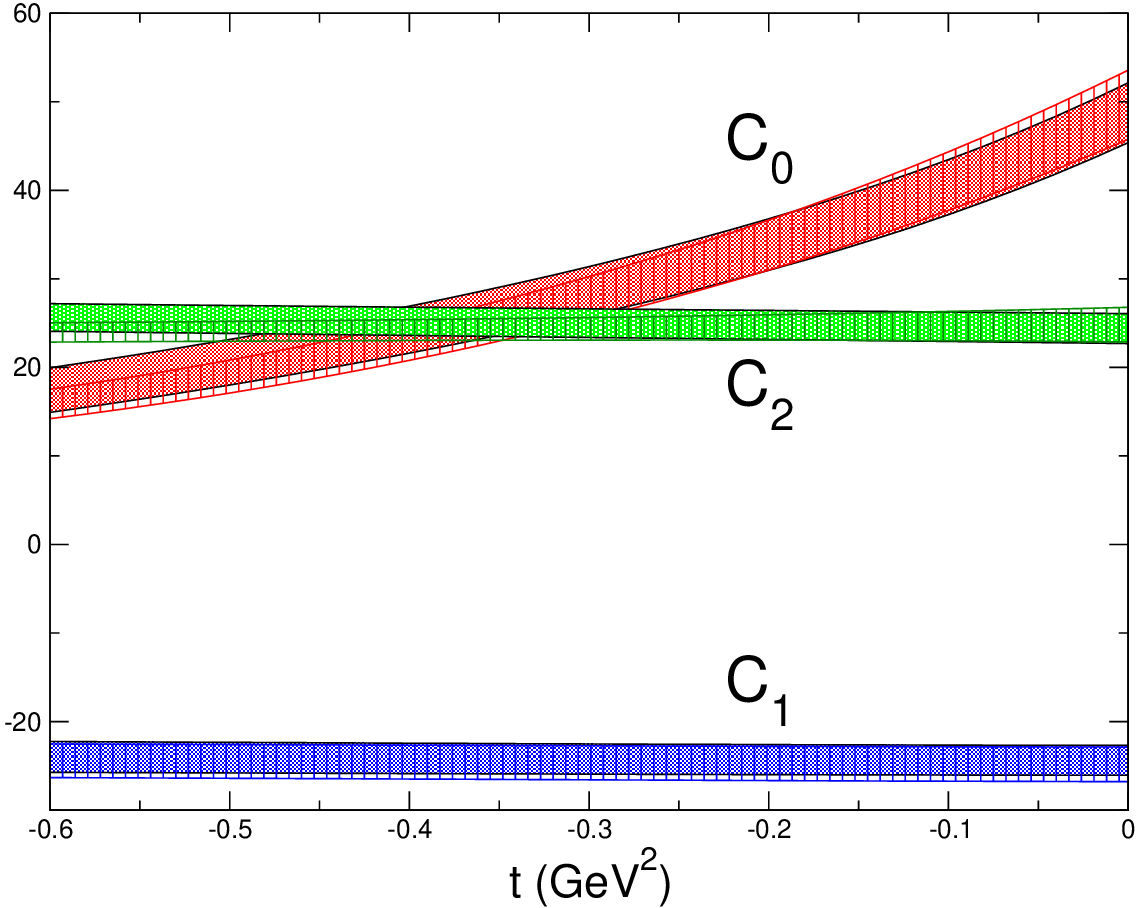} 
\caption{\label{fig:sum rule DSR} Validity of the crossing symmetry sum
  rules.} \end{center}   
\end{figure} 
\subsection{\label{sec:duality check}Checking duality} 
\begin{figure}[thb] 
\begin{center} 
\includegraphics[width=8cm]{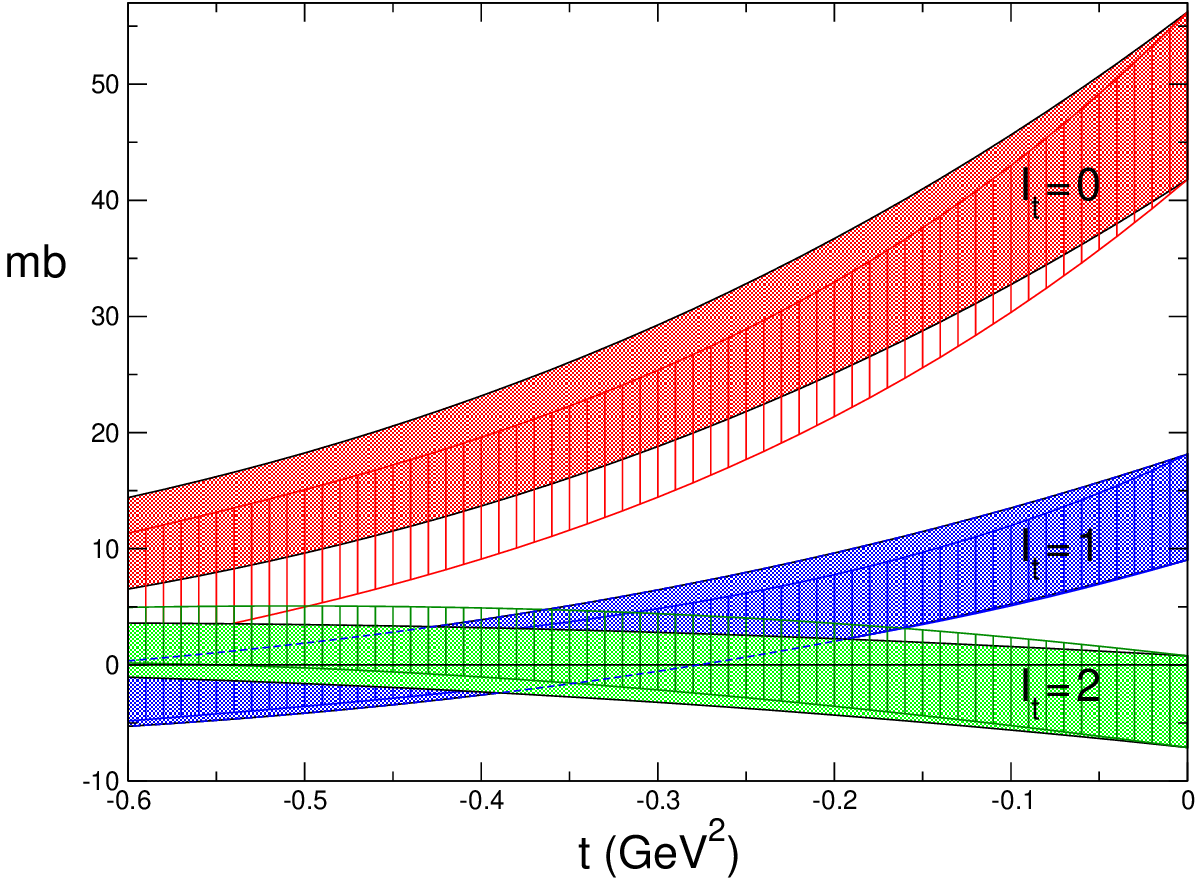} 
\caption{\label{fig:duality}Duality.  } \end{center}   
\end{figure} 
In the construction of our Regge representation, we made use of a
semi-local form of duality: we required the average over the interval
$1.5\,\mbox{GeV}\leq \sqrt{s}\leq 1.9\,\mbox{GeV}$ of the partial wave
representation for the total cross sections to agree with the average of
the Regge representation over the same interval. Moreover, we imposed the
same condition also at nonzero values of $t$, replacing the cross sections
by the quantities $\rho(s)\,\mbox{Im}T^{(I_t)}(s,t)$. In
Fig.\,\ref{fig:duality}, the full bands indicate the averages over the
partial wave representations for these quantities, while the hatched bands
represent the averages over the Regge representations (numerical values in
mb). The figure shows that the two averages agree: our partial wave and
Regge representations do satisfy the semi-local form of duality invoked in
the present paper. In fact, since we required the duality conditions for
the total cross sections to be strictly obeyed, the full and hatched bands
strictly agree at $t=0$.
 
\section{Summary and conclusions}\label{sec:summary}

In the present paper we have analyzed the high-energy $\pi \pi$ scattering
amplitude. Our main motivation is not high energy {\em per se} but rather
that this part of the amplitude plays a role in the Roy equations, even
when these are used to determine the amplitude at low energy. Close to
threshold the relevance of high energy is rather limited, but as one moves
up, towards 1 GeV and higher, a precise determination of the high-energy
contribution becomes important.

We have adopted a Regge parametrization to describe the $\pi \pi$
scattering amplitude at high energy. Some of its parameters are well known
like, {\em e.g.} the Regge trajectories. Moreover, some Regge residues can
be determined from high-energy $\pi N$ and $NN$ scattering data
\cite{DDLN-book,Cudell2002,PDG-Regge,Cudell2006} assuming factorization.
Several other parameters, however, are poorly or not at all determined. Aim
of the present paper was to pin these down. In order to do this we have
worked under the assumption that the partial wave and Regge representations
can be joined smoothly in the region between 1.7 and 2 GeV -- not only at
$t=0$ but also as a function of $t$, down to about $-0.6$ GeV$^2$. In this
setting we have imposed a set of sum rules which follow from general
principles, like crossing symmetry.

While 2 GeV can hardly be considered asymptotic, it has been already
observed in the literature (see \cite{DDLN-book} and references therein)
that, at least on average, duality between the partial wave and Regge
representations works well even at energies that low. In order to both
partly test this assumption and make our results less dependent on it, we
have supplemented the standard Regge parametrization with pre-asymptotic
terms (terms that are suppressed at asymptotic energies with respect to the
dominating Regge contributions) and have determined them by requiring that
a smooth transition between the partial wave and Regge representations does
take place.

For the partial wave representation, we use a simple parametrization which
approximately describes the data between $1.15$ and $2$ GeV. Below $1.15$
GeV, we solve the Roy equations -- in this region, the exact form of the
parametrization becomes irrelevant. In order to solve the Roy equations, we
need an input for the high-energy part, which, on the other hand, is what
we aim to determine. In summary, we have simultaneously solved a number of
constraints -- Roy equations, duality and sum rules -- which impose
relations among the different partial waves at low energy and the
high-energy amplitude. This is not an easy task, and to perform it in a way
which is as practical as possible we have adopted an iterative
scheme:
\begin{enumerate} 
\item take an input for the Regge parameters; 
\item solve the Roy equations for all partial waves up to and including the
  $G$ waves; 
\item impose the duality and crossing symmetry constraints in order to 
  determine the Regge parameters; 
\item if the change in the Regge parameters is significant, go back to 1. 
\end{enumerate} 
The outcome of this analysis is an improved Regge para\-me\-trization of
the $\pi \pi$ scattering amplitude at high energy, which: 
\begin{enumerate} 
\item is consistent with the high-energy data on hadron-hadron scattering
  under the factorization hypothesis;  
\item satisfies the Olsson sum rule and its extension to $t \neq 0$, as
  well as a set of other sum rules which follow from crossing symmetry;
\item smoothly joins the partial wave representation in the region just 
  below 2 GeV, also as far as the $t$ dependence is concerned. 
\end{enumerate}

A summary of the results of the present analysis and of the parameters
taken as input is presented in Table~\ref{tab:finalparameters}.
\begin{table}[b] 
 \begin{tabular}{|c|c|l|}\hline
{  Parameter} & {  Value } &{  Source }\\\hline
$\alpha_P(0)$ &1& PDG 2011\\\hline
$\alpha_f(0)$ &$0.54 \pm 0.05$& our estimate\\\hline
$\alpha_\rho(0)$ & $0.45\pm 0.02$ &our estimate \\\hline
$\alpha_e(0)$ & 0 &our estimate \\\hline
$\alpha_P'(0)$   & $0.25\pm 0.05 \;\mbox{GeV}^{-2}\rule{0em}{0.9em}$  & diffraction peak \\\hline
$\alpha_f'(0)$ & $0.90\pm 0.05 \;\mbox{GeV}^{-2}\rule{0em}{0.9em}$ & Chew-Frautschi plot\\\hline
$\alpha_\rho'(0)$ &$0.91\pm 0.02 \;\mbox{GeV}^{-2}\rule{0em}{0.9em}$ &Chew-Frautschi plot\\\hline
$\alpha_e'(0)$ &$0.5\pm 0.1 \;\mbox{GeV}^{-2}\rule{0em}{0.9em}$ &our estimate\\\hline
$\bar{B}$& $0.025\pm  0.001$ \rule{0em}{0.95em}& PDG 2011 \\\hline
$\sqrt{s_0}$ \rule[0cm]{0cm}{0.28cm}&$ 5.38\pm 0.50\;\mbox{GeV}$ & PDG 2011 \\\hline
$\beta_P(0)$ &$94 \pm 1$ &factorization \\\hline
$\beta_f(0)$ &$69 \pm 2$ & factorization \\\hline 
$\betabar_f(t)$\rule[-0.1cm]{0cm}{0.43cm}  &equation (\ref{eq:betaf}) & linearity, no ghosts  \\\hline
\hline
$\beta_\rho(0)$\rule[-0.15cm]{0cm}{0.47cm} & $84^{+13}_{-16}$& Olsson s.r., duality\\\hline
$\beta_e(0)$ &$-24\pm 29$ & duality \\\hline
$p_1$   &$-0.8\pm 0.6$&  duality \\\hline
$r_1$   \rule[-0.15cm]{0cm}{0.47cm}&$-0.8^{+0.8}_{-0.7}$& Olsson s.r., duality \\\hline
$b_P$& $2.5\pm 0.4\;\mbox{GeV}^{-2}\rule{0em}{0.9em}$&sum rules, duality\\\hline
$b_\rho$&$3.4\pm0.5\;\mbox{GeV}^{-2}\rule{0em}{0.9em}$&sum rules, duality\\\hline
$\beta_e'(0)$&$-120\pm 40\;\mbox{GeV}^{-2}\rule{0em}{0.9em}$&sum rules, duality\\\hline
$\betabar_P(t)$\rule[-0.1cm]{0cm}{0.43cm} & equation (\ref{eq:betaP num}) & sum rules, duality\\\hline
$\betabar_\rho(t)$\rule[-0.1cm]{0cm}{0.43cm}    &equation (\ref{eq:betarho num})&  sum rules, duality\\\hline
$\beta_e(t)$\rule[-0.1cm]{0cm}{0.43cm}    &equation (\ref{eq:betae num})&  sum rules, duality\\\hline
\end{tabular}
\caption{\label{tab:finalparameters} Regge representation of the scattering
  amplitude. The upper part lists the range used for the input parameters,
  while the lower part indicates the results of our calculation.}  
\end{table}
Besides the constants $\beta_\rho(0),p_1,r_1,\beta_e(0)$, which are
relevant for the total cross sections, our analysis also provides the
$t$-dependence of the residues, which we parametrize in terms of quadratic
polynomials. We do not repeat the explicit results for the profiles
$\betabar_X(t) \equiv \beta_X(t)/\beta_X(0)$ here. They can be found in the
equations listed in the table.  We stress that our method for  
determining the $t$-dependence of the residues is new and puts their
representation on a more solid basis.  

The main novelty of our analysis is twofold: First, the introduction of
pre-asymptotic terms, which we have shown to be needed in order to reach a
smooth transition between the Regge and partial wave representations,
especially in the channel with $I_t=0$.  Second, the determination of the
$\rho$ residue via the Olsson sum rule, which is much more robust than
factorization. These two elements explain some of the differences found
with earlier analyses. For a detailed comparison of our results for the
total cross sections, both with data and the literature, we refer the
reader to section \ref{sec:discussion sigmatot}.

As we have discussed in section \ref{sec:profile pomeron}, the outcome of
our calculation for the Pomeron profile is consistent with some of the
results available in the literature, but differs from others. Within the
accuracy of our calculation, this profile is well described by the square
of the pion electromagnetic form factor.

Our results for $\betabar_\rho(t)$ indicate that this profile has a zero in
the range $t=-0.42^{+0.09}_{-0.24}\,\mbox{GeV}^2$. This confirms the
results of Pennington \cite{Pennington-Annals}, which in fact were based on
a similar approach. On the other hand, as discussed in section
\ref{sec:profile rho}, the outcome of our calculation is not consistent
with parametrizations that explain the cross-over in $\pi N$ scattering in
terms of a zero in $\betabar_\rho(t)$.

As shown in section \ref{sec:profile e}, the sum rules studied in the
present paper also provide a good handle on the $t$-dependence of the Regge
amplitude with $I_t=2$. The main result here is that, on the interval
$0\geq t\gsim -0.6\, \mbox{GeV}^2$, this amplitude is small compared to the
contribution from the Pomeron. The size is comparable to the pre-asymptotic
terms in the non-exotic channels with $I_t=0,1$. As is the case with these,
the absolute size of the amplitude with $I_t=2$ is subject to large
uncertainties.

New solutions of the Roy equations for the $S$, $P$, $D$, $F$ and $G$ waves
based on this high-energy input have been derived and will be presented in
a forthcoming publication. These solutions extend up to 1.15 GeV and are
suitable to be used in dispersive analyses of other quantities and
processes, like the hadronic vacuum polarization contribution to
$(g-2)_\mu$, $\eta \to 3 \pi$, Kaon decays in multipion final states etc.

These extended Roy solutions also provide an improved basis for the
determination of the pole positions of resonances decaying into two pions,
like the $\sigma$ or the $\rho$. We emphasize that extrapolations based on
explicit parametrizations of the energy dependence, such as those in
\cite{Achasov:2010fh} or in \cite{Anisovich:2011rg}, cannot compete with
the analytic continuation provided by dispersion theory.  In the case of
resonances with a large width, the result for the pole position is subject
to large theoretical uncertainties associated with the choice of the
parametrization, even if those on the real axis are small
\cite{Caprini2008}.  The result for the mass and width of the $\sigma$
obtained by extrapolating the explicit parametrization in
\cite{Achasov:2010fh}, for instance, differs significantly from the pole
position calculated on the basis of the Roy equations \cite{CGL}, despite
the excellent agreement on the real axis. The dispersive method has the
advantage that it provides a complete error analysis which includes, in
particular, also the uncertainties associated with the presence of
inelastic channels.  The calculations performed in
\cite{GarciaMartin:2011jx,Moussallam:2011zg} corroborate this statement:
the result for the pole position agrees with \cite{CGL}, within the quoted
errors.

\vskip1cm 
\noindent {\bf Acknowledgments:} IC acknowledges support from CNCS,
Contract Idei Nr.464/2009. The Albert Einstein Center for Fundamental
Physics at the University of Bern is supported by the ``In\-nova\-tions-
und Kooperationsprojekt C-13'' of the ``Schwei\-zerische
Universit\"atskonferenz SUK/ CRUS''. This work was partially supported by
the Swiss National Science Foundation and by EU contract
MRTN-CT-2006-035482 (Flavianet).

\appendix 
\section{\boldmath Literature on the Regge parameters}\label{app:review}
Most of the literature on the Regge parameters deals with $NN$ and $\pi N$
scattering, for which fits of various experimental data sets were
performed. The parameters relevant for $\pi\pi$ scattering are in general
obtained only indirectly, using factorization
\cite{Rarita,Szczurek,Cudell2006,Halzen}.  Several modifications of the
standard expression (\ref{Regge}) were also considered in the literature:
for instance, an additional factor $\Gamma(\alpha(t))$ is often introduced
in the denominator of (\ref{Regge}) in order to remove the unphysical poles
due to $\sin \pi \alpha$.  Other factors dependent on the trajectory, which
remove unphysical poles, are adopted in \cite{Rarita},
\cite{PY2003,PY2004-Regge,KPY2006,KPY2008,GarciaMartin:2011cn},
\cite{Huang,Sibirtsev}, etc. 

In the present review, we mainly consider the trajectory parameters, the
residues at $t=0$ and the slope of the Pomeron profile at this point. For
simplicity, we mostly quote only the central values, omitting the errors.
We mention that, although the values reported in the literature for a
certain parameter may be quite different, correlations with different
values of other parameters may lead to a similar overall description.

\subsection{Pomeron}

In the early literature, the contribution of the Pomeron to the total cross
sections was assumed to be energy independent, which in view of
(\ref{sigmapole}) corresponds to a trajectory with intercept
$\alpha_P(0)=1$. In equation (\ref{sigmatot}), the Pomeron is described by
a triple Regge pole with $\alpha_P(0)=1$, leading to a cross section that
contains a constant term and a term which grows with $\ln^2\hspace{-0.1cm}
s$.  The ``soft Pomeron'' \cite{Donnachie-Landshoff-1992} with an intercept
slightly larger than 1 (for instance, $\alpha_P(0)=1.093$ \cite{PDG-2000}
or $\alpha_P(0)=1.08$ \cite{DDLN-book}), also yields a very good
description over a very wide range of energies, while a ``hard'' Pomeron
with $\alpha_P (0) = 1.45 $ was shown \cite{Donnachie-Landshoff-1998} to
explain the data on $ep$ scattering at HERA. In \cite {Cudell2006} both of
these variants of the Pomeron are used for the description of $pp$, $\pi p$
and $K p$ elastic scattering at small $t$.

Information on the $t$-dependence of the trajectory is obtained from the
slope of the diffraction peak in $pp$ and $\pi p$ collisions at high
energies:
\begin{equation}\label{diff}
B_{ap}(s)=  \partial_t\hspace{-0.03cm} \ln\,[ d
\sigma^{ap}(s,t)/ d t]\rule[-0.1cm]{0cm}{0cm}_{\,t\to 0} \,,
\end{equation}
where $a$ stands for $p, \bar{p}$ or $\pi^\pm$. The quantity (\ref{diff}) is expressed as
\begin{equation} \label{bb'}
B_{ap}(s)= 2 [b_P^{ap}+\alpha_P'(0)\ln (s/s_1)]\,,
\end{equation}
in terms of the slope of the trajectory, $\alpha_P'(0)$, and the slope
$b_P^{ap}$ of the Pomeron profile at $t=0$, defined in the case of $\pi\pi$
scattering in equation (\ref{eq:parabola}).

The slope of the diffraction peak in proton-proton scattering was measured
by many experiments at various energies (for references on the data sets
see \cite{Cudell2006,Oko}). Recently, the slope $B_{pp}$ was measured at
$\sqrt{s}= 7\, {\rm TeV}$ by the TOTEM Collaboration at LHC \cite{Totem}.
The trajectory slope $\alpha_P'(0)=0.25$ was shown to be consistent with
high energy data, in particular in the context of the soft Pomeron
\cite{Donnachie-Landshoff-1984}. A recent comprehensive analysis
\cite{Oko}, based on the available experimental data on $pp$ and $\bar{p}p$
scattering over a large energy scale, reports the average value
$\alpha_P'(0)= 0.24$.

Recent choices of the Pomeron trajectory are: 

\vspace{0.1cm} \hspace{-0.5cm}$\alpha_P(t)$= $1.0 + 0.20\, t$
\cite{GarciaMartin:2011cn}, $1.09 + 0.33\, t$ \cite{Cudell2006}, $1.08 +
0.25\,t$ \cite{Sibirtsev}, to be compared with the input used in the
present paper, $\alpha_P(t)=1+(0.25\pm0.05)\,t$.
 
The Pomeron residue $\beta_P(t)\equiv \beta_P^{\pi\pi}(t)$ is known with
less precision.  For the value at $t=0$, we quote: $\beta_P(0)=46\,
\mbox{\cite{ACGL}},\; 118\, \mbox{\cite{PY2003}},\; 99\,
\mbox{\cite{GarciaMartin:2011cn}},\; 66\, \mbox{\cite{Szczurek}},\; 98 \,
\mbox{\cite{Halzen}}$, to be compared with the input we are using,
$\beta_P(0)=94\pm1$.  Fig. 4a. illustrates the fact that the value used in
\cite{ACGL} is on the low side, while the one in \cite{Halzen} is high. The
value derived in \cite{Szczurek} (which is based on factorization and the
older parametrization of $NN$ and $\pi N$ data given in
\cite{Donnachie-Landshoff-1992}) is also low, but this is partly
compensated by a higher value for $\beta_f(0)$.

As concerns the $t$-dependence  near $t=0$, an exponential form
\begin{equation}\label{expo}
\beta_P^{ab}(t)=\beta_P^{ab}(0) e^{b_P^{ab} t},
\end{equation}
was often assumed for various types of elastic reactions $ab\to ab$ (early
references are \cite{Serber,Rarita,ChRi}, for more recent work see for
instance \cite{PY2003,PY2004-Regge,KPY2006,KPY2008,GarciaMartin:2011cn},
\cite{Huang,Sibirtsev}). For the slope in $\pi\pi$ scattering, $b_P\equiv
b_P^{\pi\pi}$, factorization implies
\begin{equation}\label{factoriz}
b_P= 2 b_P^{\pi p}- b_P^{p p}.
\end{equation}
Since the $\pi p$ data at $t\neq0$ are less accurate, the extraction of
$b_P$ from this relation is not very precise. The values proposed in the
literature scatter wildly: $b_P= 4\,\mbox{\cite{Pennington-Protopopescu}},
$ $3.3 \,\mbox{\cite{GarciaMartin:2011cn}},\;0.42\, \mbox{\cite{Rarita}},\;
0.79 \, \mbox{\cite{Cudell2006}}$, to be compared with the result of our
calculation, $b_P=2.5\pm 0.4$.

\subsection{\boldmath $f$  pole}
The parametrization of the $f$ Regge pole (denoted also as $P'$ in the
literature
\cite{Rarita,PY2003,PY2004-Regge,KPY2006,KPY2008,GarciaMartin:2011cn}) is
linked with the one of the Pomeron. Indeed, both contribute to the
amplitude with $I_t=0$ and are largely indiscernible. The relative weight
of the contributions from Pomeron and $f$ depends on the parametrization
chosen for the Pomeron. For one of the fits obtained by Rarita et al.
\cite{Rarita}, for instance, the Pomeron dominates over the $f$ already
below 1.8 GeV, while for the others this is the case only at much higher
energies.

Recent parametrizations of the $f$-trajectory are:   

\vspace{0.1cm}
\noindent $\alpha_f(t)=0.53 + 0.90\,t$ \cite{GarciaMartin:2011cn}, $0.69 +
0.80\,t$ \cite{Desgrolard},\, $0.96 + 0.58\,t +$\\ $0.03\,t^2$
\cite{Desgrolard}, $ 0.61 + 0.82\,t$ \cite{Cudell2006},\, $ 0.71 + 0.83\,t$
\cite{Sibirtsev}, while we are using
$\alpha_f(t)=(0.54\pm0.05)+(0.90\pm0.05)\,t$ as an input.

\vspace{0.1cm} The values of $\beta_f(0)$ proposed in the earlier
literature \cite{Pennington-Annals,Rarita} cover a large range. Some values
adopted recently are (compare section \ref{sec:discussion sigmatot}):
$\beta_f(0)= 108 $ \cite{ACGL}, 32 \cite{GarciaMartin:2011cn}, 103
\cite{Szczurek}, 153 \cite{Cudell2006}, 123 \cite{Halzen}, to be compared
with the input we are relying on, $\beta_f(0)=69\pm2$.
 
\subsection{\boldmath $\rho$  pole}

The $\rho$ trajectory is relatively well known. Some  choices made in the literature are:      

\vspace{0.1cm}\noindent $\alpha_\rho(t)= 0.53 +0.90 \,t - 0.15
\,t^2\,\mbox{\cite{GarciaMartin:2011cn}}, \; 0.44 + 0.96\,t -0.02 \,t^2\,
\mbox{\cite{Desgrolard}},\\0.48 + 0.88\,t \, \mbox{\cite{Desgrolard}},
\;0.53 + 0.8\,t\, \mbox{\cite{Huang,Sibirtsev}},\; 0.47 + 0.91\,t\,
\mbox{\cite{Cudell2006}}$, while our estimate reads
$\alpha_\rho(t)=(0.45\pm 0.02)+(0.91\pm0.02)\,t$.

\vspace{0.1cm} For the residue, the values $\beta_\rho(0) \approx 104.2$
given by Kane and Seidl \cite{Kane-Seidl}, $\beta_\rho(0) \approx 107.6$ of
Irving and Worden \cite{Irving-Worden} and $\beta_\rho(0) \approx 116$
predicted by the Lovelace-Shapiro-Veneziano model \cite{LSV} are on the
high side. Pennington \cite{Pennington-Annals} obtains $\beta_\rho(0)
\approx 72$, a value which was adopted also in \cite{ACGL}.  These numbers
are to be compared with the outcome of our calculation, which reads
$\beta_\rho(0)= 84^{+13}_{-16}$.

A smaller value, $\beta_\rho(0)=33.2$ was proposed in \cite{PY2003},
invoking consistency with data on $\pi\pi$ scattering at high energies. In
the subsequent papers \cite{PY2004-Regge,KPY2006,KPY2008,GarciaMartin:2011cn}, the
authors resorted to dispersive analyses of the pion-pion amplitude and
gradually increased their estimate: $\beta_\rho(0)=37.1$
\cite{PY2004-Regge}, $48.2$ \cite{KPY2008}, $58.4$
\cite{GarciaMartin:2011cn}--CFD solution.  The difference between the
values obtained in \cite{PY2004-Regge}-\cite{KPY2008} and our result is due
partly to a different shape of the isoscalar $S$ wave below 1 GeV (the
phase shift $\delta_0^0(s)$ given in these works displays an unphysical
``hump'' around 0.8 GeV, see the discussion in
\cite{Caprini2008,LeutwylerLisbon}). We mention that in the most recent
result of this group, quoted as CFD solution in \cite{GarciaMartin:2011cn},
this hump is no longer present.

Other recent values are $\beta_\rho(0) = 84$ (obtained in \cite{Szczurek}
with factorization, using the older parametrization of cross sections in
\cite{Donnachie-Landshoff-1992}), which is consistent with our result, and
the very small value $\beta_\rho(0)$ $ = 9.8$ found in \cite{Halzen}, which
is discussed in section \ref{sec:discussion sigmatot}.

The results given in the literature for the residue $\beta_\rho(t)$ at
$t\neq0$ are rather controversial.  The values obtained from dispersion sum
rules for the $\pi\pi$ amplitude in the seventies
\cite{Tryon,Basdevant-Schomblond,Pennington-Annals} were not conclusive,
due to the poor 
quality of the data at low energies.  The phenomenological information
provided by factorization was also rather poor: since the contribution of
the $\rho$-pole to the forward $NN$ amplitudes is small, it was neglected
altogether in some of the early fits \cite{Rarita}.  The studies of the
$\rho$-dominated charge exchange process $\pi^- p\to \pi^0\,n$ indicated
that indeed the spin-flip $\rho NN$ coupling is larger than the non-flip
coupling, which is the only one contributing to the forward $NN$ scattering
amplitude. This feature is confirmed by the recent analysis of the
pion-nucleon charge exchange amplitudes \cite{Huang}.

\subsection{Regge cuts}
The Regge cuts are introduced in phenomenological analyses especially for
extending the validity of the Regge model to larger values of $|t|$ (see
for instance the recent works \cite{Szczurek,Huang}). As we are interested
in a rather small range of $t$, we consider a Regge cut only for the
description of the exotic amplitude with $I_t=2$, where it is the dominant
contribution.

The information about this term is poor, since it does not contribute above
5 GeV, where the Regge fits to the $NN$ and $\pi N$ amplitudes are usually
done \cite{PDG-Regge,Cudell2006,Igi,Halzen}.  Alternative treatments based
on sum rules for $\pi\pi$ scattering in the seventies were not conclusive.
Pennington \cite{Pennington-Annals} notices a great sensitivity of the
finite-energy sum rule for the amplitude with $I_t=2$ with respect to the
transition point from low to high energy regimes. He assumes
$\beta_e(0)\approx 0$, this value being adopted also in \cite{ACGL}.  A
small effective residue $\beta_e(t)$ appearing in equation (\ref{Regge
  ImT2t}) can occur from cancellations of the $\rho-\rho$ and $a_2-a_2$
cuts, as argued by Worden \cite{Worden}, using symmetry arguments in the
Reggeon-calculus of Gribov \cite{Reggeon-calculus}. In
\cite{GarciaMartin:2011cn} the authors obtain $\beta_e(0)= 8$ in the UFD
fit, and $\beta_e(0)= 3$ in the CFD fit. This is consistent with our
result, which comes with a large error: $\beta_e(0)=-24\pm29$.

 The $t$-dependence of $\beta_e(t)$ is also poorly known. In
 \cite{GarciaMartin:2011cn}, the residue is taken to be proportional $ \exp
 b_e t$, with $b_e=2.4$. The ansatz prevents $\beta_e(t)$ from passing
 through zero and makes it shrink if the momentum transfer grows (compare
 Fig.\,\ref{fig:betae}).  The numerical value for the slope is
 $\beta_e'(0)= 20$ and $\beta_e'(0)= 8$ for the UFD and CFD fits,
 respectively. Our framework does not support this ansatz: the sum rules
 require a zero and a comparatively large, negative slope,
 $\beta_e'(0)=-120(40)$.

\section{Roy equations}\label{app:Roy}
\subsection{\boldmath Fixed-$t$ dispersion relation}\label{sec:fixed t}\label{app:dispersion relation}
The derivation of the Roy equations (see \cite{Roy} and appendix A of
\cite{ACGL}) is based on the fixed-$t$ dispersion relation obeyed by the
$s$-channel isospin amplitudes $\vec{T}=(T^0,T^1,T^2)$ in the interval
$-28M_\pi^2 < t < 4M_\pi^2$, written as \cite{Roy} \bea
\label{fixedt} 
\vec{T }(s,t)\al=\al(4M_\pi^2)^{-1}\,(s\,{\bf 1} + t\,
C_{st} + u\, C_{su})\,\vec{T}(4M_\pi^2,0) \no
\al+\al \int_{4M_\pi^2}^\infty
\!ds'\,g_2(s,t,s')\,\mbox{Im}\,\vec{T}(s',0)
\\\al+\al\int_{4M_\pi^2}^\infty \!ds'\,g_3(s,t,s')\,\mbox{Im}\,\vec{T}(s',t)
\,. \nonumber
\eea
The subtraction term is fixed by the $S$-wave scattering lengths: 
\bea
\vec{T}(4M_\pi^2,0)=32\,\pi\,(a_0^0,0,a_0^2)\,,
\eea  the
crossing matrices $C_{tu}=C_{ut},\,C_{su}=C_{us},\,C_{st}=C_{ts}$
are given by \bea\label{crossing} && C_{tu}=
\left(\!\begin{tabular}{rrr}
1&0&\rule{0.15cm}{0cm}0\\\rule{0em}{1em}0&--1&0\\\rule{0em}{1em}0&0&1\end{tabular}\!
\right)\hspace{0.8cm}
C_{su}=
\left(\hspace{-0.2cm}\begin{tabular}{rrr}
\mbox{$\frac{1}{3}$}&--1&\mbox{$\frac{5}{3}$}\\\rule{0em}{1em}
--\mbox{$\frac{1}{3}$}&\mbox{$\frac{1}{2}$}&\mbox{$\frac{5}{6}$}\\
\rule{0em}{1em}
\mbox{$\frac{1}{3}$}&\mbox{$\frac{1}{2}$}&\mbox{$\frac{1}{6}$}\end{tabular}
\!\right) \no&&  C_{st}=
\left(\hspace{-0.2cm}\begin{tabular}{rrr}
\mbox{$\frac{1}{3}$}&1&\hspace{-0.1cm}\mbox{$\frac{5}{3}$}\\\rule{0em}{1em}
\mbox{$\frac{1}{3}$}&\mbox{$\frac{1}{2}$}&\hspace{-0.1cm}--\mbox{$\frac{5}{6}$}\\
\rule{0em}{1em}
\mbox{$\frac{1}{3}$}&--\mbox{$\frac{1}{2}$}&\hspace{-0.1cm}\mbox{$\frac{1}{6}$}\end{tabular}
\!\right)\nonumber
\eea
and the kernels $g_2(s,t,s')$, $g_3(s,t,s')$  represent $3\times3$ matrices built with $C_{st}$, $C_{tu}$ and $C_{su}$,
\bea g_2(s,t,s')&=&-\frac{t}{\pi\, s'\,(s'-4M_\pi^2)}\,
(u\, C_{st} + s\, C_{st}\, C_{tu})\times \no
&&\left(\frac{{\bf 1}}{s'-t}
+ \frac{C_{su}}{s'-u_0}\right)\co\\
g_3(s,t,s')&=&-\frac{s\,u}{\pi\,s'(s'-u_0)}\left(\frac{{\bf 1}}{s'-s}
+ \frac{C_{su}}{s'-u}\right)\co\nonumber\eea
where $u=4M_\pi^2-s-t$ and $u_0=4M_\pi^2-t$.

We add a comment concerning the subtractions. As emphasized in
\cite{GarciaMartin:2011cn}, a single subtraction suffices: the Olsson sum
rule (\ref{Olsson}) can be used to express one of the two subtraction
constants in terms of the other. The resulting set of relations is referred
to as the GKPY-equations. Conversely, the dispersive representation for
$\vec{T}(s,t)$ obtained in that framework \cite{GarciaMartin:2011cn}
differs from the one in (\ref{fixedt}) by a multiple of the Olsson sum
rule. The advantage of using a single subtraction is that this allows a
better determination of $a_0^0$ and $a_0^2$ from experiment (see the
discussion at the end of section \ref{sec:partial waves}). The price to pay
is that the dispersive representation for the partial waves then becomes
more sensitive to the high energy part of the input used, because the
GKPY-equations converge less rapidly than the Roy equations. As we are
relying on the theoretical predictions for $a_0^0$ and $a_0^2$, retreating
to a single subtraction is no gain for us.
\subsection{Dispersive representation of the partial waves}\label{dispersive representation}
Exploiting crossing symmetry, the partial wave projection of the amplitude
in (\ref{eq:partial waves}) can be written in a form that involves a
smaller range of $t$-values: 
\be\label{pwp}
t_\ell^I(s)=\frac{1}{32\pi}\int_{0}^1 dz P_\ell(z)\, T^I(s,t_z)\,.
\ee 
This step extends the range of validity of Roy's representation \cite{Roy},
\be\label{Roy} t_\ell^I(s)= k_\ell^I(s)+ \sum_{I'=0}^2\sum_{\ell'=0}^\infty
\int_{4M_\pi^2}^\infty ds'\,K_{\ell\ell'}^{I I'}(s,s')\,\mbox{Im} \,
t_{\ell'}^{I'}(s')\,.  \ee The term $k^I_\ell(s)$ denotes the partial wave
projection of the subtraction term, which shows up only in the $S$- and
$P$-waves, \bea \label{subconst1}\hspace{-2cm} k^I_\ell(s)\al=\al a_0^I
\,\delta_\ell^0 +\frac{s-4M_\pi^2}{4M_\pi^2}\,
(2a_0^0-5a_0^2)\times\\
&&\hspace{1cm}\left(\frac{1}{3}\,\delta_0^I\,\delta_\ell^0
  +\frac{1}{18}\,\delta_1^I\,
  \delta_\ell^1-\frac{1}{6}\,\delta_2^I\,\delta^0_\ell\right).
\nonumber
\eea 
The kernels $K_{\ell\ell'}^{I I'}(s,s')$ are explicitly known
functions (see appendix A of \cite{ACGL}). They contain a diagonal,
singular Cau\-chy kernel that generates the right hand cut in the partial
wave amplitudes, as well as a logarithmically singular piece that accounts
for the left hand cut.  The validity of these equations has rigorously been
established on the interval $-\,4 M_\pi^2$ $<s< 68
M_\pi^2=(1.15\,\mbox{GeV})^2$.

In \cite{ACGL}, Roy's equations were solved for the $S$- and $P$-waves in
the region $\sqrt{s}\le 0.8~ \mbox{GeV}$. In the present paper we solve
these equations in the whole range $\sqrt{s}\le 1.15~ \mbox{GeV}$,
including also the higher partial waves, which become increasingly
important as the energy grows, because they are given more weight in the
sum and the angular momentum barrier becomes less effective: for instance,
above 1.6 GeV, the $S$-wave accounts for less than a quarter of the total
cross section with $I_s=0$. In order to have a reliable description, we
solved the Roy equations for the waves with $\ell\le 4$ below 1.15 GeV and
made use of the fact that, below 2 GeV, the masses, widths and $\pi\pi$
branching fractions of the resonances are now known quite well
\cite{PDG-2011}.  A detailed description of the solution of the Roy
equations will be given elsewhere \cite{paper-on-PW} -- in the following,
we limit ourselves to a cursory discussion.
\subsection{Input used when solving the Roy equations}\label{app:input}
The input required for the construction of the solutions of the Roy
equations was outlined in section \ref{sec:partial waves}. There, we also
reviewed the status of our knowledge of the subtraction constants. In the
present appendix, we briefly discuss the other elements of the input used
for the partial waves.
\subsubsection{Phases}\label{sec:phase input}
Our analysis relies on the following phenomenological estimates for values
of the S- and P-wave phase shifts in the low energy region
($\sqrt{s_A}=0.8\,\mbox{GeV}$, $\sqrt{s_{max}}=1.15\,\mbox{GeV}$):
\bea\label{phase input}
\delta^0_0(s_A)\al =\al 82.3^{\circ
  +10^\circ}_{\hspace{0.4em}-4^\circ}\,,\\\delta^0_0(4M_K^2)\al=\al
185^\circ \pm 10^\circ\,,\no \delta^0_0(s_{max})\al =\al 260^\circ
\pm10^\circ\,,\no \delta_1^1(s_A)\al =\al 108.9^\circ\pm2^\circ\,,\no
\delta_1^1(s_{max}) \al =\al 166.5^\circ\pm2^\circ\,.\nonumber
\eea 
We briefly comment on these estimates.
 
The phase shift analyses available at the time were used in \cite{ACGL} to
estimate the S$^0$ phase shift at 0.8 GeV, with the result
$\delta_0^0(s_A)=82.3^\circ\pm 3.4^\circ$. This estimate was criticized in
\cite{PY2003}, however, and significantly higher values were suggested. We
did reply to this objection \cite{CCGL}, but when analyzing the
consequences of the representation for the S$^0$-wave for the mass and
width of the $\sigma$ \cite{CCL}, we decided to inflate the uncertainty
range by a factor of 2, in order to make it evident that the outcome is not
tied to an input which at the time appeared to be controversial.

In the meantime, the dust settled -- the discrepancy disappeared: the
result $\delta^0_0(s_A)=85.7^\circ\pm1.6^\circ$ obtained in the most recent
paper of the Madrid-Krakow collaboration (Table XII in
\cite{GarciaMartin:2011cn}) differs from the old estimate in \cite{ACGL} by
less than one standard deviation. Moreover, the thorough analysis of the
experimental situation concerning the S$^0$ phase shift in
\cite{Achasov:2010fh} fully confirms the results of \cite{ACGL,CGL}, also
with regard to the value at 0.8 GeV. Last, not least, we point to a recent
paper by Moussallam \cite{Moussallam:2011zg}, who has analyzed the problem,
solving the Roy equations with matching point at $2M_K$ and relying on the
experimental results for the scattering lengths obtained by NA48/2
\cite{NA48Kl4} to fix the subtraction constants. He shows that, currently,
the uncertainties in the value of $\delta_0^0(s_A)$ are dominated by those
in the input used for the elasticity above $K\Kbar$ threshold: while the
shallow dip in $\eta_0^0(s)$ obtained from the data on the inelastic
channels leads to $\delta_0^0(s_A)=82.9^\circ\pm 1.7^\circ$, the deep dip
indicated by the data on elastic scattering yields
$\delta_0^0(s_A)=80.9^\circ \pm 1.4^\circ$. Both of these results are
within the range estimated in \cite{ACGL}.  Hence we might just as well
return to that estimate, but, for the time being, we prefer to stick to the
broad uncertainty range adopted in \cite{CCL} -- we plan to examine the
present experimental situation concerning the low partial waves in a
forthcoming article \cite{paper-on-PW}.

Phase shift analysis indicates that $\delta_0^0(s)$ passes through
180$^\circ$ in the vicinity of $K\Kbar$ threshold, but it is still not
known for certain whether this happens below or above that point, although
the second option appears to be more likely. The error bar attached to our
estimate allows for both possibilities. The values for $\delta_0^0(4M_K^2)$
given in \cite{Moussallam:2011zg} are consistent with our estimate in
(\ref{phase input}), albeit somewhat higher.

The P$^1$-wave is experimentally very well determined through the data on
the e.m.\,form factor of the pion, so that our estimates for the phase
shifts for $\delta_1^1(s_A)$ and $\delta_1^1(s_{max})$ in (\ref{phase
  input}) come with small errors.

The list (\ref{phase input}) does not contain entries for the exotic phase
shift $\delta_0^2(s)$. The reason is the following. As discussed in
\cite{ACGL}, the solutions of the Roy equations in general develop a cusp
at the matching point, which in the present framework is taken at
$\sqrt{s_{max}}=1.15$ GeV. We remove the unphysical phenomenon by requiring
the phase shifts as well as their derivatives to be continuous at the
matching point. This in effect imposes a condition on the behaviour of the
imaginary parts above this point: we find acceptable solutions only if we
do not prescribe the value of $\delta_0^2(s_{max})$, but let it float (for
a more thorough discussion, see \cite{ACGL}).  The same holds for the
higher partial waves: the requirement that the phase shift and its
derivative are continuous at the matching point admits solutions only if
the value of the phase at this point is not prescribed.

For the numerical evaluation of the Roy equations, we need an explicit {\it
  parametrization} of the phases. Since the solution is uniquely
determined, the choice of the parame\-tri\-za\-tion is a matter of
convenience -- the only requirement to be met is that it is flexible enough
to come close to the exact solution. In the case of the S$^0$-wave, the
fact that there is a resonance in the immediate vicinity of $K\Kbar$
threshold entails a rather intricate behaviour there and special care is
needed to find solutions of good quality, but for the other waves, we did
not encounter such problems. For details, we refer to \cite{paper-on-PW}.
\subsubsection{Elasticities}\label{sec:elasticity}
The input used for the elasticity of the partial waves in the region where
the Roy equations are solved plays a significant role, not only in
resolving the structure of the S$^0$ wave in the vicinity of $K\Kbar$
threshold, for instance, but also when analyzing the behaviour of the
partial waves off the real axis, in order to determine the position of the
resonance poles, such as the one of the $\sigma$. The elasticities ensure
that the effects on the elastic $\pi\pi$ scattering amplitude generated by
inelastic processes like $\pi\pi \leftrightarrow K\Kbar$, for instance, are
properly accounted for.

The interference between the $K\Kbar$ cut and the pole associated with the
$f_0(980)$ generates a dip in the elasticity of S$^0$, immediately above
$\sqrt{s}=2M_K$.  In the region between $2M_K$ and 1.15 GeV, where it
matters for our ana\-lysis, the uncertainties in the elasticity of S$^0$
are gradually becoming smaller. For a recent, very thorough analysis of
this field, we refer to \cite{Moussallam:2011zg}. We describe the behaviour
by means of the Flatt\'e formula \cite{Flatte} and vary the parameters
contained therein in a rather broad range, so that the uncertainty band for
$\eta_0^0(s)$ generously covers the available parametrizations. The
inelasticities of P$^1$ and S$^2$ are also estimated on the basis of
available phase shift analyses. For the higher partial waves with $I_s=0$
or 1, we use a Breit-Wigner parametrization and calculate the corresponding
inelasticities from mass, width and branching fraction for the decay into
$\pi\pi$ quoted in the data tables. The exotic waves D$^2$ and G$^2$ are
driven by the other partial waves. Since their inelasticities are very
small, these barely affect our results. For a more thorough discussion, we
refer to \cite{paper-on-PW}.
\subsubsection{Imaginary parts}\label{sec:imaginary part}
Although phenomenology does reduce the range allowed by unitarity, $0\leq
\mbox{Im}\,t^I_\ell(s)\leq 1/\rho(s)$, the uncertainties in the behaviour
of the partial wave imaginary parts above the matching point are
considerable.

We use a crude parametrization involving Breit-Wigner terms for the
resonances and polynomials for the background, impose continuity of
function and derivative at the matching point and allow for one free
parameter in the representation used, for each one of the partial waves
with $\ell\leq 4$. It is convenient to identify the free parameter with the
value of the corresponding imaginary part at $\sqrt{s_b}=2\,\mbox{GeV}$.
Our estimates for the latter read: \bea\label{input
  imt}\mbox{Im}\,t^0_0(s_b)\al =\al 0.4\pm0.25\,,\hspace{0.5cm}
\mbox{Im}\,t^1_1(s_b) = 0.26\pm0.13\,,\rule{0.5cm}{0cm}\\
\mbox{Im}\,t^2_0(s_b)\al =\al 0.3\pm0.15\,,\hspace{0.5cm}
\mbox{Im}\,t^0_2(s_b) = 0.2\pm0.075\,,\no \mbox{Im}\,t^1_3(s_b)\al =\al
0.23\pm0.07\,,\hspace{0.32cm} \mbox{Im}\,t^2_2(s_b) = 0.12\pm0.06\,,\no
\mbox{Im}\,t^0_4(s_b)\al = \al 0.22\pm0.03\,,\hspace{0.31cm}
\mbox{Im}\,t^2_4(s_b) = 0.01\pm0.01\,.\nonumber\eea We plan to describe the
parametrization used in more detail \cite{paper-on-PW}.
\section{Sum rules from crossing symmetry}\label{app:SR}
A set of sum rules for fixed $t$ is obtained by taking the first derivative
with respect to $s$ of the l.h.s.  of (\ref{st}) and then setting $s=0$.
Since the resulting expression has a zero both at $t=0$ and at
$t=4M_\pi^2$, it is convenient to divide by the factor $t(t-4M_\pi^2)$. The
sum rules then take the form \be\label{stSR} C_k(t)=0\,,\quad k=0,1,2\,,\ee
where $C_0(t)$, $C_1(t)$, $C_2(t)$ are integrals over the imaginary part of
the scattering amplitude: \bea\label{eq:Ck} C_k(t)\al\equiv\al
\int_{4M_\pi^2}^\infty ds\,
n_k(s,t)\,\mbox{Im}\,\Tbar^{\,(k)}(s,t)\\
&+&\int_{4M_\pi^2}^\infty ds\,f_k(s,t)\,\mbox{Im}\,T^1(s,0)\no &+&
\sum_{\ell=0}^2 \int_{4M_\pi^2}^\infty ds\,h_{k\ell}(s,t)\
\partial_{t'}\mbox{Im}\,T^{\ell}(s,t')\,\rule[-0.15cm]{0.015cm}{0.45cm}_{\;t'
  \rightarrow 0}\,. \nonumber\eea The functions
$\mbox{Im}\,\Tbar^{\,(0)}(s,t)$, $\mbox{Im}\,\Tbar^{\,(1)}(s,t)$,
$\mbox{Im}\,\Tbar^{\,(2)}(s,t)$ are defined in (\ref{ImTbar}). The first
line shows that the sum rules decouple, in the sense that the $t$-dependent
part of the scattering amplitude enters with definite isospin in the
$t$-channel. The explicit expressions for the corresponding kernels read
\bea &&n_0(s,t)= \frac{2s+t-4M_\pi^2}{\pi\,s^2(s+t-4M_\pi^2)^2}\,,\\
&&n_1(s,t)=\frac{t-4M_\pi^2}{\pi\,s^2(s+t-4M_\pi^2)^2}\,,\no &&
n_2(s,t)=\frac{2s+t-4M_\pi^2}{\pi\,s^2(s+t-4M_\pi^2)^2}\,.\nonumber\eea The
diagonal elements of $h_{k\ell}(s,t)$ are of the form
\bea h_{00}(s,t)\al=\al \frac{-4s-2t+12M_\pi^2}{3\pi s(s-4M_\pi^2)(s-t)(s+t-4M_\pi^2)}\,,\\
h_{11}(s,t)\al=\al \frac{-3s-t+8M_\pi^2}{2\pi
  s(s-4M_\pi^2)(s-t)(s+t-4M_\pi^2)}\,,\no h_{22}(s,t)\al=\al
\frac{-7s-5t+24M_\pi^2}{6\pi
  s(s-4M_\pi^2)(s-t)(s+t-4M_\pi^2)}\,.\nonumber\eea The off-diagonal ones
are given by
\bea &&h_{k\ell}(s,t) =\frac{c_{k\ell}}{6\pi s (s-4M_\pi^2)(s+t-4M_\pi^2)}\,,\\
&&c_{01}=6,\hspace{1.3cm}c_{02}=-10,\no&&c_{10}=2,\hspace{1.3cm}c_{12}=-5,\no
&&c_{20}=-2,\hspace{1cm}c_{21}=-3.\nonumber\eea The kernels in the second
line of (\ref{eq:Ck}) can be expressed in terms of those listed above:
\bea\label{eq:f}&& f_k(s,t)=-\frac{c_k\, n_k(s,t)+2\,h_{k1}(s,t)}{s-4M_\pi^2}\,,\\
&& c_0=2\,,\hspace{0.5cm}c_1=1\,,\hspace{0.5cm}c_2=-1\,.\nonumber\eea

As in the case of $S(t)$, the integrands in (\ref{eq:Ck}) display a
fictitious double pole for $s = 4M_\pi^2- t$, but the residue of the
singularity vanishes for a crossing symmetric integrand.  At $t=0$, the
three sum rules (\ref{stSR}) reduce to a single one, in fact one that had
been made use of already earlier.\footnote{See appendix B.2 in \cite{ACGL},
  equation (B.7).}

Since the contribution from the S- and P-waves is manifestly crossing
symmetric, these waves do not show up in the above sum rules. For the
S-waves, which generate a $t$-independent contribution, that is manifest:
these waves drop out, in $\mbox{Im}\,\Tbar(s,t)$, in $\mbox{Im}\,T^1(s,0)$,
as well as in the terms involving the partial derivative with respect to
$t$. The P$^1$-wave, on the other hand does generate a contribution to
these quantities -- it drops out only in the sum, on account of the
relation (\ref{eq:f}) between the kernels.

\section{Details of minimization procedure}\label{sec:details}
In the present appendix, we specify the discrepancy function that enters
the procedure used to solve the sum rules and duality conditions, as
described in section \ref{sec:analysis}. For a given input, these amount to
constraints imposed on the Regge output variables, \be
v^{\indR}_{out}=\{p_1, b_P,c_P,\beta_\rho(0),
r_1,b_\rho,c_\rho,\beta_e(0),b_e,c_e\}.\ee We evaluate the $t$-dependent
conditions on a finite number of points, using the lattice \be t_n=
n\,t_1\,,\quad t_1=-0.1\,\mbox{GeV}^2\,,\quad n=0,\ldots\,, 6\,.\ee The sum
rules and duality conditions then amount to a finite number of relations
relations of the form \bdm
S_r^{\indPW}+S_r^{\indR}(v^{\indR}_{out})=0\,,\quad r=1,2,\ldots\edm The
part $S_r^{\indPW}$ represents the contribution from the region where we
use the partial wave representation, while the remainder is calculated with
the Regge parametrization.

The duality conditions for the total cross sections are solved as algebraic
constraints for the variables $p_1,r_1,\beta_e(0)$. The remaining variables
in $v^R_{out}$ are determined by minimizing the discrepancy function
\be\label{eq:chi} \chi^2(v^R_{out})=\sum_r
\lambda_r\,[S_r^{\indPW}+S_r(v^{\indR}_{out})]^2\,,\ee where the sum
extends over all constraints except the duality conditions at $t=0$. The
factors $\lambda_r$ specify the weights of the remaining conditions in the
minimization process. We determine these with the response of the term
$S_r^{\indPW}\!$ to variations of the partial wave input: vary the input
parameters within the estimated uncertainty range, calculate the responses
generated by these variations in $S_r^{\indPW}\!$, evaluate the sum
$\Delta_r$ over the squares of these responses and set $\lambda_r=
1/\Delta_r$.  The choice of the weights is not of crucial importance,
because the various sum rules and duality conditions are consistent with
one another. The particular choice made homogenizes the set of relations,
giving little weight to those constraints that are sensitive to the
uncertainties in our partial wave representation and thus contain little
information about the Regge output parameters, which we determine by
minimizing $\chi^2(v^R_{out})$.

\end{document}